\documentclass[11pt,letterpaper]{article}
\usepackage[T1]{fontenc}
\usepackage[utf8]{inputenc}
\usepackage{lmodern}
\usepackage[margin=1in]{geometry}
\usepackage{amsmath,amssymb,amsthm,mathtools}
\usepackage{microtype}
\usepackage{booktabs,array,graphicx}
\usepackage{pgfplots}
\pgfplotsset{compat=1.18}
\usepackage[font=small,labelfont=bf,labelsep=period,skip=6pt]{caption}
\usepackage{enumitem}
\usepackage{xcolor}
\usepackage[colorlinks=true,linkcolor=blue!50!black,citecolor=blue!50!black,urlcolor=blue!50!black]{hyperref}
\usepackage{bookmark}
\pdftrailerid{}
\hypersetup{pdfcreator={},pdfproducer={},pdftitle={},pdfauthor={},pdfsubject={},pdfkeywords={},bookmarksdepth=2}
\numberwithin{equation}{section}
\newtheorem{theorem}{Theorem}[section]
\newtheorem{lemma}[theorem]{Lemma}
\newtheorem{proposition}[theorem]{Proposition}
\newtheorem{corollary}[theorem]{Corollary}
\theoremstyle{definition}

\theoremstyle{remark}

\DeclareMathOperator{\TV}{TV}
\DeclareMathOperator{\TC}{TC}
\DeclareMathOperator{\Unif}{Unif}
\DeclareMathOperator{\supp}{supp}
\DeclareMathOperator{\Var}{Var}
\DeclareMathOperator{\conv}{conv}
\DeclareMathOperator{\cav}{cav}
\DeclareMathOperator{\per}{per}
\newcommand{\E}{\mathbb E}
\newcommand{\one}{\mathbf 1}
\setlist{topsep=3pt,itemsep=2pt,parsep=0pt}
\allowdisplaybreaks[1]
\pdfmapline{+LibertinusSerif-Regular-tlf-t1--base LibertinusSerif-Regular "AutoEnc_75t36gtznqppeufjbancttaqqf ReEncodeFont" <[lbn1_75t36g.enc <LibertinusSerif-Regular.pfb}
\begin{document}
\hypersetup{pageanchor=false}
\begin{titlepage}
\centering
\vspace*{-0.15in}
{\fontencoding{T1}\fontfamily{LibertinusSerif-TLF}\fontseries{m}\fontshape{n}\fontsize{21.5}{26}\selectfont Shared Randomness and Parallel Rounds\\[0.15em]in Permutation Sampling\par}
\vspace{0.18in}
{\large Haoxiang Huang\footnotemark[1]\par}
\footnotetext[1]{University of Science and Technology of China.}
\vspace{0.3in}
\begin{abstract}
Sampling a uniformly random permutation of $[n]$ in parallel is a
coordination problem: $n$ processors, one per coordinate, must output
distinct values. A single round requires a shared seed of $\log n!$
nats, while $n$ rounds require no shared randomness. We determine the
tradeoff in between for $R$ rounds, in which the values of each round
are written independently given the shared randomness and the earlier
output, and the output must be a uniform permutation, exactly or
within total-variation error $\varepsilon$. Shared randomness is
measured by its support size or by its entropy, and it is either
fixed in advance or drawn adaptively during the execution.

For exact sampling, all four costs equal one explicit
combinatorial quantity up to an absolute additive constant. A fixed
error $0<\varepsilon<1$ shortens the seed by $O(1)$ nats in one
round, where the least number of seed values, an approximate
nonnegative tensor rank, is $(1-3\varepsilon/2)\,n!$ up to an
additive two for $\varepsilon\le1/2$. The saving is
$\Theta((\log n)^2)$ nats in two rounds, with a sharp constant, and of
lower order for $R=o(n)$; for $R\sim\rho n$ with $0<\rho<1$ it is
linear in $n$, with the optimal rate given by a variational formula.
Here, errors above an explicit threshold need only $o(n)$ nats, and
at every positive error below it, shared randomness drawn adaptively
has strictly smaller entropy than a seed fixed in advance.
A central tool is a change of measure that turns lower bounds with
error into entropy bounds for a single batch of distinct, nearly
independent values. The results for sublinearly and linearly many
rounds extend to sampling without replacement.
\end{abstract}
\vspace{\stretch{1}}
\begin{minipage}{\textwidth}
\setcounter{tocdepth}{1}
\tableofcontents
\end{minipage}
\vspace{\stretch{1}}
\end{titlepage}
\setcounter{tocdepth}{1}
\setcounter{page}{1}
\hypersetup{pageanchor=true}
\setlength{\abovedisplayskip}{10.5pt plus 3pt minus 3pt}
\setlength{\belowdisplayskip}{10.5pt plus 3pt minus 3pt}
\setlength{\abovedisplayshortskip}{4pt plus 3pt}
\setlength{\belowdisplayshortskip}{8pt plus 3pt minus 2pt}
\makeatletter
\renewcommand\section{\@startsection{section}{1}{\z@}%
  {-3.5ex \@plus -.3ex \@minus -.2ex}{2.3ex \@plus .1ex}%
  {\normalfont\Large\bfseries}}
\renewcommand\paragraph{\@startsection{paragraph}{4}{\z@}%
  {3.25ex \@plus .3ex \@minus .2ex}{-1em}%
  {\normalfont\normalsize\bfseries}}
\makeatother
\flushbottom
\clubpenalty=10000
\widowpenalty=10000
\displaywidowpenalty=10000
\postdisplaypenalty=10000
\section{Introduction}
\label{sec:introduction}

Every coordinate of a uniformly random permutation of $[n]$ is
uniform on $[n]$, yet $n$ independent ones are distinct only with
probability $n!/n^n=e^{-n+o(n)}$. Sampling a permutation is therefore
a problem of coordination, with two classical solutions. A sequential
sampler writes the coordinates one at a time, each uniform over the
unused values; it takes $n$ rounds and no shared randomness. A
one-shot sampler writes all coordinates in one round as functions of
a shared seed; since a product distribution supported on permutations
is a point mass, the seed must carry the entropy $\log n!$.

One round is the classical setting. A sampler that draws a shared
seed and then writes the coordinates independently outputs a mixture
of product distributions. The least number of components is the
nonnegative tensor rank of the target, which for two parties governs
sampling without communication \cite{JainEtAl2013}. The least entropy
of the mixing variable is the $n$-variable form of the common entropy
\cite{KumarLiElGamal2014}, a relative of Wyner's common information
\cite{Wyner1975}. Approximate versions underlie lower bounds for
sampling in communication complexity
\cite{AmbainisEtAl2003,GoosWatson2019,KolEtAl2019}. The uniform
permutation is also a standard target in the complexity of sampling
\cite{Viola2020,AlekseevEtAl2026}, where lower bounds
concern local, low-depth, and cell-probe samplers
\cite{Viola2012,FilmusEtAl2023,KaneOstuniWu2024,ByramjiEtAl2026}.
In those models every output has restricted access to an unlimited random
input; here access is unrestricted, and the amount of shared
randomness is the resource.

Rounds are the second resource, long studied in the parallel
generation of random permutations and other distributions
\cite{Hagerup1991,AnariGaoRubinstein2024}. Masked diffusion models
generate sequences in rounds, drawing each round's batch of positions
independently given those already revealed, and recent work bounds
the resulting error
\cite{BenHamuEtAl2025,ChenCongLi2026,LavenantZanella2025,WenEtAl2026,Wu2026}.
Shared randomness reduces this error: two shared bits can outperform
all deterministic schedules \cite{LiJinWei2026}, and randomness
shared within a batch restores dependence within it
\cite{DraxlerEtAl2026}. Sequential and one-shot sampling are thus the
ends of a tradeoff between rounds and shared randomness, which we
determine:
\begin{quote}
\emph{How much shared randomness is needed to sample a uniform
permutation of $[n]$ in $R$ rounds, and does it matter whether it is
fixed in advance or drawn during the execution?}
\end{quote}

In our model (Section~\ref{sec:model}), $n$ processors, one per
coordinate, see a public random string and all earlier output, and
processors writing in the same round use independent private coins.
After at most $R$ rounds the output is a uniform permutation, exactly
or within total-variation distance $\varepsilon$. The conditional
distributions are arbitrary, so our lower bounds hold for every
sampler that writes each round as a product distribution. Shared
randomness is measured in nats, by the logarithm of its support size
(the \emph{support cost}) or by its entropy (the \emph{entropy
cost}), and it is drawn once, before the first round (the
\emph{static} model), or at any time (the \emph{adaptive} model).

\paragraph{Results.}
We answer the question for every number of rounds
(Table~\ref{tab:depth}), and one principle organizes the answer. Of
the entropy $\log n!$ of the output, the private randomness of the
rounds supplies a part that we compute, and shared randomness must
supply the rest. On the part of the output that agrees with the
target, an error $\varepsilon$ leaves room for at most
$\log\frac1{1-\varepsilon}$ nats of total correlation, summed over
the rounds. The entropy that this correlation buys grows with the
number of rounds, from $O(1)$ in one round and order $(\log n)^2$ in
two to order $n$ when $R=\Theta(n)$, where the entropy of the shared
randomness and the time at which it is drawn begin to matter.

\emph{Exact sampling.} Let $F(n,R)$ be the largest number of
permutations that an $R$-round protocol without shared randomness can
output if it never repeats a value. All four costs of exact sampling
equal $\Delta(n,R)=\log n!-\log F(n,R)$ up to an absolute additive
constant, with equality for $R\le2$ and $R\ge n/2$
(Theorem~\ref{thm:exact}). The upper bound reuses one random
partition of the values in all rounds, and a batch of $b$ coordinates
costs about $\log b$ nats per coordinate.

\emph{What an error saves.} With error $\varepsilon$ a protocol can
\emph{abort}: with probability $\varepsilon$ it writes a fixed
non-permutation, and otherwise it runs an exact protocol. Up to
additive constants, aborting multiplies the entropy cost by
$1-\varepsilon$ and leaves the support cost unchanged. We call the
decrease in support cost that an error permits the \emph{saving}. In
one round the support cost is the logarithm of the least number of
product distributions whose mixture is within $\varepsilon$ of the
target, an approximate nonnegative tensor rank; we determine this
number up to an additive two as $(1-3\varepsilon/2)\,n!$ for
$\varepsilon\le1/2$ (Theorem~\ref{thm:oneround}). In two rounds the
saving is of order $(\log n)^2$, with a sharp constant
(Theorem~\ref{thm:fewround}). For every $R$ it is $O_\varepsilon(n)$,
hence of lower order when $R=o(n)$, where exact protocols and
aborting are optimal to leading order (Theorem~\ref{thm:sublinear}).
For $R\sim\rho n$ the saving is of leading order: the support cost is
$B_\rho(1-\varepsilon)\,n+o(n)$, with $B_\rho$ given by a variational
formula over single batches (Theorem~\ref{thm:frontiers}). It
vanishes exactly when $1-\varepsilon\le p_0(\rho)$, an explicit
threshold equal to the accuracy that Wen et al.\ \cite{WenEtAl2026}
obtain without any shared randomness, by drawing batches from
conditional marginals.

\emph{Static versus adaptive shared randomness.} For $R\sim\rho n$,
the optimal static entropy cost per coordinate is the lower convex
envelope of $B_\rho$, and the optimal adaptive one is strictly
smaller if and only if $p_0(\rho)<1-\varepsilon<1$, that is, at every
positive error for which shared randomness is still needed
(Theorem~\ref{thm:shannon-frontiers}). The reason is that a repeated
value cannot be undone: an adaptive protocol draws the shared
randomness for a round only while the values written so far are
distinct. For $\rho\ge2/3$ all curves are explicit
(Figure~\ref{fig:pair-frontier}). Finally, the results for $R=o(n)$
and $R\sim\rho n$ extend to sampling without replacement
(Theorem~\ref{thm:injections}), and the one-round lower bound holds
for the uniform distribution on any set of permutations, such as the
perfect matchings of a bipartite graph.

\begin{table}[t]
\caption{Shared randomness in nats for $R$ rounds, where
$\lambda_2=(\log3)/3$ and $d_\varepsilon=\log\frac1{1-\varepsilon}$.
The exact costs agree up to $O(1)$ for both measures and both models;
the other columns are at a fixed error $\varepsilon$, with
$\varepsilon\le1/2$ in the first row. The last row gives the entropy
cost of the static model, with $\operatorname{conv}$ the lower convex
envelope; the adaptive cost is strictly smaller precisely when
$p_0(\rho)<1-\varepsilon<1$ (Theorem~\ref{thm:shannon-frontiers}).}
\label{tab:depth}
\small
\renewcommand{\arraystretch}{1.3}
\begin{tabular*}{\textwidth}{@{\extracolsep{\fill}}lllll@{}}
\toprule
Rounds $R$ & Exact cost & Saving in support cost & Entropy cost
  & Theorem\\
\midrule
$1$ & $\log n!$ & $\log\tfrac{1}{1-3\varepsilon/2}+o(1)$
  & $(1-\varepsilon)\log n!+O(1)$ & \ref{thm:oneround}\\
$2$ & $\log n!-\lambda_2n+O(1)$
  & $(1+o(1))\tfrac{d_\varepsilon}{2\lambda_2}(\log n)^2$
  & $(1-\varepsilon)\,\Delta(n,2)+O_\varepsilon((\log n)^2)$
  & \ref{thm:fewround}\\
$o(n)$ & $(1+o(1))\,n\log(n/R)$ & $O_\varepsilon(n)$
  & $(1+o(1))(1-\varepsilon)\,n\log(n/R)$ & \ref{thm:sublinear}\\
$\sim\rho n$ & $\Phi(\rho)\,n+o(n)$
  & $[\Phi(\rho)-B_\rho(1-\varepsilon)]\,n+o(n)$
  & $(\operatorname{conv}B_\rho)(1-\varepsilon)\,n$
  & \ref{thm:frontiers}, \ref{thm:shannon-frontiers}\\
\bottomrule
\end{tabular*}
\end{table}

\paragraph{Techniques.}
Most lower bounds with error start from one change of measure
(Lemma~\ref{lem:overlap-main}), which reweights the executions so
that the output follows the common part of the output distribution
and the target. Each regime then asks how much entropy the
coordinates written in one round can have, given that they are
distinct and nearly independent: in one round the answer is a sharp
stability inequality for permanents, in two rounds a birthday bound,
and with linearly many rounds a variational problem of mean-field
Potts type; the adaptive lower bound follows a likelihood ratio round
by round. The upper bounds reuse a random partition, mix
relabelings of one protocol, and use random coding as in the theory
of channel resolvability \cite{HanVerdu1993,Cuff2013}.

\section{Model and preliminaries}
\label{sec:model}

Logarithms are natural, so randomness is measured in nats. We regard
the set $S_n$ of permutations of $[n]$ as a subset of the words
$[n]^n$ and let $P_n=\Unif(S_n)$. The values in $[n]$ are
\emph{labels}, and a word or partial word is \emph{collision-free} if
its labels are distinct. We write $H$ for entropy,
${\TV(P,Q)=\frac12\sum_x|P(x)-Q(x)|}$ for total variation distance,
$D(P\Vert Q)=\sum_xP(x)\log\frac{P(x)}{Q(x)}$ for relative entropy,
$h_2$ for binary entropy,
$\TC(\nu)=D(\nu\Vert\bigotimes_i\nu_i)$ for the total correlation of
$\nu$ with marginals $\nu_i$,
$\langle\nu_i,\nu_j\rangle=\sum_x\nu_i(x)\nu_j(x)$, and
$(m)_b=m!/(m-b)!$ for the falling factorial.

\paragraph{Protocols and costs.}
A protocol proceeds in stages and writes each coordinate once. The
public history $\mathcal H_t$ before stage $t$ records the shared
draws and the coordinates written so far. At stage $t$ the protocol
draws a shared random variable $Z_t$ with finite range, whose
distribution may depend on $\mathcal H_t$, selects a batch
$A_t=A_t(\mathcal H_t,Z_t)$ of unwritten coordinates, and writes them
independently, coordinate $i$ according to a \emph{kernel}
$q_{t,i}(\,\cdot\mid\mathcal H_t,Z_t)$. An \emph{$R$-round} protocol
writes at most $R$ nonempty batches in every execution; a stage with
an empty batch only draws shared randomness.

The draws within a batch are the private coins of the processors;
they are free, as is all computation, and every other random choice,
including that of the next batch, is part of the public coin $Z_t$.
In the \emph{static} model all shared randomness is drawn before the
first output, as a single seed; in the \emph{adaptive} model it may
be drawn at every stage. In both models the batches and kernels may
depend on the entire history. A \emph{private-coin} protocol uses no
shared randomness. Protocols are finite; those terminating almost
surely have the same optimal costs (Appendix~\ref{app:information}).

We write $Q$ also for the joint distribution of a protocol $Q$ and
for the distribution of its output $X$. The \emph{support cost} and
the \emph{entropy cost} of $Q$, the two measures of its shared
randomness, are
\begin{equation}
 B(Q)=\max\sum_t\log\bigl|\supp Q(Z_t\mid\mathcal H_t)\bigr|,
 \qquad C_H(Q)=\sum_tH_Q(Z_t\mid\mathcal H_t),
 \label{eq:costs}
\end{equation}
where the maximum is over the executions of positive probability. In
the static model these are the logarithm of the support size of the
seed and its entropy $C_H$; the least expected number of fair coins
that generate the seed is $C_H/\log2+O(1)$ \cite{KnuthYao1976}.
For $\tau\in\{\mathrm{st},\mathrm{ad}\}$ and $0\le\varepsilon<1$, let
$B^\tau_\varepsilon(n,R)$ and $C^\tau_\varepsilon(n,R)$ be the infima
of $B(Q)$ and $C_H(Q)$ over $R$-round protocols $Q$ in model $\tau$
with $\TV(P_n,Q)\le\varepsilon$. We also express accuracy through the
\emph{overlap}
\begin{equation}
 p(Q)=\sum_x\min\{P_n(x),Q(x)\}=1-\TV(P_n,Q),
 \label{eq:overlap}
\end{equation}
the probability that a target sample and the output agree under a
maximal coupling.

\begin{proposition}[Basic comparisons]
\label{prop:resource-comparisons}
In either model, $C^\tau_\varepsilon\le B^\tau_\varepsilon$; moreover
$C^{\rm ad}_\varepsilon\le C^{\rm st}_\varepsilon$ and
$B^{\rm ad}_\varepsilon\le B^{\rm st}_\varepsilon$. For $n\ge2$,
$p=1-\varepsilon$, and every $\delta>0$ such that $\varepsilon+\delta<1$,
\begin{equation}
 C^\tau_\varepsilon(n,R)\le p\,C^\tau_0(n,R)+h_2(p),
 \qquad
 B^{\rm st}_{\varepsilon+\delta}(n,R)
 \le B^{\rm ad}_\varepsilon(n,R)+\log(2/\delta).
 \label{eq:time-sharing}
\end{equation}
\end{proposition}

The first inequality in~\eqref{eq:time-sharing} comes from aborting.
The second follows from the interval algorithm \cite{HanHoshi1997},
by which a uniform seed with $M$ values simulates an adaptive protocol
$Q$ with error at most $e^{B(Q)}/M$
(Proposition~\ref{info:static-simulation}); the two models can thus
differ substantially only in entropy cost. Our lower bounds work with
the common part $\min\{P_n,Q\}$ of the output distribution and the
target.

\begin{lemma}[Reweighting by the common part]
\label{lem:overlap-main}
Let $Q$ be a protocol with $p=p(Q)>0$, and let $V$ be the law of its
transcript $T$, comprising all shared random variables and outputs,
given acceptance, when each execution is accepted with probability
$\min\{1,P_n(X)/Q(X)\}$. Then
\begin{equation}
 \log n!-H_V(X)\le\log\frac1p,\qquad
 D(V\Vert Q)\le\log\frac1p,\qquad
 \sum_t\E_V\TC(V_t)\le\log\frac1p,
 \label{eq:overlap-accounting}
\end{equation}
where $V_t$ is the conditional law of batch $t$ under $V$ given
$\mathcal H_t$ and $Z_t$, and the shared randomness satisfies
$\sum_tH_V(Z_t\mid\mathcal H_t)\le\min\{B(Q),C_H(Q)/p\}$. Explicitly,
$V(T)=V_X(X)\,Q(T\mid X)$ with $V_X(x)=\min\{P_n(x),Q(x)\}/p$, a
probability distribution supported on $S_n$.
\end{lemma}

The reweighted law is thus within $\log(1/p)$ nats of the uniform
distribution, of the protocol, and of independence within the rounds.
Every output under $V$ is a permutation, and the chain rule splits its
entropy into the parts supplied by shared and by private randomness:
\begin{equation}
 \log n!-\log\frac1p\;\le\;H_V(X)\;\le\;
 \underbrace{\sum_tH_V(Z_t\mid\mathcal H_t)}_{\text{shared}}
 \;+\;
 \underbrace{\sum_tH_V(X_{A_t}\mid\mathcal H_t,Z_t)}_{\text{private}}.
 \label{eq:split}
\end{equation}
Each regime below therefore reduces to the \emph{single-batch
problem}: bound the entropy of a collision-free batch of $b$
coordinates on $m$ unused labels in terms of its total correlation,
of which the batches under $V$ have at most $\log(1/p)$ in total.
Total correlation controls collisions: a collision-free batch $\nu$
has $\TC(\nu)\ge\sum_{i<j}\langle\nu_i,\nu_j\rangle$, the expected
number of colliding pairs among independent draws from its marginals.
At zero error it vanishes, and the bound becomes a count. The lemma
(Appendix~\ref{fr:appendix}) extends the entropy accounting used for
parallel sampling \cite{BenHamuEtAl2025,WenEtAl2026} by a change of
measure of the kind used in strong converses in information theory
\cite{TyagiWatanabe2020}.

\section{Exact sampling}
\label{sec:exact}

We first determine the entropy that private randomness can supply.
Consider a private-coin protocol that never repeats a label, at a
history with $m$ unused labels. A batch of $b$ coordinates is
collision-free with probability one exactly when the supports of its
$b$ independent draws are pairwise disjoint sets of unused labels.
The number of its outcomes, the product of the support sizes, is
largest for balanced sizes:
$L(m,b)=\lfloor m/b\rfloor^{\,b-s}\lceil m/b\rceil^{\,s}$ with
$s=m\bmod b$. Let $F(n,R)$ be the largest number of permutations that
such a protocol can output in $R$ rounds. Optimizing over the first
batch gives
\begin{equation}
 F(n,R)=\max_{1\le b\le n}L(n,b)\,F(n-b,R-1),\qquad
 F(0,R)=1,\quad F(n,0)=0\ (n>0),
 \label{eq:capacity}
\end{equation}
and uniform draws from balanced supports attain $F(n,R)$ with a
uniform output. By the chain rule, $\log F(n,R)$ is also the largest
entropy of such a protocol, even with adaptive batches.

\begin{theorem}[Exact sampling]
\label{thm:exact}
For $1\le R\le n$, put $\Delta(n,R)=\log n!-\log F(n,R)$. Then
\[
 \Delta(n,R)\le C^{\rm ad}_0(n,R)\le C^{\rm st}_0(n,R)
 \le B^{\rm st}_0(n,R)\le\Delta(n,R)+O(1),
\]
and $C^{\rm ad}_0\le B^{\rm ad}_0\le B^{\rm st}_0$, so all four exact
costs lie between $\Delta(n,R)$ and $\Delta(n,R)+O(1)$ uniformly in
$n$ and $R$. All four equal $\log n!$ for $R=1$, $\log(n!/G(n))$ for
$R=2$, where $G(n)$ is the largest product of positive integers with
sum $n$, and $\log\binom{2(n-R)}{n-R}$ for every $R\ge\lceil n/2\rceil$.
\end{theorem}

The exact cost is thus known within an absolute constant over its
whole range, from $\log n!$ at $R=1$ to zero at $R=n$. For the lower
bound, every batch of an exact protocol is collision-free with
probability one given the history and the shared variables, so its
conditional entropy is at most $\log L(m_t,b_t)$ for $m_t$ unused
labels and batch size $b_t$. By~\eqref{eq:capacity} these terms sum
to at most $\log F(n,R)$ along every execution, and the chain rule
gives $\log n!\le C_H(Q)+\log F(n,R)$.

\paragraph{Upper bound: reusing a random partition.}
Suppose first that all batches have the same size $b$. The seed is a
uniformly random balanced partition of the labels into $b$ parts, and
each coordinate of the first batch draws a uniform label from its own
part. The batch is then a uniform injection, and given the labels
written, the partition of the remaining labels is again uniform, so
the same seed serves every later batch. Each round thus consumes
exactly $\log(m)_b-\log L(m,b)$ nats of the seed, which is what the
lower bound charges to it. When the batch size increases, the
residual partition is refined by a finite form of the interval
algorithm \cite{HanHoshi1997,Li2021}. The conversion losses form a
convergent series and account for the additive constant
(Appendix~\ref{ex:section}).

\paragraph{Explicit values.}
With two rounds, the second batch can complete a collision-free first
batch deterministically, so $F(n,2)=\max_bL(n,b)=G(n)$, attained by
supports of sizes two and three. The matching seed is a random coset
of a product of cyclic groups acting on blocks of coordinates, as in
the subgroup algorithm \cite{DiaconisShahshahani1987}. For
$R\ge\lceil n/2\rceil$, single draws followed by $n-R$ pairs are
optimal, and partition reuse with two parts attains the bound. For
$R=o(n)$, $R$ batches of equal size give $\Delta(n,R)\sim n\log(n/R)$;
for $R/n\to\rho\in(0,1]$ two consecutive batch sizes are mixed, and
$\Delta(n,R)/n\to\Phi(\rho)$, where $\Phi$ is the piecewise linear
function through the points $(1/j,\log j)$, $j\ge1$.

\section{One and two rounds}
\label{sec:oneround}

We now ask what an error $\varepsilon$ saves. In one round the least
support cost is $\log r_\varepsilon(P_n)$, where $r_\varepsilon(P)$
is the least number of product distributions in a mixture within
total variation $\varepsilon$ of $P$, the approximate nonnegative
rank of $P$ with probability distributions as approximants
\cite{JainEtAl2013,GoosWatson2019}. A product distribution on $[n]^n$
is given by the row-stochastic matrix $A$ of its marginals, and its
mass on $S_n$ is $\per(A)$. Since a row-stochastic matrix with
permanent one is a permutation matrix, $r_0(P_n)=n!$. With error, a
component can place mass on several permutations; by the next
theorem, its mass on all but the most likely one is at most half its
mass outside $S_n$ (cf.\ \cite[Appendix~A]{AaronsonNguyen2016}).

\begin{theorem}[Stability for product distributions]
\label{thm:stability}
Let $n\ge2$, let $\Pi$ be a product distribution on $[n]^n$, and put
$q=\Pi(S_n)$ and $q_{\max}=\max_{\sigma\in S_n}\Pi(\sigma)$. Then the
following two inequalities hold:
\begin{equation}
 q-q_{\max}\le\frac{1-q}{2},\qquad
 q\,H(\Pi\mid S_n)\le(1-q)\log2,
 \label{eq:permanent-stability-main}
\end{equation}
where $H(\Pi\mid S_n)$ is the conditional entropy given $S_n$ (zero if
$q=0$). Both constants are optimal for every $n$. In matrix terms,
the first inequality reads
$3\per(A)-2\max_{\sigma}\prod_iA_{i\sigma(i)}\le1$.
\end{theorem}

\begin{theorem}[One round]
\label{thm:oneround}
Let $n\ge2$. For every error $0\le\varepsilon\le1/2$, the approximate
rank satisfies
\begin{equation}
 \left\lceil(1-3\varepsilon/2)\,n!\right\rceil
 \le r_\varepsilon(P_n)
 \le n!-3\left\lfloor\varepsilon\,n!/2\right\rfloor,
 \label{eq:rank-finite-main}
\end{equation}
so that $B^\tau_\varepsilon(n,1)=\log n!+\log(1-3\varepsilon/2)+O(1/n!)$
in both models. For $0\le\varepsilon<1$, the entropy cost is
$C^\tau_\varepsilon(n,1)=(1-\varepsilon)\log n!+O(1)$ with an absolute
constant. The lower bound in~\eqref{eq:rank-finite-main} holds for
every family of permutations:
$r_\varepsilon(\Unif(T))\ge\lceil(1-3\varepsilon/2)|T|\rceil$ for
nonempty $T\subseteq S_n$ and $0\le\varepsilon\le2/3$.
\end{theorem}

The two bounds in~\eqref{eq:rank-finite-main} differ by at most two.
For instance, $r_{1/2}(P_n)=n!/4$ for $n\ge4$, so an error of $1/2$
shortens the one-round seed by exactly $\log4$ nats, and aborting is
optimal for the entropy cost up to an additive constant. In one
round, an error thus buys only $O(1)$ nats, while a second round lets
it buy an amount of order $(\log n)^2$, as we show next.

For Theorem~\ref{thm:stability}, we expose the first coordinate $X_1$
and, when the other coordinates are distinct, their missing label
$J$; the output is a permutation exactly when $X_1=J$. With
$\alpha_j=\Pr(X_1=j)$ and $\beta_j=\Pr(J=j)$, Cauchy--Schwarz gives
$(\sum_j\sqrt{\alpha_j\beta_j})^2\le\sum_j\beta_j\le1$, which bounds
the mass and entropy of the agreement outside its most likely label;
induction on $n$ completes the proof.

The lower bound in~\eqref{eq:rank-finite-main} follows by a covering
argument. Let $Q$ be a mixture of $M$ product distributions within
$\varepsilon$ of a distribution $P$ on $S_n$, and let $\Sigma$
consist of one most likely permutation of each component.
Theorem~\ref{thm:stability} gives
$Q(S_n\setminus\Sigma)\le Q(S_n^c)/2\le\varepsilon/2$, and since the
overlap of $P$ and $Q$ is at most $P(\Sigma)+Q(S_n\setminus\Sigma)$,
we obtain $P(\Sigma)\ge1-3\varepsilon/2$; for $P=\Unif(T)$ this
forces $M\ge(1-3\varepsilon/2)|T|$. The upper bound pairs the
permutations that differ by exchanging the first two coordinates. A
product distribution in which these two coordinates choose
independently and uniformly between the two exchanged labels puts
mass $1/4$ on each permutation of the pair and mass $1/2$ outside
$S_n$, so $j$ such components of weight $4/n!$ and point masses on
the other $n!-4j$ permutations give $n!-3j$ components with
total-variation error $2j/n!$. The entropy bounds follow in the same
way from the second inequality of Theorem~\ref{thm:stability} and
Lemma~\ref{lem:overlap-main} (Theorem~\ref{rank:latent-entropy}).

\paragraph{Two rounds.}
Two rounds are the first case in which the saving grows with $n$. A
second round can complete any collision-free first batch, so the
first-round coordinates can use several labels each, on disjoint
supports; this is why $\log F(n,2)=\lambda_2n+O(1)$ with
$\lambda_2=(\log3)/3$. With error, some coordinates can occasionally
draw from a common reserve of $w$ labels. By the birthday bound,
about $\sqrt w$ such draws remain distinct with constant probability;
each adds about $\log n$ nats, and each reserved label costs
$\lambda_2$ nats. The best $w$ is of order $(\log n)^2$, and so is
the saving. The same mechanism, spending total correlation on shared
labels, operates in every later regime.

\begin{theorem}[Two rounds]
\label{thm:fewround}
Let $d_\varepsilon=\log\frac1{1-\varepsilon}$. For every fixed
$0<\varepsilon<1$ and in both models,
\begin{equation}\label{eq:two-round-main}
 \begin{gathered}
 B^\tau_\varepsilon(n,2)=\Delta(n,2)-
 \frac{d_\varepsilon}{2\lambda_2}(\log n)^2
 +o\bigl((\log n)^2\bigr),\\
 (1-\varepsilon)\,\Delta(n,2)-O_\varepsilon\bigl((\log n)^2\bigr)
 \le C^\tau_\varepsilon(n,2)\le(1-\varepsilon)\,\Delta(n,2)+h_2(\varepsilon).
 \end{gathered}
\end{equation}
\end{theorem}

For the lower bound, Lemma~\ref{lem:overlap-main} leaves the
following instance of the single-batch problem.

\begin{lemma}[Entropy of a nearly independent batch]
\label{lem:owner-main}
Let $\nu$ be a distribution on collision-free words in $[n]^b$,
$1\le b\le n$, with total correlation $\delta$. Then, with an $o(1)$
term that depends only on $n$,
\begin{equation}
 H(\nu)\le\lambda_2n+
 \Bigl(\frac1{2\lambda_2}+o(1)\Bigr)(\log n)^2\,\delta+\frac1e.
 \label{eq:owner-main}
\end{equation}
\end{lemma}

The proof (Lemma~\ref{fr:owner}) makes the balance above precise.
Let $A_{ix}=\nu(X_i=x)$. The collision bound of
Section~\ref{sec:model} gives $\sum_x\sum_{i<j}A_{ix}A_{jx}\le\delta$.
A label whose total mass $\sum_iA_{ix}$ is not small is therefore
used almost entirely by one coordinate, and such labels contribute at
most $\lambda_2$ nats each. On the $w$ remaining labels, the mass
$e_x=\sum_iA_{ix}-\max_iA_{ix}$ used outside the largest entry
satisfies $\sum_{i<j}A_{ix}A_{jx}\ge e_x^2/2$, so the $e_x$ sum to at
most $\sqrt{2w\delta}$ by the Cauchy--Schwarz inequality. Each unit
of this mass carries at most $\log n$ nats, and setting aside $w$
labels costs $\lambda_2w$. Up to lower-order terms, the entropy
beyond $\lambda_2n$ that the batch can gain is therefore at most
\begin{equation}
 \max_{w\ge0}\bigl\{\log n\sqrt{2w\delta}-\lambda_2w\bigr\}
 =\frac{\delta}{2\lambda_2}(\log n)^2.
 \label{eq:two-round-balance}
\end{equation}
The second batch contributes only $O_\varepsilon(1)$ under $V$, by
the second inequality of Theorem~\ref{thm:stability}.

The optimization~\eqref{eq:two-round-balance} also describes a
protocol. Hold $w$ labels in reserve, partition the others into
triples, and let one coordinate per triple draw a uniform label from
its triple or, with small probability, from the reserve. With $\mu$
reserve draws on average, these are distinct with probability about
$e^{-\mu^2/(2w)}$ and add about $\mu\log n$ nats, while the reserve
costs $\lambda_2w$; at $\mu^2/(2w)=\kappa$ the best choice gains
$\kappa(\log n)^2/(2\lambda_2)$ nats. Keeping the collision-free
first batches with a typical number of reserve draws, completing them
deterministically, and mixing $O(n!\,h)$ relabelings of the result,
where $h$ is its largest output probability, gives a sampler within
$\varepsilon$ of $P_n$ for every $\kappa<d_\varepsilon$
(Lemma~\ref{con:group-mixture}); letting $\kappa\uparrow d_\varepsilon$
proves the upper bound for the support cost in~\eqref{eq:two-round-main}.

\section{Sublinearly and linearly many rounds}
\label{sec:depth}

The picture found for one and two rounds persists for every
$R=o(n)$: an error changes the support cost only in lower-order
terms and scales the entropy cost by $1-\varepsilon$.

\begin{theorem}[Sublinearly many rounds]
\label{thm:sublinear}
For fixed $0\le\varepsilon<1$ and $R=o(n)$, in both models,
\begin{equation}
 B^\tau_\varepsilon(n,R)\sim n\log(n/R),\qquad
 C^\tau_\varepsilon(n,R)\sim(1-\varepsilon)\,n\log(n/R).
 \label{eq:sublinear}
\end{equation}
Moreover, $0\le B^\tau_0(n,R)-B^\tau_\varepsilon(n,R)=O_\varepsilon(n)$
uniformly in $1\le R\le n$.
\end{theorem}

The proof solves the single-batch problem uniformly in the batch
size: a distribution on collision-free words of $b$ coordinates on
$m$ labels with total correlation $\delta$ has entropy at most
$b\log(m/b)+tm\delta+b\Psi(t)$ for every $t>0$, with an explicit
$\Psi$ decreasing to zero (Lemma~\ref{fr:coarse-modulus}). Summed
over the batches, it gives a finite lower bound, uniform in $R$ and
in the overlap~\eqref{fr:eq-sparse-finite}. Exact protocols attain
the support cost in~\eqref{eq:sublinear}, and aborting attains the
entropy cost.

\paragraph{Linearly many rounds.}
When $R\sim\rho n$ with $0<\rho<1$, the exact cost is itself of order
$n$, so the saving enters the leading term and has to be determined
exactly. A batch of $b$ independent uniform draws from $m$ unused
labels needs no shared randomness and has a collision with
probability about $\binom b2/m$, whereas uniform draws from disjoint
blocks of $m/b$ labels never collide and give up $\log b$ nats of
entropy per coordinate. For a general batch with marginals
$\nu_1,\ldots,\nu_b$, draw a label $J$ uniformly and set
$a_i=m\,\nu_i(J)$. Then $a=(a_1,\ldots,a_b)$ is a nonnegative random
vector with $\E a_i=1$. Both quantities become averages over the
labels, so the batch enters only through the law of $a$: the entropy
deficit of the marginals and $m$ times the expected number of
colliding pairs are
\[
 d_b=\E\sum_ia_i\log a_i=\sum_iD(\nu_i\Vert\Unif([m])),\qquad
 c_b=\E\sum_{i<j}a_ia_j=m\sum_{i<j}\langle\nu_i,\nu_j\rangle.
\]
Let $\mathcal M_b$ be the set of laws
$\mu$ of nonnegative random vectors $(a_1,\ldots,a_b)$ with
$\E a_i=1$ (\emph{density laws}), define $d_b(\mu)$ and $c_b(\mu)$ by
the same expectations, and put, for every multiplier $t\ge0$,
\begin{equation}
 K_b(t)=\inf_{\mu\in\mathcal M_b}\bigl\{d_b(\mu)+t\,c_b(\mu)\bigr\}.
 \label{eq:local-action}
\end{equation}
Thus $K_b(t)$ is the best local trade between information and
collisions for a batch of size $b$ at collision multiplier $t$.
Uniform marginals give $(d_b,c_b)=(0,\binom b2)$, and uniform ones on
disjoint blocks give $(b\log b,0)$. A single batch can have a
positive deficit without shared randomness, as fixed disjoint blocks
show; the deficits are paid through the entropy accounting of the
whole protocol.

To state the result we pass to rates per coordinate. For
$\rho\in(0,1)$, let $B^\tau_\rho(p)$ be the least $\beta$ for which
there are $R_n$-round protocols $Q_n$ on $[n]$ in model $\tau$ with
$R_n/n\to\rho$, $\limsup_nB(Q_n)/n\le\beta$, and
$\liminf_np(Q_n)\ge p$, and define $C^\tau_\rho(p)$ in the same way
with $C_H$ in place of $B$.

\begin{theorem}[Linearly many rounds]
\label{thm:frontiers}
Let $p\in(0,1]$. Then $B^\tau_\rho(p)=B_\rho(p)$ in both models,
where
\begin{equation}
 B_\rho(p)=\sup_{\eta,\xi\ge0}
 \left\{\int_0^1\inf_{b\in\mathbb N}
       \frac{K_b(\xi/u)+\eta}{b}\,du-\eta\rho-\xi\log\frac1p\right\}.
 \label{eq:main-dual}
\end{equation}
Moreover, for every sequence $R_n/n\to\rho$ and every fixed
$0\le\varepsilon<1$, $B^\tau_\varepsilon(n,R_n)/n\to
B_\rho(1-\varepsilon)$.
\end{theorem}

In~\eqref{eq:main-dual}, $u$ is the fraction of unused labels,
and $\eta$ and $\xi$ are Lagrange multipliers for the number of
rounds and for the log-probability of a collision-free output. A
collision at $u$ has multiplier~$\xi/u$, since collisions become more
likely as labels are used up, and the integrand is the least cost per
coordinate of a batch. The cost of $n$ coordinates in $\rho n$ rounds
is thus determined by $K_b$.

This function has a critical multiplier. For a fixed sum $s=\sum_ia_i$,
minimizing over $v=a/s$ is the variational problem of the mean-field
Potts model \cite{EllisWang1990,GeorgiiEtAl2006}: a minimizer is
uniform or has one large coordinate and $b-1$ equal small ones. In
particular $K_b(t)=\binom b2t$ for $t\le2/b$, where uniform marginals
are optimal. Since the multiplier at $u$ is
$\xi/u$, optimal protocols use nonuniform marginals, and hence shared
randomness, only once labels are sufficiently depleted.

\begin{proposition}[Properties of $B_\rho$]
\label{prop:frontier-properties}
$B_\rho(p)$ is jointly continuous on $(0,1)\times[0,1]$ and
nondecreasing in $p$, with $B_\rho(1)=\Phi(\rho)$. There is a
threshold $p_0(\rho)\in(0,1)$ such that $B_\rho(p)=0$ if and only if
$p\le p_0(\rho)$, and there are $c_\rho>0$ and $c_p<\infty$
independent of $\rho$ such that
\begin{equation}
 c_\rho\,\varepsilon\log\frac1\varepsilon\;\le\;
 \Phi(\rho)-B_\rho(1-\varepsilon)\;\le\;c_{1-\varepsilon},
 \label{eq:bounded-gap}
\end{equation}
the first inequality holding for small $\varepsilon$, and
$p_0(\rho)=\sqrt{2\rho-1}$ for $\rho\ge2/3$
(Proposition~\ref{front:zero}).
\end{proposition}

Hence an overlap up to the threshold costs $o(n)$ nats, while the
saving in support cost is $\Theta_{\rho,\varepsilon}(n)$ and stays
bounded per coordinate as $\rho\to0$, in agreement with
Theorem~\ref{thm:sublinear}. The curve $B_\rho$ has infinite slope at
$p=1$, because moving a small mass $\alpha$ between disjoint supports
gains entropy of order $\alpha\log(1/\alpha)$ while $c_b$ increases
by order $\alpha$. The threshold is attained by uniform draws in
batches that shrink as labels are used up, and equals the accuracy of
the optimal schedules of Wen et al.\ under a hard bound on the number
of rounds \cite[Section~4.2 and Appendix~C]{WenEtAl2026}, which draw
each batch from the conditional marginals of the target. A fixed
schedule of this kind gives every permutation the same probability,
so its overlap $q$ also determines its relative entropy
$D(P_n\Vert Q)=\log(1/q)$. No protocol of support or entropy cost
$o(n)$ attains a larger overlap, whatever its kernels, so these
schedules remain optimal among all such protocols
(Proposition~\ref{front:zero}).

The lower bound in Theorem~\ref{thm:frontiers} uses that a
collision-free batch $\nu$ of $b$ coordinates on $m$ labels satisfies
$H(\nu)=b\log m-d_b-\TC(\nu)$ and $c_b\le m\,\TC(\nu)$, with $d_b$
and $c_b$ computed from its marginals. After splitting large batches,
the terms $b_t\log m_t$ sum to $\log n!+o(n)$, so
by~\eqref{eq:split} the shared randomness has
entropy at least the expected sum of the $d_{b_t}$ under $V$, while
the expected sum of the $c_{b_t}/m_t$ is at most $\log(1/p)$;
bounding each batch by $K_b$ gives~\eqref{eq:main-dual}. For the
upper bound, a batch with law $\mu$ is realized by assigning the
values of $a$ to the unused labels at random. These assignments are
equally likely to be collision-free, their average is uniform on
collision-free batches, and
a mixture of $e^{n(B_\rho(p)+o(1))}$ of them approximates it, by a
resolvability argument \cite{HanVerdu1993,Cuff2013}.

\section{Entropy cost in the static and adaptive models}
\label{sec:sharing-costs}

For $R=o(n)$ the entropy cost is, to leading order, the support cost
scaled by $1-\varepsilon$, and the two models agree. With linearly
many rounds neither statement holds. A static protocol can randomize
between samplers of different overlaps, and an adaptive protocol can
also postpone each shared draw until the stage that uses it and omit
it after a collision. Randomizing helps exactly where $B_\rho$ lies
above its lower convex envelope, while postponing helps at every
overlap strictly between the threshold and one. Both effects are
stated through profiles, which
describe how a protocol spends rounds, information, and collisions as
labels are used up. A \emph{profile} $\pi$ prescribes, for each fraction $u\in(0,1]$ of
unused labels, a probability distribution over pairs $(b,\mu)$ with
$\mu\in\mathcal M_b$. Let $r(u)$, $d(u)$, and $c(u)$ be the averages
of $1/b$, $d_b(\mu)/b$, and $c_b(\mu)/b$ under $\pi$ at $u$: the
rounds, the information, and the collision rate that the profile
spends per coordinate there, so that a profile is the continuum limit
of a schedule of batch sizes and density laws. The profile uses
$R(\pi)$ rounds per coordinate, and when a fraction $u$
of the labels remains, its output is still collision-free with
probability $s_\pi(u)$. In terms of the averages $r$ and $c$ above,
these two quantities are given by
\begin{equation}
 R(\pi)=\int_0^1r(u)\,du,\qquad
 s_\pi(u)=\exp\Bigl(-\int_u^1\frac{c(v)}{v}\,dv\Bigr).
 \label{eq:profile}
\end{equation}
Profiles may be taken piecewise constant, ending with single draws
(Appendix~\ref{app:frontiers}). Then $B_\rho(p)$ is the least value
approached by $\int_0^1d(u)\,du$ along profiles whose
limits satisfy $R(\pi)\le\rho$ and $s_\pi(0)\ge p$.

\begin{theorem}[Entropy cost with linearly many rounds]
\label{thm:shannon-frontiers}
Let $\rho\in(0,1)$.
\begin{enumerate}[label=\textup{(\roman*)},leftmargin=*]
\item In the static model, $C^{\rm st}_\rho$ is the lower convex
envelope of $p\mapsto B_\rho(p)$ on $[0,1]$, and it is attained by
randomizing between two suitable protocols of different overlaps.
\item In the adaptive model, the region $\beta\ge C^{\rm ad}_\rho(p)$
is the closed convex hull of the pairs
\begin{equation}
 \Bigl(\int_0^1s_\pi(u)\,d(u)\,du,\ s_\pi(0)\Bigr),
 \qquad R(\pi)\le\rho,
 \label{eq:adaptive-region}
\end{equation}
together with all pairs of larger cost or smaller overlap.
\item For $p\in(0,1]$, both entropy rates lie between the lower bound
$p\,B_\rho(p)$ and $B_\rho(p)$:
\[
 p\,B_\rho(p)\le C^{\rm ad}_\rho(p)\le C^{\rm st}_\rho(p)\le B_\rho(p),
\]
and $C^{\rm ad}_\rho(p)<C^{\rm st}_\rho(p)$ if and only if
$p_0(\rho)<p<1$. Moreover, $C^{\rm st}_\rho(p)<p\,\Phi(\rho)$ for
$0<p<1$.
\end{enumerate}
\end{theorem}

Parts (i) and (ii) differ in one factor: a static protocol that
follows a profile $\pi$ spends $\int_0^1d(u)\,du$ nats per
coordinate, and an adaptive one spends $\int_0^1s_\pi(u)\,d(u)\,du$,
because the information for the batch at $u$ is drawn only while the
output is collision-free. By part~(iii) the difference is strict
exactly between the threshold and exact sampling. Under the
reweighted law $V$ the shared randomness has the same optimal rate
$B_\rho(p)$ in both models, the support rate
(Proposition~\ref{info:overlap-cost}).

For (i), we group the seeds of a static protocol by probability and
apply the support bound to each group. For the
lower bound in (ii), at a collision-free history we compare the collision-free completions
with the uniform distribution on the remaining permutations through
their largest likelihood ratio; a shared draw increases its logarithm
by at most the surprisal of the draw, and the single-batch bound
controls the effect of a batch, which gives a recursion with
continuum limit~\eqref{eq:adaptive-region}. Part~(iii) is strict
because, at an overlap in $(p_0(\rho),1)$, the multipliers
in~\eqref{eq:main-dual} are positive and finite, and then
exactly disjoint supports are suboptimal, so a nearly optimal
profile spends information where $s_\pi<1$ and postponing it saves a
positive amount (Theorems~\ref{front:static-entropy},
\ref{ad:frontier}, \ref{strict:universal}).

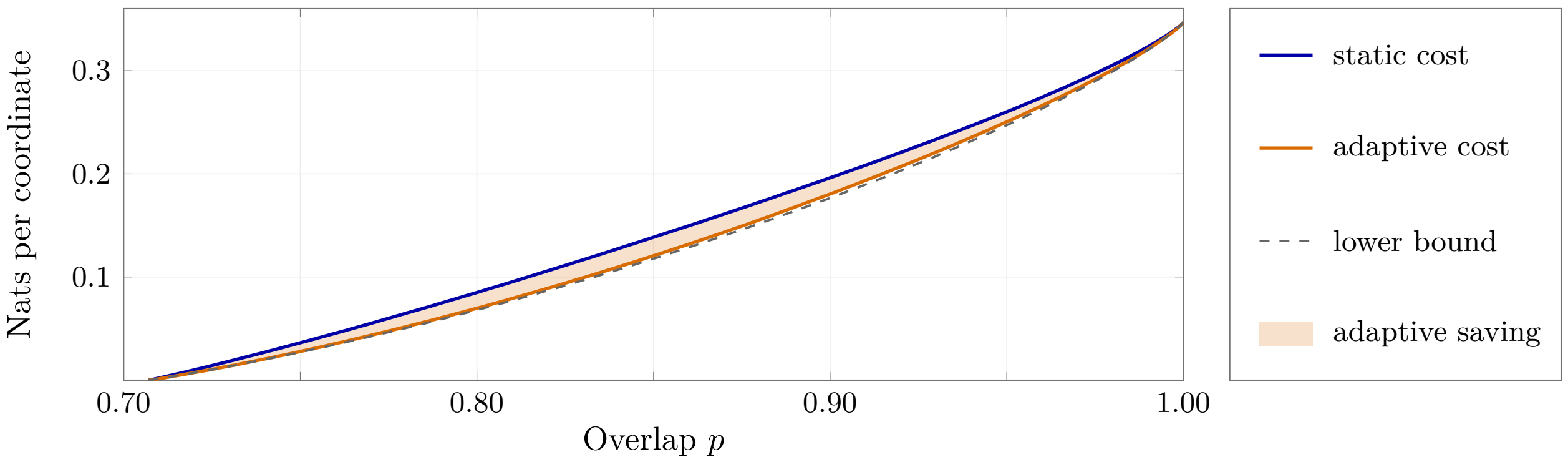
\begin{figure}[t]
\pgfmathsetlengthmacro{\figL}{36pt}%
\pgfmathsetlengthmacro{\figG}{14pt}%
\pgfmathsetlengthmacro{\figK}{100pt}%
\pgfmathsetlengthmacro{\figW}{\textwidth-\figL-\figG-\figK}%
\pgfmathsetlengthmacro{\figH}{112pt}%
\pgfmathsetmacro{\figYL}{-22.5/(\figW/1pt)}%
\noindent
\begin{tikzpicture}
\useasboundingbox (-\figL,-26pt) rectangle (\figW+\figG+\figK,\figH+3pt);
\begin{axis}[
 at={(0,0)},anchor=south west,
 scale only axis,width=\figW,height=\figH,
 xmin=.7,xmax=1,xtick={.7,.75,.8,.85,.9,.95,1},
 xticklabels={0.70,,0.80,,0.90,,1.00},
 ymin=0,ymax=.36,ytick={0,.1,.2,.3},yticklabels={,0.1,0.2,0.3},
 tick label style={font=\footnotesize},label style={font=\small},
 axis line style={black!55,line width=.4pt},tick style={black!55},
 major tick length=2.5pt,tick align=inside,
 grid=major,grid style={black!7,line width=.3pt},
 xlabel={Overlap $p$},xlabel style={yshift=3pt},
 ylabel={Nats per coordinate},
 ylabel style={at={(axis description cs:\figYL,.5)},anchor=south},
 clip=true]
\addplot[fill=orange!85!black,fill opacity=.20,draw=none] coordinates {
(0.7071,0.0000)
(0.7106,0.0026)
(0.7140,0.0053)
(0.7175,0.0080)
(0.7209,0.0108)
(0.7244,0.0137)
(0.7278,0.0166)
(0.7313,0.0196)
(0.7347,0.0226)
(0.7382,0.0256)
(0.7417,0.0287)
(0.7451,0.0318)
(0.7486,0.0350)
(0.7520,0.0381)
(0.7555,0.0414)
(0.7589,0.0446)
(0.7624,0.0478)
(0.7658,0.0511)
(0.7693,0.0544)
(0.7727,0.0578)
(0.7762,0.0612)
(0.7796,0.0646)
(0.7831,0.0680)
(0.7866,0.0714)
(0.7900,0.0749)
(0.7935,0.0784)
(0.7969,0.0819)
(0.8004,0.0854)
(0.8038,0.0890)
(0.8073,0.0925)
(0.8107,0.0961)
(0.8142,0.0998)
(0.8176,0.1034)
(0.8211,0.1071)
(0.8246,0.1108)
(0.8280,0.1145)
(0.8315,0.1182)
(0.8349,0.1219)
(0.8384,0.1257)
(0.8418,0.1295)
(0.8453,0.1333)
(0.8487,0.1371)
(0.8522,0.1410)
(0.8556,0.1448)
(0.8591,0.1487)
(0.8626,0.1526)
(0.8660,0.1565)
(0.8695,0.1605)
(0.8729,0.1644)
(0.8764,0.1684)
(0.8798,0.1724)
(0.8833,0.1764)
(0.8867,0.1805)
(0.8902,0.1845)
(0.8936,0.1886)
(0.8971,0.1927)
(0.9006,0.1969)
(0.9040,0.2010)
(0.9075,0.2052)
(0.9109,0.2094)
(0.9144,0.2137)
(0.9178,0.2180)
(0.9213,0.2223)
(0.9247,0.2267)
(0.9282,0.2311)
(0.9316,0.2356)
(0.9351,0.2401)
(0.9385,0.2446)
(0.9420,0.2492)
(0.9455,0.2539)
(0.9489,0.2586)
(0.9524,0.2634)
(0.9558,0.2683)
(0.9593,0.2732)
(0.9627,0.2783)
(0.9662,0.2834)
(0.9696,0.2887)
(0.9731,0.2941)
(0.9765,0.2996)
(0.9800,0.3053)
(0.9805,0.3062)
(0.9810,0.3070)
(0.9815,0.3079)
(0.9820,0.3087)
(0.9825,0.3096)
(0.9830,0.3104)
(0.9835,0.3113)
(0.9840,0.3122)
(0.9845,0.3131)
(0.9850,0.3140)
(0.9855,0.3149)
(0.9860,0.3158)
(0.9865,0.3167)
(0.9870,0.3176)
(0.9875,0.3185)
(0.9880,0.3194)
(0.9885,0.3204)
(0.9890,0.3213)
(0.9895,0.3223)
(0.9900,0.3232)
(0.9905,0.3242)
(0.9910,0.3252)
(0.9915,0.3262)
(0.9920,0.3272)
(0.9925,0.3282)
(0.9930,0.3292)
(0.9935,0.3303)
(0.9940,0.3313)
(0.9945,0.3324)
(0.9950,0.3335)
(0.9955,0.3346)
(0.9960,0.3357)
(0.9965,0.3369)
(0.9970,0.3381)
(0.9975,0.3393)
(0.9980,0.3406)
(0.9985,0.3419)
(0.9990,0.3433)
(0.9995,0.3448)
(1.0000,0.3466)
(1.0000,0.3466)
(0.9995,0.3446)
(0.9990,0.3430)
(0.9985,0.3415)
(0.9980,0.3400)
(0.9975,0.3387)
(0.9970,0.3373)
(0.9965,0.3360)
(0.9960,0.3347)
(0.9955,0.3335)
(0.9950,0.3322)
(0.9945,0.3310)
(0.9940,0.3298)
(0.9935,0.3287)
(0.9930,0.3275)
(0.9925,0.3264)
(0.9920,0.3252)
(0.9915,0.3241)
(0.9910,0.3230)
(0.9905,0.3219)
(0.9900,0.3208)
(0.9895,0.3198)
(0.9890,0.3187)
(0.9885,0.3176)
(0.9880,0.3166)
(0.9875,0.3156)
(0.9870,0.3145)
(0.9865,0.3135)
(0.9860,0.3125)
(0.9855,0.3115)
(0.9850,0.3105)
(0.9845,0.3095)
(0.9840,0.3085)
(0.9835,0.3075)
(0.9830,0.3066)
(0.9825,0.3056)
(0.9820,0.3046)
(0.9815,0.3037)
(0.9810,0.3027)
(0.9805,0.3018)
(0.9800,0.3008)
(0.9765,0.2944)
(0.9731,0.2882)
(0.9696,0.2822)
(0.9662,0.2763)
(0.9627,0.2706)
(0.9593,0.2650)
(0.9558,0.2595)
(0.9524,0.2540)
(0.9489,0.2487)
(0.9455,0.2435)
(0.9420,0.2383)
(0.9385,0.2332)
(0.9351,0.2282)
(0.9316,0.2232)
(0.9282,0.2184)
(0.9247,0.2135)
(0.9213,0.2088)
(0.9178,0.2041)
(0.9144,0.1994)
(0.9109,0.1948)
(0.9075,0.1902)
(0.9040,0.1857)
(0.9006,0.1812)
(0.8971,0.1768)
(0.8936,0.1724)
(0.8902,0.1680)
(0.8867,0.1637)
(0.8833,0.1594)
(0.8798,0.1552)
(0.8764,0.1510)
(0.8729,0.1469)
(0.8695,0.1428)
(0.8660,0.1388)
(0.8626,0.1348)
(0.8591,0.1308)
(0.8556,0.1269)
(0.8522,0.1230)
(0.8487,0.1192)
(0.8453,0.1154)
(0.8418,0.1117)
(0.8384,0.1080)
(0.8349,0.1043)
(0.8315,0.1007)
(0.8280,0.0972)
(0.8246,0.0936)
(0.8211,0.0901)
(0.8176,0.0867)
(0.8142,0.0833)
(0.8107,0.0799)
(0.8073,0.0766)
(0.8038,0.0733)
(0.8004,0.0701)
(0.7969,0.0669)
(0.7935,0.0638)
(0.7900,0.0607)
(0.7866,0.0576)
(0.7831,0.0546)
(0.7796,0.0516)
(0.7762,0.0487)
(0.7727,0.0458)
(0.7693,0.0430)
(0.7658,0.0402)
(0.7624,0.0374)
(0.7589,0.0347)
(0.7555,0.0320)
(0.7520,0.0294)
(0.7486,0.0268)
(0.7451,0.0243)
(0.7417,0.0218)
(0.7382,0.0194)
(0.7347,0.0170)
(0.7313,0.0146)
(0.7278,0.0124)
(0.7244,0.0101)
(0.7209,0.0080)
(0.7175,0.0059)
(0.7140,0.0038)
(0.7106,0.0018)
(0.7071,0.0000)
} -- cycle;
\addplot[blue!65!black,line width=1pt] coordinates {
(0.7071,0.0000)
(0.7106,0.0026)
(0.7140,0.0053)
(0.7175,0.0080)
(0.7209,0.0108)
(0.7244,0.0137)
(0.7278,0.0166)
(0.7313,0.0196)
(0.7347,0.0226)
(0.7382,0.0256)
(0.7417,0.0287)
(0.7451,0.0318)
(0.7486,0.0350)
(0.7520,0.0381)
(0.7555,0.0414)
(0.7589,0.0446)
(0.7624,0.0478)
(0.7658,0.0511)
(0.7693,0.0544)
(0.7727,0.0578)
(0.7762,0.0612)
(0.7796,0.0646)
(0.7831,0.0680)
(0.7866,0.0714)
(0.7900,0.0749)
(0.7935,0.0784)
(0.7969,0.0819)
(0.8004,0.0854)
(0.8038,0.0890)
(0.8073,0.0925)
(0.8107,0.0961)
(0.8142,0.0998)
(0.8176,0.1034)
(0.8211,0.1071)
(0.8246,0.1108)
(0.8280,0.1145)
(0.8315,0.1182)
(0.8349,0.1219)
(0.8384,0.1257)
(0.8418,0.1295)
(0.8453,0.1333)
(0.8487,0.1371)
(0.8522,0.1410)
(0.8556,0.1448)
(0.8591,0.1487)
(0.8626,0.1526)
(0.8660,0.1565)
(0.8695,0.1605)
(0.8729,0.1644)
(0.8764,0.1684)
(0.8798,0.1724)
(0.8833,0.1764)
(0.8867,0.1805)
(0.8902,0.1845)
(0.8936,0.1886)
(0.8971,0.1927)
(0.9006,0.1969)
(0.9040,0.2010)
(0.9075,0.2052)
(0.9109,0.2094)
(0.9144,0.2137)
(0.9178,0.2180)
(0.9213,0.2223)
(0.9247,0.2267)
(0.9282,0.2311)
(0.9316,0.2356)
(0.9351,0.2401)
(0.9385,0.2446)
(0.9420,0.2492)
(0.9455,0.2539)
(0.9489,0.2586)
(0.9524,0.2634)
(0.9558,0.2683)
(0.9593,0.2732)
(0.9627,0.2783)
(0.9662,0.2834)
(0.9696,0.2887)
(0.9731,0.2941)
(0.9765,0.2996)
(0.9800,0.3053)
(0.9805,0.3062)
(0.9810,0.3070)
(0.9815,0.3079)
(0.9820,0.3087)
(0.9825,0.3096)
(0.9830,0.3104)
(0.9835,0.3113)
(0.9840,0.3122)
(0.9845,0.3131)
(0.9850,0.3140)
(0.9855,0.3149)
(0.9860,0.3158)
(0.9865,0.3167)
(0.9870,0.3176)
(0.9875,0.3185)
(0.9880,0.3194)
(0.9885,0.3204)
(0.9890,0.3213)
(0.9895,0.3223)
(0.9900,0.3232)
(0.9905,0.3242)
(0.9910,0.3252)
(0.9915,0.3262)
(0.9920,0.3272)
(0.9925,0.3282)
(0.9930,0.3292)
(0.9935,0.3303)
(0.9940,0.3313)
(0.9945,0.3324)
(0.9950,0.3335)
(0.9955,0.3346)
(0.9960,0.3357)
(0.9965,0.3369)
(0.9970,0.3381)
(0.9975,0.3393)
(0.9980,0.3406)
(0.9985,0.3419)
(0.9990,0.3433)
(0.9995,0.3448)
(1.0000,0.3466)
};
\addplot[orange!85!black,line width=1pt] coordinates {
(0.7071,0.0000)
(0.7106,0.0018)
(0.7140,0.0038)
(0.7175,0.0059)
(0.7209,0.0080)
(0.7244,0.0101)
(0.7278,0.0124)
(0.7313,0.0146)
(0.7347,0.0170)
(0.7382,0.0194)
(0.7417,0.0218)
(0.7451,0.0243)
(0.7486,0.0268)
(0.7520,0.0294)
(0.7555,0.0320)
(0.7589,0.0347)
(0.7624,0.0374)
(0.7658,0.0402)
(0.7693,0.0430)
(0.7727,0.0458)
(0.7762,0.0487)
(0.7796,0.0516)
(0.7831,0.0546)
(0.7866,0.0576)
(0.7900,0.0607)
(0.7935,0.0638)
(0.7969,0.0669)
(0.8004,0.0701)
(0.8038,0.0733)
(0.8073,0.0766)
(0.8107,0.0799)
(0.8142,0.0833)
(0.8176,0.0867)
(0.8211,0.0901)
(0.8246,0.0936)
(0.8280,0.0972)
(0.8315,0.1007)
(0.8349,0.1043)
(0.8384,0.1080)
(0.8418,0.1117)
(0.8453,0.1154)
(0.8487,0.1192)
(0.8522,0.1230)
(0.8556,0.1269)
(0.8591,0.1308)
(0.8626,0.1348)
(0.8660,0.1388)
(0.8695,0.1428)
(0.8729,0.1469)
(0.8764,0.1510)
(0.8798,0.1552)
(0.8833,0.1594)
(0.8867,0.1637)
(0.8902,0.1680)
(0.8936,0.1724)
(0.8971,0.1768)
(0.9006,0.1812)
(0.9040,0.1857)
(0.9075,0.1902)
(0.9109,0.1948)
(0.9144,0.1994)
(0.9178,0.2041)
(0.9213,0.2088)
(0.9247,0.2135)
(0.9282,0.2184)
(0.9316,0.2232)
(0.9351,0.2282)
(0.9385,0.2332)
(0.9420,0.2383)
(0.9455,0.2435)
(0.9489,0.2487)
(0.9524,0.2540)
(0.9558,0.2595)
(0.9593,0.2650)
(0.9627,0.2706)
(0.9662,0.2763)
(0.9696,0.2822)
(0.9731,0.2882)
(0.9765,0.2944)
(0.9800,0.3008)
(0.9805,0.3018)
(0.9810,0.3027)
(0.9815,0.3037)
(0.9820,0.3046)
(0.9825,0.3056)
(0.9830,0.3066)
(0.9835,0.3075)
(0.9840,0.3085)
(0.9845,0.3095)
(0.9850,0.3105)
(0.9855,0.3115)
(0.9860,0.3125)
(0.9865,0.3135)
(0.9870,0.3145)
(0.9875,0.3156)
(0.9880,0.3166)
(0.9885,0.3176)
(0.9890,0.3187)
(0.9895,0.3198)
(0.9900,0.3208)
(0.9905,0.3219)
(0.9910,0.3230)
(0.9915,0.3241)
(0.9920,0.3252)
(0.9925,0.3264)
(0.9930,0.3275)
(0.9935,0.3287)
(0.9940,0.3298)
(0.9945,0.3310)
(0.9950,0.3322)
(0.9955,0.3335)
(0.9960,0.3347)
(0.9965,0.3360)
(0.9970,0.3373)
(0.9975,0.3387)
(0.9980,0.3400)
(0.9985,0.3415)
(0.9990,0.3430)
(0.9995,0.3446)
(1.0000,0.3466)
};
\addplot[black!60,dashed,line width=.7pt] coordinates {
(0.7071,0.0000)
(0.7106,0.0018)
(0.7140,0.0038)
(0.7175,0.0058)
(0.7209,0.0078)
(0.7244,0.0099)
(0.7278,0.0121)
(0.7313,0.0143)
(0.7347,0.0166)
(0.7382,0.0189)
(0.7417,0.0213)
(0.7451,0.0237)
(0.7486,0.0262)
(0.7520,0.0287)
(0.7555,0.0312)
(0.7589,0.0338)
(0.7624,0.0365)
(0.7658,0.0392)
(0.7693,0.0419)
(0.7727,0.0447)
(0.7762,0.0475)
(0.7796,0.0503)
(0.7831,0.0532)
(0.7866,0.0562)
(0.7900,0.0592)
(0.7935,0.0622)
(0.7969,0.0652)
(0.8004,0.0684)
(0.8038,0.0715)
(0.8073,0.0747)
(0.8107,0.0780)
(0.8142,0.0812)
(0.8176,0.0846)
(0.8211,0.0879)
(0.8246,0.0913)
(0.8280,0.0948)
(0.8315,0.0983)
(0.8349,0.1018)
(0.8384,0.1054)
(0.8418,0.1090)
(0.8453,0.1127)
(0.8487,0.1164)
(0.8522,0.1201)
(0.8556,0.1239)
(0.8591,0.1278)
(0.8626,0.1316)
(0.8660,0.1356)
(0.8695,0.1395)
(0.8729,0.1435)
(0.8764,0.1476)
(0.8798,0.1517)
(0.8833,0.1558)
(0.8867,0.1601)
(0.8902,0.1643)
(0.8936,0.1686)
(0.8971,0.1729)
(0.9006,0.1773)
(0.9040,0.1818)
(0.9075,0.1863)
(0.9109,0.1908)
(0.9144,0.1954)
(0.9178,0.2001)
(0.9213,0.2049)
(0.9247,0.2097)
(0.9282,0.2145)
(0.9316,0.2195)
(0.9351,0.2245)
(0.9385,0.2296)
(0.9420,0.2348)
(0.9455,0.2400)
(0.9489,0.2454)
(0.9524,0.2509)
(0.9558,0.2564)
(0.9593,0.2621)
(0.9627,0.2679)
(0.9662,0.2739)
(0.9696,0.2799)
(0.9731,0.2862)
(0.9765,0.2926)
(0.9800,0.2992)
(0.9805,0.3002)
(0.9810,0.3012)
(0.9815,0.3022)
(0.9820,0.3032)
(0.9825,0.3042)
(0.9830,0.3052)
(0.9835,0.3062)
(0.9840,0.3072)
(0.9845,0.3082)
(0.9850,0.3093)
(0.9855,0.3103)
(0.9860,0.3114)
(0.9865,0.3124)
(0.9870,0.3135)
(0.9875,0.3145)
(0.9880,0.3156)
(0.9885,0.3167)
(0.9890,0.3178)
(0.9895,0.3189)
(0.9900,0.3200)
(0.9905,0.3211)
(0.9910,0.3223)
(0.9915,0.3234)
(0.9920,0.3246)
(0.9925,0.3257)
(0.9930,0.3269)
(0.9935,0.3281)
(0.9940,0.3293)
(0.9945,0.3306)
(0.9950,0.3318)
(0.9955,0.3331)
(0.9960,0.3344)
(0.9965,0.3357)
(0.9970,0.3371)
(0.9975,0.3385)
(0.9980,0.3399)
(0.9985,0.3414)
(0.9990,0.3429)
(0.9995,0.3446)
(1.0000,0.3466)
};
\end{axis}
\begin{scope}[shift={(\figW+\figG,0)},
  every node/.style={anchor=west,font=\footnotesize,inner sep=0pt}]
\draw[black!55,line width=.4pt] (0,0) rectangle (\figK,\figH);
\draw[blue!65!black,line width=1pt] (9pt,.875*\figH) -- (25pt,.875*\figH);
\node at (31pt,.875*\figH) {static cost};
\draw[orange!85!black,line width=1pt] (9pt,.625*\figH) -- (25pt,.625*\figH);
\node at (31pt,.625*\figH) {adaptive cost};
\draw[black!60,dashed,line width=.7pt] (9pt,.375*\figH) -- (25pt,.375*\figH);
\node at (31pt,.375*\figH) {lower bound};
\fill[orange!85!black,fill opacity=.20] (9pt,.125*\figH-3.5pt) rectangle (25pt,.125*\figH+3.5pt);
\node at (31pt,.125*\figH) {adaptive saving};
\end{scope}
\end{tikzpicture}
\caption{Shared randomness per coordinate at $\rho=3/4$, from the
threshold $p_0=\sqrt{1/2}$, where it vanishes, to exact sampling. The
static entropy cost equals the support cost; the adaptive entropy cost
lies below it by the shaded band and stays close to the lower bound
$p\,B_\rho(p)$ of Theorem~\ref{thm:shannon-frontiers}.}
\label{fig:pair-frontier}
\end{figure}
\paragraph{Explicit curves for $\rho\ge2/3$.}
When $2/3\le\rho<1$, batches of size one and two are optimal: pairs
while the fraction $u$ of unused labels exceeds $2\rho-1$, and single
draws afterwards. A pair splits the unused labels into two halves,
and each of its coordinates prefers one half with bias
$\theta(u)\in[0,1]$, which costs
$I(\theta)=\log2-h_2\bigl(\tfrac{1+\theta}2\bigr)$ nats per
coordinate and gives $c_2=1-\theta^2$. With $\beta$ nats per
coordinate, the optimal error for the support and static entropy
costs is $1-e^{-J_\rho(\beta)}$, where
\begin{equation}
 J_\rho(\beta)=\inf_\theta\biggl\{\,\frac12\int_{2\rho-1}^1
 \frac{1-\theta(u)^2}{u}\,du\;:\;
 \int_{2\rho-1}^1I\bigl(\theta(u)\bigr)\,du\le\beta\biggr\}
 \label{eq:pair-main}
\end{equation}
is the least collision exponent within the information budget
(Theorem~\ref{pair:domain}). The optimal bias solves the Curie--Weiss
equation $\operatorname{arctanh}\theta=t\theta$ at the local multiplier
$t$, so pairs are uncoordinated early and increasingly coordinated as
labels are used up. Optimality among all batch sizes follows from
$K_b(t)\ge(b-1)K_2(bt/2)$, and Theorem~\ref{pair:adaptive} gives the
adaptive curve. For $\sqrt{2\rho-1}<p<1$ we have
$p\,B_\rho(p)<C^{\rm ad}_\rho(p)<C^{\rm st}_\rho(p)=B_\rho(p)$: here
$B_\rho$ is convex, so randomizing between overlaps does not help,
while postponing shared draws does. At $\rho=3/4$ the relative saving
of the adaptive model falls from about $29\%$ near the threshold to
$18\%$ at overlap $0.8$ and $8\%$ at $0.9$
(Figure~\ref{fig:pair-frontier}).

\section{Sampling without replacement and open problems}
\label{sec:injections}

Since the analysis tracks only the fraction of unused labels, it
extends to injections of $[k]$ into $[N]$. Let
$F_{N,k,R}=\max_{b\le k}L(N,b)\,F_{N-b,k-b,R-1}$ as
in~\eqref{eq:capacity}, and $\Delta_{N,k,R}=\log(N)_k-\log
F_{N,k,R}$.

\begin{theorem}[Injections]\label{thm:injections}
Let $1\le R\le k\le N$, with target $P_{N,k}$ uniform on injections.
\begin{enumerate}[label=\textup{(\roman*)},leftmargin=*]
\item The four exact costs lie between $\Delta_{N,k,R}$ and
$\log\binom Nk+\Delta(k,R)+O(1)$. If $k/N\to1$ and
$R/N\to\rho\in(0,1)$, they are $\Phi(\rho)N+o(N)$, exactly as for
permutations.
\item If $k/N\to\alpha\in(0,1]$ and $R=o(k)$, then for fixed
$0\le\varepsilon<1$ the support and entropy costs are asymptotic to
$k\log(k/R)$ and $(1-\varepsilon)\,k\log(k/R)$.
\item Let $k/N\to\alpha\in(0,1)$ and $R/N\to\rho\in(0,\alpha)$. Then
the support cost per label in both models is $B_{\alpha,\rho}(p)$ for
$0<p<1$, given by~\eqref{eq:main-dual} with the integral over
$u\in[1-\alpha,1]$, and
$B^\tau_\varepsilon(N,k,R)/N\to B_{\alpha,\rho}(1-\varepsilon)$ for
fixed $0<\varepsilon<1$; at $p=1$, the least costs per label
attainable along sequences with error tending to zero are
$\alpha\,\Phi(\rho/\alpha)$. Parts \textup{(i)} and \textup{(ii)} of
Theorem~\ref{thm:shannon-frontiers} hold with $B_{\alpha,\rho}$, $[1-\alpha,1]$, and $s_\pi(1-\alpha)$
in place of $B_\rho$, $[0,1]$, and $s_\pi(0)$; the adaptive entropy cost is strictly smaller than the
static one for $p_0(\alpha,\rho)<p<1$, where the threshold
$p_0(\alpha,\rho)\in(0,1)$ takes the value $\sqrt{1-2\alpha+2\rho}$ for every
$\rho\ge\max\{\alpha/2,\alpha-1/3\}$.
\end{enumerate}
\end{theorem}

The function $K_b$ thus governs sampling without replacement, with
permutations as the case $\alpha=1$
(Appendix~\ref{app:partial-injection-frontiers}). In each regime
the cost is the entropy of the target minus what conditionally
independent rounds can supply, and an error enters only through a
bound on total correlation. Both notions make sense for other systems
of constraints, such as perfect matchings of bipartite graphs, where
the one-round bound already applies, proper colorings, and Latin
squares.

The change of measure applies to any finite uniform target, and the
question there is which single-batch bounds admit matching
constructions without the symmetry of $S_n$. In the permutation
model, open questions are the order of the saving for fixed $R\ge3$,
the second-order term of the entropy cost in two rounds, the values
of $\rho$ for which $B_\rho$ is convex, and explicit adaptive curves
for $\rho<2/3$; for injections at fixed density $0<\alpha<1$, the
first-order exact cost.
\label{main:end}\par\pagebreak
\clearpage
\phantomsection
\addcontentsline{toc}{section}{References}
\newcommand{\etalchar}[1]{$^{#1}$}

\clearpage
\phantomsection
\addcontentsline{toc}{section}{Appendices}
\addtocontents{toc}{\protect\setcounter{tocdepth}{0}}
\appendix
\setlength{\abovedisplayskip}{9pt plus 3pt minus 5pt}
\setlength{\belowdisplayskip}{9pt plus 3pt minus 5pt}
\section{Exact sampling and maximal private support}
\label{ex:section}

This appendix proves Theorem~\ref{thm:exact}. Both the lower bound
and the sampler rest on one counting fact. The coordinates of a batch
are written independently given the history and the shared
randomness, so a batch that never produces a collision assigns
pairwise disjoint sets of unused labels to its coordinates. The
largest possible support of such a batch is therefore an integer
partition problem. For integers $m\ge b\ge1$, write $m=ub+v$, where
$0\le v<b$, and define
\[
 L(m,b)=u^{b-v}(u+1)^v.
\]
This is the maximum product of $b$ positive integers with sum at most
$m$. The maximizing parts use all $m$ units and differ by at most one:
transferring a unit between two parts that differ by at least two
strictly increases their product.

\subsection{Maximal private support and exact lower bounds}

A private-coin protocol uses no shared randomness. The largest support of
a private-coin protocol that never produces a collision obeys a one-batch
recurrence, and its logarithm bounds the entropy that an exact
sampler generates privately. Shared randomness must supply the rest
of the entropy of a uniform permutation, which gives the lower bound
in Theorem~\ref{thm:exact}.

\begin{lemma}[Maximal private support]
\label{ex:capacity}
Let $F(n,R)$ be the maximum support size of a private-coin protocol that
generates a permutation of $[n]$ in at most $R$ rounds and whose
output is always collision-free.
For $n\ge1$ and $R\ge1$,
\begin{equation}
F(n,R)=\max_{1\le b\le n}L(n,b)F(n-b,R-1),
\label{ex:capacity-recursion}
\end{equation}
with $F(0,R)=1$ and $F(n,0)=0$ for $n>0$.
A protocol attaining $F(n,R)$ can be uniform on its support. For
$1\le R\le n$,
\begin{equation}
F(n,R)=\max_{\substack{b_1+\cdots+b_R=n\\b_i\ge1}}
\prod_{i=1}^RL(m_i,b_i),\qquad m_i=\sum_{j=i}^Rb_j.
\label{ex:schedule-capacity}
\end{equation}
\end{lemma}
\begin{proof}
At a history with $n$ unused labels, the coordinates of every batch
that is almost surely collision-free have pairwise disjoint supports
contained in those labels. A batch of size $b$ consequently has at
most $L(n,b)$ possible outputs, and each output has at most
$F(n-b,R-1)$ continuations. This bound applies to every continuation,
including those whose batch sizes depend on earlier outputs.
Conversely, split the unused labels into $b$ parts whose sizes differ
by at most one. Independent uniform draws from these parts, followed
by a relabeled maximizing protocol on the remaining labels, attain
this product and are uniform on their support.

The recurrence thus optimizes over compositions with at most $R$
parts. To obtain exactly $R$ parts, observe that
\begin{equation}
L(m,b)\le mL(m-1,b-1),\qquad b\ge2,
\label{ex:split}
\end{equation}
because deleting one part from a maximizing partition leaves $b-1$
parts of total size at most $m-1$, while the deleted part is at most
$m$. Replacing a batch of size $b\ge2$ by a singleton and a batch of
size $b-1$ therefore preserves or increases the support bound.
Repeating this operation yields exactly $R$ parts.
\end{proof}

A sequence of batch sizes is called a schedule. The next lemma
compares the product of the batch bounds along a schedule with a
falling factorial; its error terms enter the asymptotic computations
later in this appendix.

\begin{lemma}[Finite schedule corrections]
\label{ex:schedule-corrections}
For integers \(N\ge k\ge1\) and \(L\ge1\), let positive integers
\(b_1,\ldots,b_J\le L\) sum to \(k\), and put
\(m_j=N-\sum_{i<j}b_i\). Define
\[
 e(m,b)=b\log(m/b)-\log L(m,b),\quad
 E=\sum_j e(m_j,b_j),\quad
 \Gamma=\sum_j[b_j\log m_j-\log(m_j)_{b_j}].
\]
Then
\begin{equation}
 \log\frac{(N)_k}{\prod_j L(m_j,b_j)}
 =\sum_j b_j\log b_j-\Gamma+E,
 \qquad 0\le e(m,b)\le\frac{b^3}{2m^2},
 \label{ex:schedule-correction-identity}
\end{equation}
\[
 0\le\Gamma\le\frac{(L-1)k}{N-k+1},\qquad
 0\le E\le\frac{L^2}{2}\sum_{h=N-k+1}^{N}h^{-2}<L^2.
\]
\end{lemma}

\begin{proof}
The identity follows by telescoping the falling factorials. If
$m/b=q+\theta$, with $q\ge1$ and $0\le\theta<1$, concavity of the
logarithm and its second derivative give
\[
 0\le e(m,b)\le \frac{b\theta(1-\theta)}{2q^2}
 \le\frac{b}{8q^2}\le\frac{b^3}{2m^2}.
\]
Number the emitted labels downward from $N$ in order of emission, so
that batch $j$ receives the integers in $(m_j-b_j,m_j]$. These
intervals partition the numbers used. Hence
\[
 \sum_j\frac{b_j^3}{m_j^2}
 \le L^2\sum_{h=N-k+1}^{N}h^{-2},
\]
and the bound on $E$ follows from $\sum_{h\ge1}h^{-2}<2$.
Finally, each term in the expansion of $\Gamma$ satisfies
\[
 -\log(1-i/m_j)\le\frac{i}{m_j-i}
 \le\frac{L-1}{N-k+1},\qquad 0\le i<b_j.
\]
There are $k$ such terms, which proves the remaining estimate.
\end{proof}

An exact sampler is a protocol whose output is a uniform permutation.
As in Section~\ref{sec:model}, the support cost of a protocol is the
largest sum, over its executions, of the logarithmic support sizes of
its shared draws, and its entropy cost $C_H$ is the sum of the
conditional entropies of these draws given the history. In the static
model all shared randomness is drawn before the first round; in the
adaptive model it may be drawn at every stage.

\begin{proposition}[Exact resource lower bounds]
\label{ex:exact-lower}
For $1\le R\le n$, put $\Delta(n,R)=\log(n!/F(n,R))$. In both the
static and the adaptive model, every $R$-round exact sampler has
entropy cost at least $\Delta(n,R)$ and support cost at least
$\log\lceil n!/F(n,R)\rceil$.
\end{proposition}
\begin{proof}
Condition on the history and the shared draw preceding each private
batch, of size $b_t$ on $m_t$ unused labels. Exactness forces the
supports of its coordinates to be pairwise disjoint subsets of the
unused labels, so its entropy is at most $\log L(m_t,b_t)$. Along
every execution, the recurrence for $F$ bounds the sum of these terms
by $\log F(n,R)$. The chain rule for the entropy of the transcript
therefore gives
\[
\log n!\le C_H+\E\sum_t\log L(m_t,b_t)
\le C_H+\log F(n,R).
\]
The expected shared entropy is at most the largest sum of
logarithmic support sizes along an execution. For a finite protocol,
the exponential of this maximum is an integer. Thus the same lower
bound yields the stated ceiling for the support cost in both models.
\end{proof}

The upper bound is constructive.
For a composition $\mathbf b=(b_1,\ldots,b_R)$, write
\[
 P(\mathbf b)=\prod_{i=1}^R L(m_i,b_i),\qquad
 m_i=\sum_{j=i}^R b_j,
\]
and let $F_\uparrow(n,R)$ be the maximum of $P(\mathbf b)$ over
nondecreasing compositions of $n$ into $R$ positive parts. We first
construct exact samplers for such schedules, and afterwards compare
$F_\uparrow(n,R)$ with $F(n,R)$.

\subsection{Balanced partitions and finite common sources}

The exact sampler shares a uniform partition of the unused labels
into parts of nearly equal size, and each coordinate of a batch draws
privately from its own part. The partition of the remaining labels is
again uniform and serves the next batch; when the batch size grows,
a small additional source converts it into a finer partition.

For integers $M\ge0$ and $b\ge1$, write $M=ub+v$, where
$0\le v<b$, and put
\[
 D(M,b)=(u!)^b(u+1)^v,\qquad
 A(M,b)=\frac{M!}{D(M,b)}.
\]
Here $A(M,b)$ counts the ordered partitions of $M$ labels into $b$
parts in which the first $v$ parts have size $u+1$ and the remaining
parts have size $u$; such partitions are called balanced. This
convention fixes the positions of the larger parts and gives
$A(0,b)=1$.

\begin{lemma}[Partition recycling]
\label{ex:partitions}
Let $m\ge b\ge1$. Choose a uniform balanced partition of $m$ labels
into $b$ parts and draw one independent uniform label from each
part. The resulting tuple is a uniform ordered injection.
Conditional on this tuple, the residual partition is uniform among
its $A(m-b,b)$ compatible balanced partitions, and
\begin{equation}
\frac{A(m,b)}{A(m-b,b)}
=\frac{(m)_b}{L(m,b)}.
\label{ex:partition-identity}
\end{equation}
\end{lemma}

\begin{proof}
Each injective tuple is compatible with exactly $A(m-b,b)$ initial
partitions. Each such partition has probability $1/A(m,b)$ and
assigns the tuple conditional probability $1/L(m,b)$.
Cancelling the factorials of the part sizes proves
\eqref{ex:partition-identity}; the tuple consequently has probability
$1/(m)_b$. The same calculation shows that all compatible residual
partitions have equal conditional probability.
\end{proof}

When the number of parts increases, the residual partition must be
converted into a new uniform partition. We use an interval construction
for random generation \cite{HanHoshi1997}, together with a finite
common source: one random variable from which every conditional
distribution arising in that construction is generated by a
deterministic map. The common-source viewpoint is also central to
minimum entropy coupling \cite{Li2021}.

\begin{lemma}[A finite common source]
\label{ex:common-source}
Let $S\ge1$ and $d\ge2$ be integers. Every finite family of
probability laws, each supported on at most $d$ outcomes with masses
in $S^{-1}\mathbb Z$, can be realized as deterministic functions of one
finite random variable whose support has size at most
$2+\log S/\log(d/(d-1))$.
\end{lemma}
\begin{proof}
The source is built greedily, one atom at a time.
Suppose that the remaining mass in each law is $r$. Choose an atom
of largest remaining mass in each law, and let $w$ be the smallest of
these masses. Introduce a source atom of mass $w$, mapping it to the
chosen atom in each law, and subtract $w$ from those atoms.
Since $w\ge r/d$, the remaining mass contracts by a factor at most
$1-1/d$. At least one outcome is exhausted at every step, so the
procedure terminates. All residual masses remain multiples of $1/S$;
after $j$ steps, their integer total is at most $S(1-1/d)^j$.
The total vanishes once this bound is below one, which gives the
support bound.
\end{proof}

\begin{lemma}[Uniform conversion]
\label{ex:converter}
Let $K\ge A\ge1$ be integers, and set $\chi=K/A$ and
$a=A/\gcd(A,K)$. There is a deterministic map that converts a uniform
variable on $[A]$, together with an independent finite source, into a
uniform variable on $[K]$. The additional source can have support at
most $\chi+1+\log(2a)/\log(3/2)$. If $\chi$ is an integer, a uniform
source on $[\chi]$ suffices.
\end{lemma}
\begin{proof}
Write $g=\gcd(A,K)$ and $k=K/g$. Identify the input with a uniform
pair in $[g]\times[a]$ and retain its first coordinate. It remains to
convert the second coordinate from a uniform variable on $[a]$ to a
uniform variable on $[k]$.

Partition $[0,ak)$ into $a$ rows of length $k$ and, separately,
into $k$ cells of length $a$. The intersection lengths, divided by
$k$, give a transition kernel with the required uniform marginals.
When $\chi=k/a$ is an integer, each row consists of $\chi$ complete
cells. Otherwise, put $q=\lfloor\chi\rfloor$ and remove $q-1$
complete cells from every row. Each row then meets at most three
remaining cells, with total integer length
\[
 S=k-(q-1)a\in(a,2a).
\]
Lemma~\ref{ex:common-source}, with $d=3$, realizes all the residual
row distributions from one source of support at most
$2+\log S/\log(3/2)$. Add $q-1$ source atoms of mass $a/k=1/\chi$
for the removed cells and give the residual source total mass $S/k$.
The combined support has size at most $q+1+\log S/\log(3/2)$, which
is bounded as claimed.
Combining the output with the retained $[g]$ coordinate completes the
conversion.
\end{proof}

\subsection{Arithmetic of balanced partition counts}

Adding one part multiplies the number of balanced partitions by a
ratio that is exponentially large and has a small reduced
denominator. The excess support of all conversions is therefore
bounded by a convergent series, which yields a static exact sampler
on every nondecreasing schedule.

\begin{lemma}[Adjacent partition ratios]
\label{ex:factorial-ratio}
Let \(M\ge d+1\), \(s=\lfloor M/(d+1)\rfloor\), and
\(\chi_d=A(M,d+1)/A(M,d)\).
Then
\[
\chi_d\ge\frac{(s+1)^s}{s!}\ge2^s,
\]
and the denominator of \(\chi_d\) in lowest terms divides \(s!\).
\end{lemma}

\begin{proof}
The identity
\[
 \frac{A(M+1,d)}{A(M,d)}
 =\frac{M+1}{\lfloor M/d\rfloor+1}
\]
shows that $\chi_d$ is nondecreasing in $M$. For the lower bound it
therefore suffices to take $M=s(d+1)$. In that case the part sizes of
a balanced partition into $d$ parts have the form $s+t_i$, where
$\sum_i t_i=s$. Consequently,
\[
 \chi_d=\frac{\prod_i(s+t_i)!}{(s!)^{d+1}}
 \ge\frac{(s+1)^s}{s!}\ge2^s,
\]
where the last inequality follows from
$s!\le((s+1)/2)^s$.
For the denominator assertion, write
$M=s(d+1)+r=vd+r_0$, with $0\le r\le d$ and $0\le r_0<d$.
Cancelling factorials gives $(s+1)^s/s!$ when $v=s$, and otherwise gives
\[
 \chi_d=\frac{(s+1)^{d-r}(\prod_{j=s+2}^{v}j)^d(v+1)^{r_0}}{s!}.
\]
In either case the reduced denominator divides $s!$.
\end{proof}

\begin{theorem}[Exact sampling on a nondecreasing schedule]
\label{ex:partition-realization}
For a nondecreasing positive integer composition $\mathbf b$ of $n$,
put $\beta(\mathbf b)=\log(n!/P(\mathbf b))$. There is a static exact
sampler $Q$ on this schedule whose support cost satisfies
\[
B(Q)<\beta(\mathbf b)+c_{\mathrm{conv}},
\qquad c_{\mathrm{conv}}=\sum_{s=2}^\infty\log\left(1+\frac{4+3s^2}{2^s}\right)<10.
\]
\end{theorem}
\begin{proof}
We describe the sampler, verify that it is exact, and then bound the
support of its seed.
The shared seed, drawn before the first round, consists of a uniform
balanced $b_1$-partition of the labels and independent sources for
all subsequent increases in batch size. Each batch draws one
independent uniform label from each part of the current partition,
and the residual partition is retained. At an increase $b\to c$,
Lemma~\ref{ex:converter} converts the uniform index of this partition
in $[A(M,b)]$ to a uniform index in $[A(M,c)]$, where $M$ is the
number of remaining labels.

By Lemma~\ref{ex:partitions}, averaged over the seed, the residual
partition remains uniform conditional on the output prefix. The
source for the next conversion is independent of that prefix and of
the current partition, because earlier batches use only the preceding
sources. For each fixed seed and output prefix, the parts are
fixed and the batch law is a product. Thus the construction is a
static exact sampler.

For the support bound, set $K_t=A(m_t,b_t)$ and
$\Lambda_t=A(m_t-b_t,b_t)$, and call a transition between
consecutive batches strict when the batch size increases. Equal-size
transitions satisfy $K_{t+1}=\Lambda_t$, while $\Lambda_R=1$.
Multiplying \eqref{ex:partition-identity} over the schedule gives
\begin{equation}\label{ex:partition-realization-telescope}
\log K_1+\sum_{\text{strict transitions }t}
\log(K_{t+1}/\Lambda_t)=\beta(\mathbf b).
\end{equation}
The seed has logarithmic support $\log K_1$ plus the logarithmic
supports of the conversion sources, so it remains to bound the excess
of each source over the logarithm of its ratio $K_{t+1}/\Lambda_t$.
At a strict transition $b\to c$, put
\[
 s_d=\left\lfloor\frac{M}{d+1}\right\rfloor,\qquad
 S=\sum_{d=b}^{c-1}s_d,\qquad
 \chi=\frac{A(M,c)}{A(M,b)}.
\]
By Lemma~\ref{ex:factorial-ratio}, the reduced denominator $a$ of
$\chi$ divides $\prod_ds_d!$, and hence divides $S!$, while
$\chi\ge2^S$. In particular, $\log a\le S^2$.
Using $\log(3/2)>1/3$ and $\log2<1$ in
Lemma~\ref{ex:converter}, the logarithmic support of the conversion
source exceeds $\log\chi$ by less than
\[
 \log\left(1+\frac{4+3S^2}{2^S}\right).
\]
A transition into the final batch has ratio $D(M,b)$, an integer, and
therefore has zero excess. At every other strict transition, at least
two batches of size at least $c$ remain. Thus
$S\ge s_*:=\lfloor M/c\rfloor\ge2$. Before the next strict transition,
the number of remaining labels decreases to $M'\le M-c$ and the new
batch size increases to $c'>c$, so
\[
 \left\lfloor M'/c'\right\rfloor\le s_*-1.
\]
The function $(4+3s^2)/2^s$ decreases at integers $s\ge2$, since
$3s^2-6s+1>0$. Each excess is therefore bounded by the summand of
the series indexed by $s_*$, and the indices of distinct strict
transitions are distinct. The excesses sum to at most
$c_{\mathrm{conv}}$; combining this with
\eqref{ex:partition-realization-telescope} proves the support bound.

For the numerical bound, set
\[
 h_s=\frac{4+3s^2}{2^s},\qquad
 P_N=\prod_{s=2}^N(1+h_s),\qquad
 T_N=\sum_{s>N}h_s=\frac{4+3(N^2+4N+6)}{2^N}.
\]
If $T_N<1$, then $e^{c_{\mathrm{conv}}}\le P_N/(1-T_N)$.
The exact rational estimates
$T_{12}=299/2048$ and $P_{12}/(1-T_{12})<(8/3)^{10}<e^{10}$
give $c_{\mathrm{conv}}<10$.
\end{proof}

\subsection{A uniform comparison with nondecreasing schedules}

The partition construction applies to nondecreasing batch sizes,
whereas the maximal private support is a maximum over all
compositions. Restricting to nondecreasing schedules changes the
maximal private entropy by less than $28$, uniformly in the number of
labels and rounds. The argument combines an ordering property of
optimal integer schedules with a rounding procedure for a continuous
optimization problem.

Let $\psi(x)$ denote the piecewise linear interpolation of $\log u$
at the positive integers, and set
\[
 g(m,b)=b\psi(m/b),\qquad
 c_u=\log(1+1/u),\qquad p_u=\log u-u c_u.
\]
The function $g$ is continuous, jointly concave, and homogeneous, with
$g(m,b)=\log L(m,b)$ at integer arguments. It is linear on each range
$u\le m/b\le u+1$ between consecutive positive integers, where
\begin{equation}
g(m,b)=c_um+p_ub.
\label{ex:linear-piece}
\end{equation}
Both $p_u$ and $h(u)=u\log(1+1/u)$ are increasing. The
\emph{remaining mass} at a batch is the total size of that batch and
of all later batches. The \emph{reward} of a composition with
positive real parts is the sum of $g(m,b)$ over its batches, where
$b$ is the size of a batch and $m$ its remaining mass. For an integer
composition, the reward is the logarithm of \(P(\mathbf b)\).

\begin{lemma}[Ordering and two-size sorting]
\label{ex:ordering}
Every integer composition maximizing \(P(\mathbf b)\) satisfies
\[
b_i\le b_j+1\qquad(i<j).
\]
For a sequence $W$ with entries in $\{q,q+1\}$ and a remaining
tail of mass $T$, let $W_\uparrow$ be its increasing rearrangement
and let $P(W;T)$ be the product of the maximal supports of its
batches. If \(T=(q+1)t+a\), \(0\le a\le q\), then
\[
\log P(W;T)-\log P(W_\uparrow;T)\le\frac{a}{q+1}<1.
\]
\end{lemma}

\begin{proof}
The ordering follows from an exchange argument on adjacent parts, and
the two-size bound from a telescoping potential.

Consider adjacent real parts $a>b$ of sum $S$, the first of which
has remaining mass $m$, and let
$J(a)=g(m,a)+g(m-a,S-a)$ be the reward of the pair as a function of
$a$, with $m$ and $S$ fixed. With $u=\lfloor m/a\rfloor$ and
$v=\lfloor(m-a)/b\rfloor$, the identity
$m/a-(m-a)/b=1-(a-b)(m-a)/(ab)\le1$ gives $u\le v+1$, hence
\[
 J'(a)=p_u-p_v-c_v\le p_{v+1}-p_v-c_v=h(v)-h(v+1)<0.
\]
Continuity extends this comparison across breakpoints. Transferring
one unit across an adjacent drop of at least two therefore increases
the reward. If the asserted ordering fails for a maximizing
composition, a violating pair of minimum separation must delimit a
block $(q+1,q,\ldots,q,q-1)$. Average adjacent entries of this block in
repeated left-to-right sweeps. Each averaging preserves the
nonincreasing order and mean $q$, and strictly increases the reward
whenever the two entries differ. It also decreases the sum of squared
deviations from $q$ by half the squared difference of that pair.
The sweeps consequently converge to the constant block with all
entries equal to $q$. Continuity then gives an integer composition
with larger reward, contradicting maximality.

For the two-size comparison, define
$\Psi_q(qv+r)=q\log(v!)+r\log(v+1)$ for $0\le r<q$.
A batch of size $q$ contributes the telescoping difference
$g(m,q)=\Psi_q(m)-\Psi_q(m-q)$. The additional contribution from a
batch of size $q+1$ is
\[
 E_q(m)=g(m,q+1)-\Psi_q(m)+\Psi_q(m-q-1).
\]
Writing $m=qk+r$ with $1\le r\le q$ gives
$E_q(m)=U_{q,r}(k)-e(m,q+1)$, where $e$ is the nonnegative
interpolation error from the lemma on finite schedule corrections
and
\[
 \begin{aligned}
 U_{q,r}(k)&=-(q+1)\log(1+1/q)
 +(q+1)\log(1+r/(qk))-r\log(1+1/k),\\
 U'_{q,r}(k)&=\frac{r(r-k-q-1)}{k(k+1)(qk+r)}<0.
 \end{aligned}
\]
For $k\le r$, one has
$E_q(qk+r)=-k\log(1+1/k)$, which decreases to $U_{q,r}(r)$.
For $k>r$, the bounds $E_q(qk+r)\le U_{q,r}(k)\le U_{q,r}(r)$
show that moving a larger batch earlier also gives no gain from this
initial range.

Index the batches of size $q+1$ from the end of the sequence. In the
increasing arrangement, their remaining masses are
$m_j^0=T+j(q+1)$; every other arrangement has
$m_j=m_j^0+qz_j$ for some $z_j\ge0$. The $\Psi_q$ terms telescope,
and the monotonicity of $U_{q,r}$ bounds each increase by the
interpolation error at $m_j^0$. Write
$m_j^0=(q+1)u+a$, where $u=t+j$. The range without gain identified
above is $u+a\le q$. For $u\ge s:=q+1-a$, the second-derivative bound
for logarithmic interpolation gives an error of at most
$a(q+1-a)/[2(q+1)u^2]$. Since these values of $u$ are distinct,
the total increase is at most
\[
 \frac{as}{2(q+1)}\sum_{u\ge s}u^{-2}
 \le\frac{as}{2(q+1)}\sum_{u\ge s}\frac2{u(u+1)}
 =\frac a{q+1}.
\]
\end{proof}

When the batch sizes are allowed to be positive reals, the maximal
reward is linear in the mass, and its slope obeys a scalar recursion.
Define
\begin{equation}
\lambda_1=0,\quad u_1=1,\qquad
\lambda_r=\max_{u\ge1}
\frac{\log u+(u-1)\lambda_{r-1}}u,
\label{ex:lambda}
\end{equation}
where, for $r\ge2$, the maximum is over positive integers $u$, and
$u_r$ denotes its smallest maximizer. The expression tends to
$\lambda_{r-1}$ as $u\to\infty$ and exceeds this limit for some
finite $u$, so the maximum is attained.

\begin{lemma}[Continuous policy and quotient runs]
\label{ex:policy}
The maximum of $\sum_i g(m_i,b_i)$ over positive real compositions
of $m$ into $r$ parts is $\lambda_rm$. It is attained by choosing
$b=m/u_r$ at each stage and applying the same rule to the remaining
mass and rounds.
For \(r\ge2\),
\[
u_2=3,\qquad
p_{u_r-1}\le\lambda_{r-1}\le p_{u_r},\qquad
u_r-u_{r-1}\in\{1,2\},\qquad r\le u_r\le2r-1.
\]
The policy has nondecreasing batch sizes.

Consider a maximal consecutive segment $s+2,s+3,\ldots,U$ of the
sequence $(u_r)$, beginning after a jump from $s$ to $s+2$.
We call this segment a run. For a quotient $u$ in the run, define
\[
\eta_u=\lambda_{r(u)-1}-p_{u-1},
\qquad \theta_u=(u-1)\eta_u,
\]
where $r(u)$ is the unique index with $u_{r(u)}=u$. The first
quantity measures the distance to the lower threshold for choosing
$u$. These quantities satisfy
\begin{align}
0\le \eta_{s+2}&\le d_s,\qquad
\theta_{u+1}=\theta_u+u d_{u-1},\qquad
0\le \theta_u<1, \label{ex:run-recurrence}\\
U+1&<e^2(s+3), \label{ex:run-ratio}\\
\eta_{s+2+j}&\le\frac{j+1}{2s^2}, \label{ex:run-end}
\end{align}
where \(d_v=h(v+1)-h(v)\).
For a complete run ending before a jump \(U\to U+2\), if \(s\ge16\),
its length satisfies \(U-s-1\ge s+1\).
\end{lemma}

\begin{proof}
The continuous problem is solved by induction on the number of
rounds. For $r=1$, the only batch contains all remaining mass and the
reward is zero. For $r\ge2$, the induction hypothesis leaves the
objective $g(m,b)+\lambda_{r-1}(m-b)$ for the first batch size $b$,
which is affine between consecutive points
$b=m/u$. After normalization by $m$, the difference between the
values at $u+1$ and $u$ is
$(\lambda_{r-1}-p_u)/[u(u+1)]$. This proves the recursion and the
threshold inequalities, and $p_2<0<p_3$ gives $u_2=3$.
Put $\delta_u=c_{u-1}-c_u=\log(u^2/(u^2-1))$, so
$p_u-p_{u-1}=u\delta_u$. The elementary derivative estimates
\begin{equation}
 h'(x)=\int_0^{1/x}\frac{t}{(1+t)^2}\,dt,
 \qquad\frac1{2(x+1)^2}\le h'(x)\le\frac1{2x(x+1)}
 \label{ex:h-derivative}
\end{equation}
follow by bounding the denominator for the lower estimate and
differentiating $a^2/[2(1+a)]-\int_0^a t/(1+t)^2\,dt$ for the upper.
For $v=u_{r-1}\ge3$, substitution of its thresholds gives
\[
 p_v+d_{v-1}\le\lambda_{r-1}\le p_{v+1}+d_v<p_{v+2},
\]
since $d_v\le1/(2v^2)<1/(v+2)\le(v+2)\delta_{v+2}$.
These inequalities give $u_r-u_{r-1}\in\{1,2\}$ and hence
$r\le u_r\le2r-1$. Along the resulting schedule, the ratio of the
next batch size to the current one is
$(u_r-1)/u_{r-1}\ge1$, which proves monotonicity.

Within a run, substituting the recursion gives
$\eta_{u+1}=d_{u-1}+(1-1/u)\eta_u$. At the entry of the run,
$0\le \eta_{s+2}\le d_s$; this includes the case $s=1$, where
$\eta_3=-p_2=d_1$.
The thresholds imply $\theta_u\le u(u-1)\delta_u<u/(u+1)<1$, and
integration of \eqref{ex:h-derivative} gives
\begin{equation}
 \frac1{2(u+1)}\le u d_{u-1}\le\frac1{2(u-1)}.
 \label{ex:run-increments}
\end{equation}
Consequently runs are finite and
$\tfrac12\log((U+1)/(s+3))\le
\tfrac12\sum_{u=s+2}^{U-1}(u+1)^{-1}<1$, which proves the ratio bound.
The recurrence and $d_{s+j}\le1/(2s^2)$ give \eqref{ex:run-end}.
If a complete run had length at most $s\ge16$, then $U\le2s+1$
and continuation of the recurrence would yield
\[
 \theta_{U+1}\le\frac1{2s}+\frac12\sum_{v=s+1}^{2s}v^{-1}
 \le\frac1{2s}+\tfrac12\log2<\tfrac12.
\]
A jump to $U+2$, however, requires
$\theta_{U+1}>U(U+1)\delta_{U+1}\ge U/(U+1)>1/2$, a contradiction.
\end{proof}

The comparison theorem applies the continuous policy to the tail of
a composition, where it must become an integer schedule with parts
bounded below. Let
$V(N,R,\ell)$ denote the maximum of $\sum_i g(m_i,b_i)$ over real
compositions of $N$ into $R$ parts, each at least the positive integer
$\ell$.

\begin{lemma}[Constrained monotone rounding]
\label{ex:rounding}
Let \(N,R,\ell\) be positive integers with \(N\ge R\ell\) and
\(R\le4\ell\). There is a nondecreasing integer
composition of $N$ into $R$ parts, each at least \(\ell\), with reward
strictly greater than \(V(N,R,\ell)-25\).
\end{lemma}

\begin{proof}
For $r\ge2$, define the loss relative to the continuous optimum by
\[
 \mathcal E_r(m,b)
 =\lambda_rm-g(m,b)-\lambda_{r-1}(m-b),
\]
and set $\mathcal E_1(m,m)=0$. Along a schedule, these losses
telescope to
$\lambda_RN-\sum_i g(m_i,b_i)$. We first treat the case in which
the continuous optimum respects the lower bound $\ell$.
Rounding down suffices near the beginning and end of the schedule;
within the intervening runs, almost constant batch sizes
keep the accumulated loss bounded. The lower constraint is treated
at the end of the proof.

\emph{Rounding down.} Suppose $N\ge\ell u_R$. With one round
remaining, the construction outputs all remaining mass and incurs
zero loss. At an integer state with $r\ge2$, put
$u=u_r$, $b=\lfloor m/u\rfloor$, and $x=m-ub$.
The next mass satisfies $(u-1)b+x\ge u_{r-1}b$, so successive
rounded batch sizes are nondecreasing and remain at least $\ell$.
Writing $v=\lfloor m/b\rfloor$ and integrating the derivative of
the objective from $b$ to $m/u$ gives
\begin{equation}
 \mathcal E_r(m,b)\le\frac xu\sum_{j=u}^v j\delta_j
 \le\frac xu(1+x/b)\frac{u}{u^2-1}
 \le\frac1{u+1}+\frac1b.
 \label{ex:floor-loss}
\end{equation}
Here $v-u\le x/b$, $x\le u-1$, and $j\delta_j\le j/(j^2-1)$.
Summed over the schedule, the reciprocal-batch terms contribute at
most $R/\ell\le4$.

\emph{Complete runs.} The terms $1/(u+1)$ are controlled by treating
an entire run $U,U-1,\ldots,s+2$, with $s\ge16$, at once.
Suppose its initial mass is $m=qU+x$, where $q\ge U$ and
$0\le x<U$, and put
\[
 J=\left\lfloor\frac{q+x}{s}\right\rfloor,\qquad
 D=(q+x)\bmod s,\qquad L=\lceil D/J\rceil.
\]
Thus $L=0$ when $D=0$. Begin with batches of the common size $q$.
Distributing an additional total mass $D$ over the final $L$ batches
will make the outgoing mass divisible by $s$.
Lemma~\ref{ex:policy} gives $U\ge2s+2$, so $J\ge2$ and
$L\le s-1<U-s-1$. For $L>0$, give these last batches the
nondecreasing increments
$D-(L-1)J,J,\ldots,J$. The outgoing mass is then
\begin{equation}
 q(s+1)+x-D=s(q+J),
 \label{ex:buffer-output}
\end{equation}
so the next run begins with zero remainder and the larger common
batch size $q+J\ge s$.

\emph{Run losses.} At the stage of the run with quotient $u$, write
$m=qu+x-E$, $b=q+c$, where $c$ is the increment of the current batch
and $E$ the sum of the preceding increments. The sign of $m-ub$
determines the linear piece of $g$.
If $m-ub\ge0$, the ratio of mass to batch size lies in $[u,u+1]$
since $m-ub\le x<U\le q\le b$, and the loss of the stage is
$(p_u-\lambda_{r-1})(m-ub)/u\le x/(u^2-1)$.
Equation~\eqref{ex:buffer-output} makes $x=0$ at every subsequent
processed run. The positive-remainder losses therefore total at most
$x/(s+1)<U/(s+1)<e^2(s+3)/(s+1)<9$.
If $m-ub<0$, the stage is among the last $L$. Writing $u=s+2+j$,
where $j$ incremented stages follow, the identities
$D=E+c+jJ$ and $q+x=sJ+D$ give
\begin{equation}
 m-(u-1)b=(s+j)(J-c)\ge0.
 \label{ex:buffer-ratio}
\end{equation}
The loss of the stage is therefore $\eta_u(b-m/u)$. By
\eqref{ex:run-end} and $(b-m/u)_+\le c+E/u$, the total of these
losses over the run is at most
\[
 \frac{LD}{2s^2}(1+L/s)
 <\frac12(1/J+1/s)(1+1/J+1/s).
\]
Suppose the first processed run has upper quotient $U_0\le\ell$.
All common batch sizes then remain at least $\ell$. The lower
boundaries of successive processed runs decrease by factors greater
than two, giving
\[
 \sum s<U_0\le\ell,\qquad
 \sum\frac1s\le\frac18,\qquad
 \sum\frac1J\le\sum\frac{2s}{q}\le2.
\]
Since $J\ge2$ and $s\ge16$, the total negative-remainder loss is
less than $(25/32)(2+1/8)=425/256<2$.

\emph{Joining the constructions.} If $u_R\le2^{14}$, rounding down
throughout gives total loss less than
$\log((2^{14}+1)/3)+4<14$.
For $u_R>2^{14}$, the inequalities
$u_R\le2R-1\le8\ell-1$ give $\ell>2048$.
Round down through the initial, possibly incomplete run and continue
until the first complete run with upper quotient $U_0\le\ell$.
The ratio bound for the preceding run gives
\[
 U_0+3>\min\{u_R+1,\ell+1\}/e^2.
\]
It follows that $U_0>253$ and that the lower boundary of this run
exceeds $28$, using $e^2<8$. The harmonic loss in the initial
rounded segment is at most
$\log((u_R+1)/(U_0+2))<\log(9e^2)<5$.
When the preceding run is complete, this follows from
$U_0+2>(\ell+1-e^2)/e^2$ and $u_R+1\le8\ell$;
when it is the initial run, $u_R+1$ takes the place of $\ell+1$.
At the start of this run, $q\ge\ell\ge U_0$, so the complete-run
construction applies. Continue it for all runs whose lower boundary
is at least $16$, then finish by rounding down. The largest quotient
in this final segment is below $18e^2$, so its harmonic loss is less
than $\log(18e^2)<5$. Equation~\eqref{ex:buffer-output} ensures that
the batch sizes remain nondecreasing at the transition.

The five contributions are the initial harmonic loss, the reciprocal
batch-size terms, the positive and negative remainders within runs,
and the final harmonic loss. Their total is less than
$5+4+9+2+5=25$. Every step preserves the total mass and the prescribed
number of rounds.

\emph{The lower constraint.} Joint concavity of $g$ gives concavity
of $V(m,r,\ell)$ in $m$. The nondecreasing continuous policy gives
$V(m,r,\ell)=\lambda_rm$ for $m\ge\ell u_r$; concavity then makes
its one-sided slopes below this threshold at least $\lambda_r$.
If $m<\ell u_r$, which forces $r\ge2$, the derivative in $b$ of
$g(m,b)+V(m-b,r-1,\ell)$ is at most
$p_{\lfloor m/b\rfloor}-\lambda_{r-1}
\le p_{u_r-1}-\lambda_{r-1}\le0$, including one-sided derivatives
at kinks. The constrained optimum consequently begins with batches
of size $\ell$ until the remaining continuous problem satisfies its
lower bound. These initial batches are already integral. Applying the
preceding construction to the remaining problem, which still has
at most $4\ell$ rounds, proves the lemma.
\end{proof}

\begin{theorem}[Comparison with nondecreasing schedules]
\label{ex:monotone}
For every \(1\le R\le n\),
\[
0\le\log F(n,R)-\log F_\uparrow(n,R)<28.
\]
\end{theorem}

\begin{proof}
The proof splits a maximizing composition into a prefix, in which
every batch is small relative to its remaining mass, and a tail. The
prefix is sorted directly, and the tail is replaced by a rounded
continuous schedule.

Choose an integer maximizing composition, and let $k$ be the first
index for which $m_k\le b_k^2$; the final batch guarantees that such
an index exists. Put $T=m_k$.

If $k>1$, set $q=b_{k-1}$. Lemma~\ref{ex:ordering} bounds every
part before $k$ by $q+1$, while the choice of $k$ gives
$T\ge q^2-q+1$. Since $g(m,b)=\log L(m,b)$ at integer arguments,
Lemma~\ref{ex:schedule-corrections} bounds the interpolation error
of every rearrangement of this prefix by
\[
\frac{(q+1)^2}{2}\sum\frac{b}{m^2}
\le\frac{(q+1)^2}{2}\int_T^n\frac{dx}{x^2}
\le\frac{(q+1)^2}{2T}\le2.
\]
The smooth reward \(b\log(m/b)\) increases under sorting:
for adjacent \(a>b\), the gain from putting $b$ first is
\[
a\log(1-b/m)-b\log(1-a/m)\ge0,
\]
because \(\log(1-x)/x\) decreases on \((0,1)\).
Sorting the prefix therefore decreases its exact reward by at most
two. The empty prefix contributes zero.

Let $\ell$ be the minimum batch size in the tail before rounding
and $r$ the number of batches in the tail.
By Lemma~\ref{ex:ordering}, \(b_k\le\ell+1\), so
\[
r\ell\le T\le b_k^2\le(\ell+1)^2\le4\ell^2.
\]
Lemma~\ref{ex:rounding} replaces the tail by a nondecreasing integer
composition with the same mass and number of batches, with all parts
at least $\ell$ and loss less than $25$.
Every prefix part is at most $\ell+1$. Thus any remaining inversion
lies in the block formed by the final prefix parts of size $\ell+1$
and the initial tail parts of size $\ell$.
The two-size bound in Lemma~\ref{ex:ordering} sorts this block at
additional loss less than one. The total loss is less than
$2+25+1=28$.
\end{proof}

Combined with the partition construction, the comparison determines
all four exact costs up to an explicit additive constant, which is
the uniform statement of Theorem~\ref{thm:exact}.

\begin{theorem}[Exact costs up to an additive constant]
\label{ex:exact-resources}
For every $1\le R\le n$, each of the four exact costs belongs to
$[\Delta(n,R),\Delta(n,R)+38)$. More precisely,
\[
\begin{gathered}
 \Delta(n,R)\le C^{\mathrm{ad}}_0(n,R)
 \le C^{\mathrm{st}}_0(n,R)
 \le B^{\mathrm{st}}_0(n,R)<\Delta(n,R)+38,\\
 C^{\mathrm{ad}}_0(n,R)\le B^{\mathrm{ad}}_0(n,R)
 \le B^{\mathrm{st}}_0(n,R).
\end{gathered}
\]
\end{theorem}
\begin{proof}
Proposition~\ref{ex:exact-lower} gives the lower bound. For the upper
bound, Theorem~\ref{ex:monotone} gives
$\log F(n,R)-\log F_\uparrow(n,R)<28$.
Theorem~\ref{ex:partition-realization} realizes a maximizing
nondecreasing schedule
with static support cost less than $\log(n!/F_\uparrow(n,R))+10$.
The remaining comparisons hold because entropy is bounded by
logarithmic support and because every static protocol is an
adaptive protocol.
\end{proof}

\subsection{Fixed-round asymptotics}

For a fixed number of rounds, the continuous optimum determines the
maximal private entropy, and with it the four exact costs, up to a
bounded error. This is the expansion stated after
Theorem~\ref{thm:exact}.

\begin{corollary}[Fixed-round expansion]
\label{ex:fixed-round}
For each fixed positive integer $R$, let \(\lambda_R\) be defined by
\eqref{ex:lambda}. Then
\[
\log F(n,R)=\lambda_Rn+O_R(1),
\]
and each of the four exact resource costs equals
\[
\log(n!)-\lambda_Rn+O_R(1).
\]
If \(D_1=1\) and \(D_r=u_rD_{r-1}\), then
\[
\lambda_Rn-\lambda_R(D_R-1)
\le\log F(n,R)\le\lambda_Rn,
\]
and equality in the upper bound holds whenever \(D_R\mid n\).
\end{corollary}

\begin{proof}
The continuous optimum bounds every integer schedule from above.
If $D_R\mid n$, the recursive policy is integral: its first batch
has size $n/u_R$, and the number of remaining labels is divisible by
$D_{R-1}$. By induction on the number of rounds, the policy
therefore attains the support size $e^{\lambda_Rn}$.

For general $n$, put $n_0=D_R\lfloor n/D_R\rfloor$.
When $n_0>0$, apply this construction to $n_0$ labels and assign the
remaining labels deterministically to additional positions in the
last batch. This yields the support size $e^{\lambda_Rn_0}$.
When $n_0=0$, the desired lower bound follows from $F(n,R)\ge1$.
Since $n-n_0\le D_R-1$, the stated inequalities follow.
Theorem~\ref{ex:exact-resources} gives the corresponding expansion
of the four costs.
\end{proof}

\subsection{The exact rate at linear round density}

When the number of rounds is proportional to the number of labels,
the exact cost per label converges to an explicit function of the
ratio. This limit is the zero-error case of
Theorem~\ref{thm:frontiers}.
Let $\Phi$ be the piecewise-affine function through the points
$(1/j,\log j)$, $j\ge1$.
\begin{theorem}[Linear-density exact rate]\label{ex:linear-density}
If $R/n\to\rho\in(0,1]$, then $\Delta(n,R)/n\to\Phi(\rho)$.
All four exact costs have the same limit after normalization by $n$.
\end{theorem}
\begin{proof}
The lower bound reduces to batches of bounded size and then uses
convexity; the upper bound uses two consecutive batch sizes.

Fix an integer $L\ge2$. Subdivide every batch of an exact sampler
into consecutive groups of size at most $L$, keeping the private
distribution of every coordinate. By conditional independence, the
subdivision preserves the output law and the shared randomness,
while the number of rounds increases to at most $R+n/L$.
Each new batch has disjoint row supports and entropy at most
$b\log(m/b)$, in terms of its size and of the number of labels
still unused when it is written.
Write $\mathsf H_n=\sum_{j=1}^n1/j$. Along every execution,
\[
\sum_t\{b_t\log m_t-\log(m_t)_{b_t}\}
\le (L-1)\mathsf H_n,
\]
because the term for a batch is at most
$(L-1)\sum_{j=m_t-b_t+1}^{m_t}1/j$.
The chain rule for entropy and the telescoping of the factorials
give
\[
 C_H\ge\E\sum_t b_t\log b_t-(L-1)\mathsf H_n.
\]
The convex envelope of the points $(1/b,\log b)$ is $\Phi$.
Applying convexity with weights $b_t/n$ and using the bound on the
number of rounds yields
\[
 \sum_t b_t\log b_t
 \ge n\Phi\bigl(\min\{1,R/n+1/L\}\bigr).
\]
Apply this inequality to the static exact sampler of cost below
$\Delta(n,R)+38$, then let $n\to\infty$ followed by $L\to\infty$.
This proves the lower bound.

For the upper bound, take $R$ batches of sizes
$q=\lfloor n/R\rfloor$ and $q+1$, with total size $n$. They satisfy
$\sum_t b_t\log b_t=n\Phi(R/n)$.
Lemma~\ref{ex:schedule-corrections} gives
$\Delta(n,R)\le n\Phi(R/n)+O_\rho(1)$.
The uniform bound of Theorem~\ref{thm:exact} then transfers this
limit to all four exact costs.
\end{proof}

\subsection{Finite exact optima}

The explicit values in Theorem~\ref{thm:exact} concern one round,
two rounds, and a number of rounds at least half the number of
labels. In each case an exact sampler with a uniform seed meets the
lower bound given by the maximal private support.

\begin{proposition}
\label{ex:one-round}
For one output round, all four exact costs are \(\log(n!)\).
\end{proposition}

\begin{proof}
Lemma~\ref{ex:capacity} gives \(F(n,1)=1\).
A uniform permutation drawn as the seed and output
deterministically attains the resulting lower bound.
\end{proof}

For two rounds, and again later, we use the following subgroup
construction, based on sequential stabilizer sampling
\cite{DiaconisShahshahani1987}.

\begin{lemma}[A subgroup sampler]
\label{ex:group-sampler}
Partition the positions into blocks $C_j$, and let
$H_j\le S_{C_j}$ have a base of length at most $R-1$; that is, a
sequence of positions whose images determine every element of $H_j$.
For $H=\prod_j H_j\le S_n$, there is an exact sampler using at most
$R$ rounds and one uniform initial seed of size $n!/|H|$.
\end{lemma}

\begin{proof}
Fix a representative of each left coset of $H$, and choose one of
these representatives, denoted by $\pi$, uniformly; this choice is
the seed.
Within each block, the group elements compatible with the base
images sampled so far form a coset $gK$ of the corresponding
pointwise stabilizer. For the next base point $b$, sample its image
uniformly from $gK(b)$. The fibers of the image map have equal size,
so this is the conditional law of a uniform element of $H_j$.

In each round, sample one base image from every block whose base is
not yet exhausted. These samples are independent across blocks, and
the labels written in that round are their images under $\pi$.
After at most $R-1$ rounds, the base images determine an element
$h\in H$. Output all remaining coordinates of $\pi h$
deterministically in the last round. The output is uniform within
the chosen coset, and since the coset is chosen uniformly, it is
uniform on $S_n$.
\end{proof}

\begin{theorem}[Two rounds]
\label{ex:two-round}
For \(n\ge2\), let
\[
G(n)=
\begin{cases}
3^{n/3},&n\equiv0\pmod3,\\
4\cdot3^{(n-4)/3},&n\equiv1\pmod3,\\
2\cdot3^{(n-2)/3},&n\equiv2\pmod3.
\end{cases}
\]
Then \(F(n,2)=G(n)\), and all four exact costs equal
\[
\log\frac{n!}{G(n)}.
\]
\end{theorem}

\begin{proof}
By the recurrence for the maximal private support,
\[
 F(n,2)=\max_b L(n,b)=\max_{\sum a_i=n}\prod_i a_i.
\]
A maximizing partition has no part equal to one. Splitting a part
$a\ge5$ into $3$ and $a-3$ increases its product; replacing three
twos by two threes also increases the product, while a four may be
replaced by two twos without changing it. Hence some maximizing
partition consists of threes and at most two twos, which gives
$G(n)$.

Partition the positions into blocks of the resulting sizes and use
the regular cyclic group on each block. Each group has a one-point
base, and their direct product has order $G(n)$.
Lemma~\ref{ex:group-sampler} therefore meets the lower bound given
by the maximal private support with a uniform seed of size
$n!/G(n)$. This proves all four cost identities.
\end{proof}

\begin{theorem}[The high-round finite optimum]
\label{ex:high-density}
For \(\lceil n/2\rceil\le R\le n\), put \(d=n-R\). Then
\[
F(n,R)=\frac{n!}{(2d)!}(d!)^2
=\frac{n!}{\binom{2d}{d}},
\]
and all four exact costs equal \(\log\binom{2d}{d}\).
In particular, \(F(2L,L)=(L!)^2\).
\end{theorem}

\begin{proof}
Three exchanges of batch sizes, none of which decreases the support
size that the recurrence assigns to a schedule, turn a maximizing
schedule into singletons followed by pairs. A uniform partition of
the labels left for the pairs into two equal parts, chosen by the
seed, then attains the resulting value.

The first exchange is the splitting inequality \eqref{ex:split}.
The second moves a singleton before a batch of size $b$:
\[
 mL(m-1,b)\ge(m-b)L(m,b),\qquad m\ge b+1.
\]
To verify it, write $m=ub+s$. The ratio $L(m-1,b)/L(m,b)$ is
$(u-1)/u$ when $s=0$ and $u/(u+1)$ otherwise.

The third exchange replaces the segment $1,2,\ldots,2,b$, containing
$k$ pairs and a batch of size $b\ge3$, by $k+1$ pairs followed by a
batch of size $b-1$.
We first show that, for $M\ge4$ and $3\le b\le M$,
\begin{equation}
 L(M,b)\le\frac M4L(M-1,b-1).
 \label{ex:large-batch-exchange}
\end{equation}
Let $x,y$ be the two smallest parts of a balanced partition attaining
$L(M,b)$. The other parts have total at least two: if there were only
one further part of size one, all three parts would be one, giving
$M=3$. Hence
\[
 4xy\le(x+y+2)(x+y-1)\le M(x+y-1).
\]
Replacing $x,y$ by the single part $x+y-1$ gives a partition of
$M-1$ into $b-1$ parts and proves
\eqref{ex:large-batch-exchange}.

For the third exchange starting with $m$ remaining labels, put
$M=m-1-2k$ and $P_2(m,j)=\prod_{i=0}^{j-1}L(m-2i,2)$.
The ratio of the new support size to the old one is
\[
 \frac{P_2(m,k+1)}{mP_2(m-1,k)}\frac{L(M-1,b-1)}{L(M,b)}.
\]
The first factor is $(M+1)/4$ for even $m$ and $M(m+1)/(4m)$
for odd $m$, using $L(2h,2)=h^2$ and $L(2h+1,2)=h(h+1)$.
It is at least $M/4$, which proves the exchange. In the remaining
case $M=3$, necessarily $b=3$ and $m$ is even, and both factors
equal one.

Begin with a maximizing schedule and split it into exactly $R$
batches by the first exchange. Their mean size is at most two, so
the presence of a batch of size at least three guarantees a
singleton. Move all singletons to the beginning by the second
exchange, and apply the third to the segment from the last
singleton to the first batch of size at least three.
Each such step decreases $\sum_i(b_i-2)_+$ while preserving the
number of labels and the number of rounds. Iterating yields $n-2d$
singletons followed by $d$ pairs, whose support size is
\[
 \prod_{m=2d+1}^n m\prod_{j=1}^d j^2
 =\frac{n!(d!)^2}{(2d)!}.
\]

To attain this value exactly, draw an independent uniform seed in
$[\binom{2d}{d}]$. Draw the first $n-2d$ coordinates sequentially
without replacement, then use the seed to choose a uniform ordered
partition of the remaining labels into two parts of equal size.
In each of $d$ further rounds, draw one independent uniform label
from each part. Lemma~\ref{ex:partitions} proves uniformity, and the
seed attains the lower bounds for all four costs. When $d=0$,
sequential sampling uses no shared randomness.
\end{proof}

\section{Finite mixtures under transitive group actions}
\label{app:group-mixtures}

The following averaging argument turns a private-coin protocol into an
approximate sampler with a uniform seed. Averaging the private
output law over finitely many of its images under a symmetry group
brings it close to the uniform law on the target, and the number of
images needed depends on the concentration of the private law on
that target.

\begin{lemma}[Finite approximation by group averaging]
\label{con:group-mixture}
Let a finite group act on a finite output space, preserve a target set
$\mathcal T$ of size $M$, and act transitively on $\mathcal T$.
Suppose its action preserves admissibility and the round bound of a
private-coin protocol with output law $Q$.
Let $Q_0\le Q$ have mass $s$, be supported on $\mathcal T$, and satisfy
$\max_x Q_0(x)\le e^A/M$.
For every positive integer $J$, there are fixed group elements
$g_1,\ldots,g_J$ such that the mixture
$Q_J=J^{-1}\sum_{j=1}^J g_jQ$ satisfies
\begin{equation}
 \TV(\Unif(\mathcal T),Q_J)
 \le1-s+\frac12\sqrt{\frac{s e^A}{J}}.
 \label{con:group-mixture-typical}
\end{equation}
Here $g_jQ$ denotes the pushforward of $Q$ under $g_j$.
If $Q$ is uniform on a subset of $\mathcal T$ of size $F$, the
sharper bound is
\begin{equation}
 \TV(\Unif(\mathcal T),Q_J)
 \le\frac12\sqrt{\frac{M/F-1}{J}}.
 \label{con:group-mixture-uniform}
\end{equation}
The mixture is implemented by one uniform initial shared variable
with support at most $J$.
\end{lemma}

\begin{proof}
Choose $g_1,\ldots,g_J$ independently and uniformly from the group.
Write $U=\Unif(\mathcal T)$ and
$\overline Q_0=J^{-1}\sum_{j=1}^J g_jQ_0$.
Transitivity gives $\E(g_jQ_0)(x)=s/M$ for every $x\in\mathcal T$,
while
\[
 \sum_x Q_0(x)^2\le\frac{s e^A}{M}.
\]
Independence and Cauchy--Schwarz therefore imply
\[
 \E\|\overline Q_0-sU\|_1
 \le\sqrt{M\,\E\|\overline Q_0-sU\|_2^2}
 \le\sqrt{\frac{s e^A}{J}}.
\]
The remaining measure $J^{-1}\sum_j g_j(Q-Q_0)$ has total mass
$1-s$. By the triangle inequality, the expected total variation
distance is at most the right side of
\eqref{con:group-mixture-typical}, and some fixed choice of the group
elements does at least as well.

For a uniform law on $F$ target points, take $Q_0=Q$ and $s=1$.
The sum of the coordinate variances is then exactly
$(1/F-1/M)/J$, which gives
\eqref{con:group-mixture-uniform} by the same argument.
Drawing a uniform component index as the shared seed and executing
the corresponding private-coin protocol, which is admissible by
hypothesis, implements the mixture with the stated shared support.
\end{proof}

\section{Permanent stability and approximate nonnegative tensor rank}
\label{rank:section}

This appendix proves the one-round results of
Section~\ref{sec:oneround}. A one-round protocol is a mixture of
product distributions; its support cost is the logarithm of the
number of components, and its entropy cost is the entropy of the
mixing variable.

Let $P_n=\Unif(S_n)$, viewed as a distribution on $[n]^n$.
For a distribution $P$ on this product space, define
\[
r_\varepsilon(P)=
\min\left\{M:
Q=\sum_{t=1}^{M}w_t\Pi_t,\quad
\Pi_t=\bigotimes_{i=1}^{n}\pi_{t,i},\quad
\sum_{t=1}^{M}w_t=1,\quad w_t\ge0,\quad
\TV(P,Q)\le\varepsilon
\right\}.
\]
Thus $r_\varepsilon(P)$ is the minimum nonnegative tensor rank of a
probability distribution within total variation $\varepsilon$ of $P$.
Let $G_\varepsilon(P)$ be the infimum of the entropy of the mixing
variable $T$ over the same finite product mixtures.
We use $h_2(t)=-t\log t-(1-t)\log(1-t)$, with $0\log0=0$.

\subsection{Two sharp inequalities for product measures}

A product distribution supported on permutations is a point mass.
The following theorem, which is Theorem~\ref{thm:stability} of the
main text, is the quantitative form of this fact.

\begin{theorem}[Permanent stability]
\label{rank:entropy}
Let $X_1,\ldots,X_n\in[n]$ be independent, with joint law $\Pi$, and put
\[
q=\Pi(S_n),\qquad m=\max_{\sigma\in S_n}\Pi(\sigma).
\]
Then
\begin{equation}
qH(\Pi\mid S_n)\le(1-q)\log2,
\label{rank:entropy-ineq}
\end{equation}
where the left side is defined as zero when $q=0$, and
\begin{equation}
q-m\le\frac{1-q}{2}.
\label{rank:mode-ineq}
\end{equation}
Both constants are attained for every $n\ge2$.
\end{theorem}

For a row-stochastic matrix $A=(A_{ij})_{i,j\le n}$, write
$w_\sigma=\prod_iA_{i,\sigma(i)}$ and $q=\per(A)$.
The theorem equivalently states that
\[
\sum_{\sigma\in S_n}w_\sigma\log\frac{q}{w_\sigma}
\le(1-q)\log2,
\qquad
3\per(A)-2\max_{\sigma\in S_n}w_\sigma\le1.
\]
The proof rests on a lemma about the event that two independent
random labels agree. In its application the labels are the first
coordinate and the label omitted by the remaining coordinates.

\begin{lemma}[Agreement inequalities]
\label{rank:agreement}
Suppose $p_j\ge0$, $\sum_jp_j=1$, $b_j\ge0$, and
$\sum_jb_j=r\le1$. Set $a_j=p_jb_j$, $q=\sum_ja_j$, and
$a_* =\max_ja_j$. Then
\[
qH((a_j/q)_j)\le(r-q)\log2,
\qquad
q-a_*\le\frac{r-q}{2},
\]
with the left side of the first inequality read as zero when $q=0$.
\end{lemma}

\begin{proof}
First, the binary entropy satisfies
$h_2(t)\le2\log2\sqrt{t(1-t)}$.
By symmetry it suffices to consider $0\le t\le1/2$.
The derivative of the right side minus the left side is
\[
g(u)=\log2\,(u-u^{-1})-2\log u,
\qquad u=\sqrt{(1-t)/t}.
\]
The sign of $g'(u)$ is the sign of
$(\log2)u^2-2u+\log2$. Consequently the difference first increases
and then decreases between its zero values at $t=0$ and $t=1/2$,
and so it is nonnegative on this interval.
Splitting off one atom and applying the entropy chain rule inductively
now gives, for every probability vector $v$,
\begin{equation}
H(v)\le2\log2\sum_{i<j}\sqrt{v_iv_j}.
\label{rank:multiway}
\end{equation}
At the induction step, the binary contribution is bounded using
\[
 \sqrt{v_1(1-v_1)}\le\sum_{j>1}\sqrt{v_1v_j}.
\]

The Cauchy--Schwarz inequality, followed by an expansion of the
square, yields
\[
\left(\sum_j\sqrt{a_j}\right)^2
\le\left(\sum_jp_j\right)\left(\sum_jb_j\right)=r,
\qquad
2\sum_{i<j}\sqrt{a_ia_j}\le r-q.
\]
Applying \eqref{rank:multiway} to the probability vector $a/q$ proves
the entropy bound.
For an index $j_*$ maximizing $a_j$, we also have
\[
\sum_{i<j}\sqrt{a_ia_j}
\ge\sum_{j\ne j_*}\sqrt{a_{j_*}a_j}
\ge\sum_{j\ne j_*}a_j=q-a_*.
\]
This proves the second inequality. Both conclusions are immediate when
$q=0$.
\end{proof}

\begin{proof}[Proof of Theorem~\ref{rank:entropy}]
We prove the two inequalities together. The argument is an induction
in which the agreement inequalities handle the first coordinate.
Let $(J,c)$ denote either $(H,\log2)$ or $(1-\max\nu,1/2)$, where $J$
is a functional of a distribution $\nu$: its entropy in the first
case, and the mass outside its largest atom in the second.
Both functionals satisfy
\begin{equation}
J(X,Y)\le J(X)+\E J(Y\mid X).
\label{rank:chain}
\end{equation}
For entropy this is the chain rule. For the second functional, put
\[
 a_x=\Pr(X=x),\qquad \beta_x=\max_y\Pr(Y=y\mid X=x),
\]
and choose
$x_*$ with $a_{x_*}=\max_xa_x$. Then
\[
\max_xa_x\beta_x\ge a_{x_*}\beta_{x_*}
\ge a_{x_*}+\sum_xa_x\beta_x-1,
\]
which is equivalent to \eqref{rank:chain}.

We induct on $n$; the cases $n=1$ and $q=0$ are immediate.
Let $F$ be the event that $X_2,\ldots,X_n$ are distinct, and let $G$
be the event that none of them equals $X_1$. Write
$r=\Pr(F)$, $s=\Pr(G)$, and $E=F\cap G$, so that $\Pr(E)=q$ and, by
inclusion--exclusion,
\begin{equation}
r+s-q\le1.
\label{rank:union}
\end{equation}
For each label $j$, set
\[
b_j=\Pr(F\text{ and the omitted label is }j),\quad
p_j=\Pr(X_1=j),\quad g_j=\Pr(G\mid X_1=j).
\]
By independence, the masses of $X_1$ on $E$ are $p_jb_j$, and
$\sum_jb_j=r$. Lemma~\ref{rank:agreement} therefore gives
$qJ(X_1\mid E)\le c(r-q)$.

Conditioning on $X_1=j$ and $G$ leaves $n-1$ independent coordinates
with values in $[n]\setminus\{j\}$. The probability that they form a
permutation of these labels is $b_j/g_j$ when $g_j>0$.
The induction hypothesis, multiplied by $g_j$, gives
\[
b_jJ(X_{2:n}\mid X_1=j,E)\le c(g_j-b_j).
\]
Terms on zero-probability events are taken to be zero.
Multiplying by $p_j$ and summing over all labels shows that the
remaining coordinates contribute at most $c(s-q)$. Combining the two
contributions by means of \eqref{rank:chain} and then using
\eqref{rank:union}, we obtain
\[
qJ(\Pi\mid S_n)\le c(r+s-2q)\le c(1-q).
\]
Finally, let the first two coordinates be independent and uniform on
$\{1,2\}$ and fix each remaining coordinate at a different label.
This law has
$q=1/2$, $m=1/4$, and $H(\Pi\mid S_n)=\log2$, attaining both bounds.
\end{proof}

\begin{corollary}[A covering bound for product mixtures]
\label{rank:mode}
If $Q=\sum_{t=1}^{M}w_t\Pi_t$ is a mixture of product laws on
$[n]^n$, there is a set $A\subseteq S_n$ of size at most $M$ such that
\[
Q(S_n\setminus A)\le\tfrac12Q(S_n^c).
\]
\end{corollary}

\begin{proof}
For each component choose a permutation $\sigma_t$ of maximum
$\Pi_t$-mass and put $A=\{\sigma_t:t\le M\}$.
Summing \eqref{rank:mode-ineq} with weights $w_t$ proves the claim.
\end{proof}

The covering bound allows us to compare the approximate rank of a
target with the least number of permutations that carry most of its
mass. For $P(S_n)=1$ and $0\le\eta\le1$, let
\[
K_\eta(P)=\min\{|A|:A\subseteq S_n,\ P(A)\ge1-\eta\}.
\]

\begin{corollary}
\label{rank:target-cover}
For every distribution $P$ supported on $S_n$ and
$0\le\varepsilon\le2/3$,
\[
K_{3\varepsilon/2}(P)\le r_\varepsilon(P)\le K_\varepsilon(P).
\]
Consequently, for every nonempty $T\subseteq S_n$,
\[
r_\varepsilon(\Unif(T))
\ge\left\lceil(1-3\varepsilon/2)|T|\right\rceil.
\]
This applies in particular to any nonempty family of perfect matchings
in a bipartite graph with $n$ vertices in each class, represented by the
matched label of each left vertex.
\end{corollary}

\begin{proof}
Let $Q$ be a product mixture of error $e\le\varepsilon$, and let $A$
be the cover from Corollary~\ref{rank:mode}. Since $Q(S_n^c)\le e$,
the overlap of the two distributions, that is, one minus their total
variation distance, satisfies
\[
1-e=\sum_x\min\{P(x),Q(x)\}
\le P(A)+Q(S_n\setminus A)\le P(A)+e/2.
\]
Thus $P(A)\ge1-3e/2$, proving the lower bound.
For the upper bound, condition $P$ on a minimum $(1-\varepsilon)$-cover,
that is, a set attaining $K_\varepsilon(P)$, and express the
resulting law as a mixture of point masses. Its total variation
distance from $P$ equals the omitted mass.
\end{proof}

\subsection{The sharp rank curve through error one half}

The covering bound and an explicit mixture determine the approximate
rank of the uniform distribution on permutations, up to an additive
two, at every error up to one half. The resulting bounds are the
first part of Theorem~\ref{thm:oneround}.

\begin{theorem}
\label{rank:approximate-rank}
Let $n\ge2$, $N=n!$, and $0\le\varepsilon\le1/2$. Then
\begin{equation}
\left\lceil\left(1-\frac{3\varepsilon}{2}\right)N\right\rceil
\le r_\varepsilon(P_n)
\le N-3\left\lfloor\frac{\varepsilon N}{2}\right\rfloor.
\label{rank:finite-curve}
\end{equation}
In particular, for $0\le k\le\lfloor N/4\rfloor$,
$r_{2k/N}(P_n)=N-3k$.
Uniformly over $0\le\varepsilon\le1/2$,
\[
\frac{r_\varepsilon(P_n)}{n!}
=1-\frac{3\varepsilon}{2}+O(1/n!),
\qquad
\log r_\varepsilon(P_n)
=\log n!+\log\left(1-\frac{3\varepsilon}{2}\right)+O(1/n!).
\]
For $n\ge4$, $r_{1/2}(P_n)=n!/4$.
\end{theorem}

\begin{proof}
The lower bound is Corollary~\ref{rank:target-cover}.
For the upper bound, pair the permutations by exchanging their first
two coordinates. For each of
$k\le\lfloor N/4\rfloor$ pairs, take the product law whose first two
coordinates independently choose between the two exchanged labels,
each with probability $1/2$, while all other coordinates are fixed.
Give this component weight $4/N$. It places mass $1/N$ on each of
the paired permutations and total mass $2/N$ outside $S_n$.
Add $N-4k$ distinct point masses outside these pairs, each of weight
$1/N$. The resulting mixture has total weight one, $N-3k$ components,
and total variation error $2k/N$.

Taking $k=\lfloor\varepsilon N/2\rfloor$ proves the upper bound.
The two bounds differ by at most two, and they coincide when the
argument of the floor is an integer. Since
$1-3\varepsilon/2\ge1/4$ on the stated interval, this also proves both
uniform asymptotic formulas. Finally, $4\mid n!$ for $n\ge4$, which
gives the endpoint value.
\end{proof}

\subsection{Entropy bounds for all errors below one}

The entropy inequality of permanent stability, combined with
Lemma~\ref{lem:overlap-main}, bounds the rank and the mixing entropy
at every error below one. For the set of all permutations, the
entropy bounds are the entropy-cost bounds of
Theorem~\ref{thm:oneround}.

\begin{theorem}[Entropy bounds for permutation-supported targets]
\label{rank:latent-entropy}
\label{rank:subfamily-entropy}
Let $n\ge2$, $\varnothing\ne\Omega\subseteq S_n$,
$N=|\Omega|$, $P=\Unif(\Omega)$, $0\le\varepsilon<1$, and
$p=1-\varepsilon$. Then
\[
\log r_\varepsilon(P)
\ge\log N+\log p-\frac{\varepsilon}{p}\log2,
\]
and
\[
p\log N+p\log p-\varepsilon\log2
\le G_\varepsilon(P)
\le p\log N+h_2(\varepsilon).
\]
\end{theorem}

\begin{proof}
Write a feasible product mixture as $Q(t,x)=w_t\Pi_t(x)$ and let
$s=1-\TV(P,Q_X)\ge p$ be its overlap with the target.
Lemma~\ref{fr:overlap} supplies a law $V$ supported on $\Omega$ with
$sV\le Q$ and $H_V(X)\ge\log(sN)$.
Put $q_t=\Pi_t(S_n)$. Lemma~\ref{fr:conditional-transport} and the
entropy inequality of permanent stability give
\[
H_V(X\mid T)
\le\frac1s\sum_tw_tq_tH(\Pi_t\mid S_n)
\le\frac{1-\sum_tw_tq_t}{s}\log2
\le\frac{1-s}{s}\log2.
\]
Hence the number $M$ of components satisfies
\[
\log M\ge H_V(T)
\ge\log(sN)-\frac{1-s}{s}\log2.
\]
The last expression increases with $s$, which proves the bound on the
rank.

For the mixing entropy, the marginal domination $sV_T\le w$ expresses
$w$ as a mixture with weight $s$ on $V_T$.
Concavity of entropy gives $H(w)\ge sH_V(T)$,
and hence
\[
H(w)\ge s\log(sN)-(1-s)\log2.
\]
If $pN<1$, the asserted lower bound is nonpositive. If $pN\ge1$,
the displayed expression is increasing for $s\ge p$, because its
derivative is $\log(sN)+1+\log2>0$. This proves the entropy lower
bound in both cases.
For the upper bound, take the mixture of point masses with weight
$p/N$ on each element of $\Omega$ and weight $1-p$ on a fixed word
outside $S_n$. This mixture has error $1-p$
and mixing entropy $p\log N+h_2(p)$.
\end{proof}

\section{Entropy bounds and sampling with few rounds}
\label{fr:appendix}

This appendix supplies the entropy estimates behind
Theorems~\ref{thm:fewround} and~\ref{thm:sublinear}. Three tools come
first: the law obtained by conditioning a protocol on its overlap
with the target, a lower bound for the one-batch function $K_b$ that
is uniform in the batch size, and an entropy bound for injections
with small total correlation.

\subsection{Overlap and entropy accounting}

The overlap of a protocol with a uniform target is one minus their
total variation distance. The first lemma conditions the protocol on
the part of its output law that lies below the target and bounds the
output entropy, the total correlation of the batches, and the shared
entropy of the resulting law $V$; it contains
Lemma~\ref{lem:overlap-main}. The second lemma transfers conditional
entropy bounds from the protocol law to laws that it dominates, such
as $V$.

Write $\mathcal H_t=(Z_1,Y_1,\ldots,Z_{t-1},Y_{t-1})$ for the public
history before the shared choice $Z_t$, and let $Y_t$ be the output
batch that follows. The set of coordinates written in this batch is
determined by $(\mathcal H_t,Z_t)$. Allowing empty batches, we may
represent a finite protocol with a fixed number of stages. Denote
its complete transcript by $T$ and its final output by $X$.

\begin{lemma}[Overlap and information accounting]
\label{fr:overlap}
Let $P$ be uniform on a finite set $\Omega$ of cardinality $N$, let
$Q$ be a protocol law whose overlap with $P$ is positive, and put
\[
p=\sum_x\min\{P(x),Q_X(x)\}>0,\qquad d=-\log p.
\]
There is a law $V$ on the same transcripts, supported on outputs in
$\Omega$, such that
\begin{equation}
H_V(X)\ge\log N+\log p,\qquad
D(V\Vert Q)\le d,\qquad
\frac{dV}{dQ}\le\frac1p.
\label{fr:eq-overlap}
\end{equation}
If $\delta_t$ is the total correlation of $Y_t$ under $V$, conditional
on $(\mathcal H_t,Z_t)$, then
\begin{equation}
\E_V\sum_t\delta_t\le d.
\label{fr:eq-correlation-budget}
\end{equation}
Moreover,
\begin{align}
H_Q(X)&=\sum_tI_Q(Z_t;X\mid \mathcal H_t)
       +\sum_tH_Q(Y_t\mid \mathcal H_t,Z_t),
\label{fr:eq-causal-chain}\\
\sum_tH_V(Z_t\mid \mathcal H_t)&\le B,
\qquad
\sum_tH_V(Z_t\mid \mathcal H_t)\le C_H/p.
\label{fr:eq-shared-bounds}
\end{align}
Here $B$ is the support cost of the protocol, that is, the maximum
over paths of the sum of the logarithms of the conditional support
sizes of the shared choices, and
$C_H=\sum_tH_Q(Z_t\mid \mathcal H_t)$ is its entropy cost.
\end{lemma}

\begin{proof}
For $Q_X(x)>0$, define
\[
V(T,x)=Q(T\mid x)\frac{\min\{P(x),Q_X(x)\}}p,
\]
and set the mass to zero otherwise. This is a probability law with
$V_X(x)\le1/(pN)$ and $dV/dQ\le1/p$. These bounds imply
\eqref{fr:eq-overlap}.

At each history, the private batch law under $Q$ is a product. For any
law $\nu$ and product law $\bigotimes_iq_i$ on the same coordinates,
\[
D\!\left(\nu\middle\Vert\bigotimes_iq_i\right)
=\TC(\nu)+\sum_iD(\nu_i\Vert q_i).
\]
Apply this identity to the batch terms in the chain rule for
$D(V\Vert Q)$. The remaining terms are nonnegative, which proves
\eqref{fr:eq-correlation-budget}.
The transcript determines $X$, and $(X,\mathcal H_t,Z_t)$ determines
$Y_t$. The chain rule for mutual information along the transcript
therefore proves \eqref{fr:eq-causal-chain}; the same identity holds
under $V$.

Finally, conditional entropy is bounded by the logarithm of the size
of the conditional support. Summing these bounds along each path and
averaging under $V$ gives the first bound in
\eqref{fr:eq-shared-bounds}. Conditional cross entropy and the
bound $dV/dQ\le1/p$ give the second:
\[
\sum_tH_V(Z_t\mid \mathcal H_t)
\le\E_V\sum_t-\log Q(Z_t\mid \mathcal H_t)
\le\frac1p\E_Q\sum_t-\log Q(Z_t\mid \mathcal H_t)=\frac{C_H}{p}.
\]
\end{proof}

\begin{lemma}[Conditional entropy under a dominated law]
\label{fr:conditional-transport}
Let $Q(t,x)=w_tK_t(x)$ be a law on finite random variables $(T,X)$.
For each $t$, let $E_t$ be an event in the output space, put
$q_t=K_t(E_t)$, and, when $q_t>0$, let
$\nu_t=K_t(\,\cdot\mid E_t)$.
If a probability law $V$ is supported on $X\in E_T$ and satisfies
$sV\le Q$ pointwise for some $0<s\le1$, then
\begin{equation}
H_V(X\mid T)\le\frac1s\sum_t w_tq_tH(\nu_t).
\label{rank:tilt-transport}
\end{equation}
Terms with $q_t=0$ are defined as zero.
\end{lemma}

\begin{proof}
For every $t$ of positive $V_T$-mass, the conditional law
$V_{X\mid t}$ is absolutely continuous with respect to $\nu_t$.
The cross-entropy bound and $sV\le Q$ yield
\[
\begin{aligned}
H_V(X\mid T)
&\le\sum_{t,x\in E_t}V(t,x)(-\log\nu_t(x))\\
&\le\frac1s\sum_{t,x\in E_t}w_tK_t(x)(-\log\nu_t(x))
 =\frac1s\sum_tw_tq_tH(\nu_t).
\end{aligned}
\]
All summands are nonnegative, and terms of zero mass vanish.
\end{proof}

\subsection{A uniform density inequality}

This subsection bounds the one-batch function $K_b$ of
Section~\ref{sec:depth} from below uniformly in the batch size and
deduces an entropy bound for injections, the main input to
Theorem~\ref{thm:sublinear}. Explicitly, for $b\ge1$ and $t\ge0$,
\begin{equation}
K_b(t)=\inf_{\substack{a_i\ge0\\ \E a_i=1,\ 1\le i\le b}}
\E\left[\sum_{i=1}^b a_i\log a_i+
                  t\sum_{i<j}a_i a_j\right].
\label{fr:eq-local-dual}
\end{equation}
The infimum ranges over all joint laws, on any probability space, of
nonnegative random variables $a_i$ with mean one; such a joint law
is called a mean-one density law. In particular, $K_1(t)=0$.

\begin{lemma}[Uniform density and injection-entropy bounds]
\label{fr:coarse-modulus}
Put $k_0=3\sqrt3/e$. For $t>0$, define
\[
z=\frac{t}{2k_0},\qquad
x_t=\max\left\{\sqrt2,\frac{1+\sqrt{1+z^2}}z\right\},
\qquad
\Psi(t)=\log\left(1+(x_t^2-1)e^{-zx_t}\right).
\]
The function $\Psi$ is decreasing and tends to zero as $t\to\infty$.
Every mean-one density law satisfies
\begin{equation}
\E\left[\sum_i a_i\log a_i+t\sum_{i<j}a_i a_j\right]
\ge b\log b-b\Psi(t).
\label{fr:eq-density-modulus}
\end{equation}
If $\nu$ is a law on injections of $b$ coordinates into $m$ labels and
$\delta=\TC(\nu)$, then
\begin{equation}
H(\nu)\le b\log(m/b)+tm\delta+b\Psi(t).
\label{fr:eq-coarse-entropy}
\end{equation}
For $A_0=3\sqrt6/e$ and $t\ge4A_0$,
\[
 \Psi(t)=\log\bigl(1+e^{-t/A_0}\bigr).
\]
\end{lemma}

\begin{proof}
The density inequality is proved by separating the average size $f$
of the densities from their shape $v$: the entropy term can fall
below $b\log b$ only where $f$ is bounded below, and there the
collision term compensates.

For $b=1$, the density bound follows from Jensen's inequality.
Suppose $b\ge2$, and put
\[
f=\frac1b\sum_i a_i,\qquad
v_i=\frac{a_i}{bf}\quad(f>0),\qquad
 e_v=1-\max_i v_i,
\qquad h(f)=f\log f-f+1.
\]
At $f=0$, choose any probability vector $v$; all terms involving it
are multiplied by $f$. Since $\E f=1$, direct substitution gives
\begin{align*}
\E\sum_i a_i\log a_i
 &=b\log b-b\E[fH(v)-h(f)],\\
\E\sum_{i<j}a_ia_j
 &=\frac{b^2}{2}\E[f^2(1-\|v\|_2^2)].
\end{align*}
Thus the first expectation can fall below $b\log b$ only through the
event on which the term $fH(v)$ exceeds $h(f)$.

Let $E_+=\{fH(v)>h(f)\}$. On this event,
\begin{equation}
f>1/(k_0\sqrt b).
\label{fr:eq-density-positive}
\end{equation}
Indeed, for $x=k_0\sqrt b$,
\[
h(1/x)-\frac{\log b}{x}
=1-\frac{3[1+\log(x/3)]}{x}\ge0,
\]
by the inequality $\log y\le y-1$. Because the function
$h(f)-f\log b$ decreases for $0<f\le1$ and $H(v)\le\log b$, this
proves \eqref{fr:eq-density-positive}.

Set $\bar e=\E[fe_v\one_{E_+}]$ and
$F_b(u)=h_2(u)+u\log(b-1)$. The entropy bound $H(v)\le F_b(e_v)$
and the concavity of $F_b$, applied with the weights $f\one_{E_+}$
and using $\E[f\one_{E_+}]\le1$ and $F_b(0)=0$, give
\[
\E[fH(v)-h(f)]\le F_b(\bar e).
\]
Moreover, $1-\|v\|_2^2\ge e_v$. Applying
\eqref{fr:eq-density-positive} to the collision term on $E_+$ gives
\[
\frac1b\E\sum_{i<j}a_ia_j
\ge\frac{\sqrt b}{2k_0}\bar e.
\]
By the elementary identity
\[
 \sup_u\{F_b(u)-su\}=\log\bigl(1+(b-1)e^{-s}\bigr),
\]
the left side of the
density inequality, divided by $b$, therefore falls below $\log b$
by at most
$\log(1+(b-1)e^{-t\sqrt b/(2k_0)})$.
Substituting $x=\sqrt b$ and maximizing $(x^2-1)e^{-zx}$ over
$x\ge\sqrt2$ gives $x_t$, and hence \eqref{fr:eq-density-modulus}.
The maximum decreases with $t$ and tends to zero. For $t\ge4A_0$, it
is attained at $x=\sqrt2$, yielding the stated expression for $\Psi$.

To obtain the injection bound, let $X$ have law $\nu$ with marginals
$p_i$, and put
\[
r=\sum_{i<j,x}p_i(x)p_j(x).
\]
For each $i$, compare the joint law of $(X_i,X_{<i})$ with the
independent coupling $p_i\otimes\nu_{<i}$. Under this coupling the
coordinates of $X_{<i}$ remain distinct, so the probability that $X_i$
equals one of them is
$s_i=\sum_{j<i,x}p_i(x)p_j(x)$.
That event has probability zero under $\nu$.
The data-processing inequality for relative entropy, applied to the
indicator of this event, gives
\[
I(X_i;X_{<i})\ge-\log(1-s_i)\ge s_i.
\]
These mutual informations sum to the total correlation, so summing
over $i$ proves $r\le\delta$.
Finally, apply \eqref{fr:eq-density-modulus} to the densities
$a_i(x)=mp_i(x)$ under the uniform distribution on labels. The two
sums on its left side then have expectations
$b\log m-\sum_iH(p_i)$ and $mr$, respectively.
Using $H(\nu)\le\sum_iH(p_i)$ and $r\le\delta$ gives
\eqref{fr:eq-coarse-entropy}.
\end{proof}

\subsection{Entropy bounds for nearly independent injections}

Under the reweighted law $V$, the first batch of a two-round
protocol is an injection whose expected total correlation is at most
$-\log p$, where $p$ is the overlap. This subsection bounds the
entropy of such a law, taking as benchmark the entropy available to
independent coordinates with disjoint supports.
Let $G(n)$ be the largest product of positive integers summing to
$n$, with $G(0)=G(1)=1$, and put $\lambda_2=(\log3)/3$.
The integer-product formula in Theorem~\ref{ex:two-round} gives
\begin{equation}
\log G(n)=\lambda_2n-\gamma_{n\bmod3}\quad(n\ge2),\qquad
\gamma_0=0,\quad\gamma_1=4\lambda_2-\log4,\quad
\gamma_2=2\lambda_2-\log2.
\label{fr:eq-integer-product}
\end{equation}
The next lemma relates the entropy of an injection in excess of this
benchmark to its total correlation. Its leading constant determines
the sharp two-round saving in Theorem~\ref{thm:fewround}.

\begin{lemma}[Entropy of an injection]
\label{fr:owner}
Let $\nu$ be a law on $[n]^b$, with $1\le b\le n$, supported on tuples
with distinct coordinates. Put $\delta=\TC(\nu)$ and $\ell=\log n$.
For $0<\vartheta<1$, define
$a_\vartheta=\lambda_2-\vartheta\log(2e/\vartheta)$.
If $a_\vartheta>0$, then
\begin{equation}
H(\nu)\le\lambda_2n+
\left(\frac{\ell^2}{2a_\vartheta}
      +\frac{2(2\ell+1)}{\vartheta}\right)\delta+\frac1e.
\label{fr:eq-sharp-local}
\end{equation}
There is also an integer $u\in[0,n]$ such that
\begin{equation}
\begin{split}
H(\nu)-\log G(n)
\le{}&-\log\frac{G(n)}{G(n-u)}
      +\ell\sqrt{2u\delta}
      +u\vartheta\log(2e/\vartheta)\\
&+\frac{2\delta}{\vartheta}(2\ell+1)
      +\varphi(2\delta/\vartheta),
\qquad
\varphi(t)=t\log(1+1/t),\quad\varphi(0)=0.
\end{split}
\label{fr:eq-refined-owner}
\end{equation}
\end{lemma}

\begin{proof}
The proof retains one largest entry in each column of substantial
mass of the matrix of marginal probabilities. The retained entries
behave as independent coordinates with disjoint supports, and the
contribution of the discarded entries is estimated through pairwise
collision probabilities, whose sum is at most the total correlation.

Write the marginal probabilities as $A_{ix}=\Pr(X_i=x)$ and set
\[
f_x=\sum_iA_{ix},\qquad a_x=\max_iA_{ix},\qquad
\rho_x=f_x-a_x,\qquad r_x=\sum_{i<j}A_{ix}A_{jx}.
\]
Each row sums to one. Since $\nu$ is supported on injections, each
column sum satisfies $f_x\le1$.
The collision argument at the end of the preceding proof gives
$r:=\sum_xr_x\le\delta$.
Moreover, $\sum_iA_{ix}^2\le a_xf_x$, so
\begin{equation}
r_x\ge f_x\rho_x/2\ge\rho_x^2/2.
\label{fr:eq-column-collision}
\end{equation}

The entropy of $\nu$ is at most the sum of the marginal entropies,
which we split into the contributions of the individual entries.
Let $u$ be the number of columns with $f_x<\vartheta$, and call these
columns small. In every other column, retain one entry of maximum
mass and discard the remaining entries. Discard all entries in the
$u$ small columns.
Suppose row $i$ retains $s_i>0$ entries with total mass $g_i$.
Their entropy contribution is at most
\[
g_i\log(s_i/g_i)\le\log s_i+1-g_i.
\]
There are $n-u$ retained entries in total, so the product of the
positive $s_i$ is at most $G(n-u)$.
Writing $\zeta$ for the total discarded mass, the entropy
contribution of the retained entries is therefore at most
$\log G(n-u)+\zeta$; it is also at most $\lambda_2(n-u)+\zeta$.
Rows with no retained entries contribute zero.

We now bound the entropy contribution of the discarded entries
together with the term $\zeta$.
In a column with $0<f_x<\vartheta$, separating its largest entry from
the others bounds its entropy contribution by
\[
\rho_x\log b+f_x\log(2/f_x).
\]
After adding its discarded mass $f_x$, this is at most
$\rho_x\ell+\vartheta\log(2e/\vartheta)$; a zero column contributes
zero. By \eqref{fr:eq-column-collision} and the Cauchy--Schwarz
inequality, the sum of these contributions over the small columns is
at most
\[
\ell\sqrt{2u\delta}+u\vartheta\log(2e/\vartheta).
\]
The columns with $f_x\ge\vartheta$ have total discarded mass
$\beta\le2r/\vartheta$. Their discarded entries occupy at most $n^2$
positions, so the entropy contribution of these entries plus their
mass is at most
\[
\beta(2\ell+1)-\beta\log\beta.
\]
Since $-\beta\log\beta\le\varphi(\beta)$ and
$\varphi'(t)=\log(1+1/t)-1/(t+1)>0$, we may bound this expression by
$2\delta(2\ell+1)/\vartheta+\varphi(2\delta/\vartheta)$.
Adding the bounds for the retained and the discarded entries proves
\eqref{fr:eq-refined-owner}.

To prove \eqref{fr:eq-sharp-local}, use the bound
$\lambda_2(n-u)+\zeta$ for the retained entries. Each small column
lowers this bound by $\lambda_2$; after this amount is subtracted,
the contribution of the column is at most
\[
\rho_x\ell-a_\vartheta
\le\frac{\ell^2\rho_x^2}{4a_\vartheta}
\le\frac{\ell^2r_x}{2a_\vartheta}.
\]
For the other columns, use $-\beta\log\beta\le1/e$.
Summing and applying $r\le\delta$ proves \eqref{fr:eq-sharp-local}.
\end{proof}

\subsection{The two-round sampling cost}

This subsection proves Theorem~\ref{thm:fewround}. Write
\[
\Delta_2(n)=\log(n!/G(n)),\qquad
 d_\varepsilon=-\log(1-\varepsilon).
\]
The exact two-round optimum is $\Delta_2(n)$ by
Theorem~\ref{ex:two-round}. At a fixed error $\varepsilon$, the
support cost falls below this optimum by a term of order
$(\log n)^2$ with a sharp constant, and the entropy cost equals
$1-\varepsilon$ times the optimum up to a term of the same order.

\begin{proposition}[Finite two-round converse]
\label{fr:two-round-converse}
For $n\ge2$, $\tau\in\{\mathrm{st},\mathrm{ad}\}$, every
$0<\varepsilon<1$, and every $0<\vartheta<1$ with
$a_\vartheta>0$,
\begin{equation}
\begin{split}
\Delta_2(n)-B^\tau_\varepsilon(n,2)
\le{}&C_n(\vartheta)d_\varepsilon+
      \lambda_2n-\log G(n)+d_\varepsilon+\frac1e
      +\frac{\varepsilon}{1-\varepsilon}\log2,\\
C_n(\vartheta)={}&
      \frac{(\log n)^2}{2a_\vartheta}
      +\frac{2(2\log n+1)}{\vartheta}.
\end{split}
\label{fr:eq-two-round-converse}
\end{equation}
\end{proposition}

\begin{proof}
Take a feasible protocol $Q$, let $p$ be its actual overlap, and put
$d=-\log p$. Let $V$ be the reweighted law of
Lemma~\ref{fr:overlap}. The proof bounds the entropy of each of the
two batches under this law.

Merge the shared draws made before each of the two output batches
into single variables $Z$ and $W$, respectively, and denote the
batches that follow them by $Y$ and $X'$.
By the chain rule for conditional entropy, the total shared entropy
is unchanged.
Conditionally on $Z$, every nonempty first batch is an injection under
$V$. Lemma~\ref{fr:owner}, together with
$\E_V\TC(V_{Y\mid Z})\le d$, gives
\[
H_V(Y\mid Z)\le\lambda_2n+C_n(\vartheta)d+1/e.
\]
An empty first batch has zero entropy and satisfies the same bound.

Consider next a history $h=(z,y,w)$ whose first batch is injective.
Let $q_h$ be the probability under $Q$ that the second batch completes
it to a collision-free output, and let $r_h$ be the probability that
every remaining coordinate chooses an unused label.
The latter is a product event. When $r_h>0$, conditioning on it leaves
a product distribution on the unused labels. Permanent stability
(Theorem~\ref{thm:stability}) therefore gives
\[
q_hH(\nu_h)\le(r_h-q_h)\log2\le(1-q_h)\log2,
\]
where $\nu_h$ is the law of the completion conditioned on being
collision-free.
When $q_h=0$ the left side is read as zero, and an empty completion
also has zero entropy. Histories with $q_h=0$ have probability zero
under $V$.
Lemma~\ref{fr:conditional-transport} now yields
\[
\begin{aligned}
H_V(X'\mid Z,Y,W)
&\le\frac1p\sum_{h:\,y\text{ injective}}Q(h)q_hH(\nu_h)\\
&\le\frac{1-p}{p}\log2.
\end{aligned}
\]
In the last step, the collision-free completions have total
probability at least $p$, whereas the histories with an injective
first batch have total probability at most one.

The transcript entropy identity and the shared-entropy bound now give
$H_V(X)\le B+H_V(Y\mid Z)+H_V(X'\mid Z,Y,W)$.
Combining this with $H_V(X)\ge\log n!-d$ and using
$p\ge1-\varepsilon$ proves the proposition.
\end{proof}

\begin{corollary}[Two-round entropy bounds]
\label{fr:two-round-shannon}
Put $p=1-\varepsilon$, and let $E_n(\vartheta,\varepsilon)$ denote
the right-hand side of \eqref{fr:eq-two-round-converse}. For either
model and every admissible $\vartheta$,
\[
p\,[\Delta_2(n)-E_n(\vartheta,\varepsilon)]
\le C^\tau_\varepsilon(n,2)
\le p\Delta_2(n)+h_2(\varepsilon).
\]
Consequently, for each fixed $0<\varepsilon<1$,
\begin{equation}
\label{fr:eq-two-round-shannon}
\begin{split}
p\Delta_2(n)
-\left(\frac{p d_\varepsilon}{2\lambda_2}+o_\varepsilon(1)\right)
(\log n)^2-O_\varepsilon(1)
&\le C^\tau_\varepsilon(n,2)\\
&\le p\Delta_2(n)+h_2(\varepsilon).
\end{split}
\end{equation}
\end{corollary}

\begin{proof}
Let a feasible protocol have actual overlap $s\ge p$, and repeat the
preceding proof with the bound
$\sum_tH_V(Z_t\mid\mathcal H_t)\le C_H/s$ on the shared entropy.
Since $-\log s\le d_\varepsilon$ and $(1-s)/s\le\varepsilon/p$,
\[
C_H/s\ge\Delta_2(n)-E_n(\vartheta,\varepsilon).
\]
When the right side is nonnegative, multiplication by $s\ge p$ gives
the lower bound; when it is negative, the bound follows from
$C_H\ge0$.
The upper bound is obtained by aborting: run an exact two-round
sampler with probability $p$ and output a fixed word outside $S_n$
otherwise.
Finally, take $\vartheta=(\log n)^{-1/2}$. Then
\[
C_n(\vartheta)
=\left(\frac1{2\lambda_2}+o(1)\right)(\log n)^2,
\qquad
\lambda_2n-\log G(n)=O(1),
\]
which proves \eqref{fr:eq-two-round-shannon}.
\end{proof}

The matching upper bound on the support cost uses the following
construction.

\begin{lemma}[A two-round sampling construction]
\label{fr:reservoir}
Let $n,u,j_0$ be integers with $u\equiv n\pmod3$,
$m=(n-u)/3\ge1$, and $0\le j_0\le\min(m,u)$.
Let $0<\varpi<1$ satisfy $3\varpi<(1-\varpi)u$.
There is a private two-round law whose collision-free probability is
\[
q=\sum_{k=j_0}^{\min(m,u)}
\binom mk\varpi^k(1-\varpi)^{m-k}\frac{(u)_k}{u^k}
\]
and whose collision-free outputs each have probability at most
\[
h=\left(\frac{1-\varpi}{3}\right)^{m-j_0}
  \left(\frac\varpi u\right)^{j_0}.
\]
For $0<\varepsilon<1$ with $q>1-\varepsilon$, a static shared seed gives
\begin{equation}
B^{\rm st}_\varepsilon(n,2)
\le\log n!+\log h+\log q+
     2\log\frac1{q-1+\varepsilon}+\log2.
\label{fr:eq-mixture-upper}
\end{equation}
\end{lemma}

\begin{proof}
Set aside $u$ labels as a reserve and partition the remaining labels
into $m$ triples.
In the first round, one coordinate per triple chooses uniformly from
its triple with probability $1-\varpi$, and uniformly from the reserve
with probability $\varpi$.
If at least $j_0$ choices use the reserve and all of those labels are
distinct, the prefix is retained and is completed deterministically
to a permutation.
Every other prefix is completed to a word with a repeated label.
Different retained prefixes give different final permutations.

The number $K$ of reserve choices has the distribution
$\operatorname{Bin}(m,\varpi)$.
Conditioned on $K=k$, the chosen reserve labels are distinct with
probability $(u)_k/u^k$, which proves the formula for $q$.
A prefix with $k$ reserve choices has probability
$((1-\varpi)/3)^{m-k}(\varpi/u)^k$.
The assumed inequality makes this probability decrease with $k$, so
every retained prefix has probability at most $h$.

Apply Lemma~\ref{con:group-mixture} to the action of label permutations,
using a mixture of
\[
M=\left\lceil\frac{n!qh}{(q-1+\varepsilon)^2}\right\rceil
\]
fixed relabelings. The total variation distance of the resulting law
from the uniform law on permutations is at most
\[
1-q+\tfrac12\sqrt{n!qh/M}\le\varepsilon.
\]
Since $n!h\ge q$, the expression inside the ceiling exceeds one.
The ceiling therefore increases it by at most a factor of two, which
proves \eqref{fr:eq-mixture-upper}.
\end{proof}

\begin{theorem}[Sharp two-round approximation]
\label{fr:sharp-two-round}
For every fixed $0<\varepsilon<1$ and
$\tau\in\{\mathrm{st},\mathrm{ad}\}$,
\begin{equation}
B^\tau_\varepsilon(n,2)
=\Delta_2(n)-\frac{d_\varepsilon}{2\lambda_2}(\log n)^2
 +o_\varepsilon((\log n)^2).
\label{fr:eq-fixed-error}
\end{equation}
\end{theorem}

\begin{proof}
For the converse, take $\vartheta=(\log n)^{-1/2}$ in
Proposition~\ref{fr:two-round-converse}. Then
$C_n(\vartheta)/(\log n)^2\to1/(2\lambda_2)$ and
$\lambda_2n-\log G(n)=O(1)$.

For achievability, fix $0<d<d_\varepsilon$, write $\ell=\log n$, and
choose
\[
\begin{gathered}
u=\frac{d\ell^2}{2\lambda_2^2}+O(1),\qquad
u\equiv n\pmod3,\qquad m=(n-u)/3,\\
\mu=\frac{d\ell}{\lambda_2},\qquad
\varpi=\mu/m,\qquad j_0=\lfloor\mu-\mu^{2/3}\rfloor.
\end{gathered}
\]
These parameters satisfy the hypotheses of
Lemma~\ref{fr:reservoir} for all sufficiently large $n$.
For $K\sim\operatorname{Bin}(m,\varpi)$, we have $K/\mu\to1$ in
probability and
\[
\log\frac{(u)_K}{u^K}
=-\frac{K(K-1)}{2u}+O(K^3/u^2)\longrightarrow-d
\]
in probability on an event whose probability tends to one.
The cutoff $K\ge j_0$ discards probability $o(1)$.
The conditional probability that the reserve labels are distinct is
bounded by one, so convergence in probability implies convergence of
the expectations; thus $q\to e^{-d}>1-\varepsilon$.

The bound $h$ on the probability of a collision-free output
satisfies
\[
\begin{aligned}
-\log h
&=m\log3-m\log(1-\varpi)
  +j_0\log\frac{(1-\varpi)u}{3\varpi}\\
&=\lambda_2n+
  \left(\frac{d}{2\lambda_2}+o(1)\right)\ell^2.
\end{aligned}
\]
Here the logarithm multiplying $j_0$ is
$\ell+\log\ell+O(1)$, while $m\log3=\lambda_2(n-u)$.
The lemma now gives the matching support bound with
coefficient $d/(2\lambda_2)$. Letting $d$ increase to
$d_\varepsilon$ proves the asserted coefficient. The construction uses
static sharing and therefore applies to the adaptive model as well.
\end{proof}

\subsection{All sublinear round counts}

For a sublinear number of rounds both costs are determined to first
order, which gives Theorem~\ref{thm:sublinear}. The lower bound
applies the injection-entropy bound to each batch, and the upper
bound is obtained by aborting a subgroup sampler.

\begin{theorem}[Sublinear-round costs]
\label{fr:sparse}
Let $1\le R\le n$, $0<p\le1$, $d=-\log p$, and set
\[
K_0=\frac{16k_0^2}{e^2},\qquad
A(d)=2+\log(1+K_0)+2\log(1+d).
\]
Every permutation protocol with overlap $p$ satisfies
\begin{equation}
\begin{split}
B&\ge n\log(n/R)-nA(d)-d,\\
C_H&\ge p\,[n\log(n/R)-nA(d)-d].
\end{split}
\label{fr:eq-sparse-finite}
\end{equation}
For every fixed $0\le\varepsilon<1$, every integer sequence
$1\le R_n=o(n)$, and both models
$\tau\in\{\mathrm{st},\mathrm{ad}\}$,
\begin{align}
B^\tau_\varepsilon(n,R_n)
&=(1+o(1))n\log(n/R_n),
\label{fr:eq-sparse-support}\\
C^\tau_\varepsilon(n,R_n)
&=(1-\varepsilon+o(1))n\log(n/R_n).
\label{fr:eq-sparse-shannon}
\end{align}
Moreover, writing $d_\varepsilon=-\log(1-\varepsilon)$, the support
saving satisfies, uniformly over $1\le R\le n$,
\[
0\le B^\tau_0(n,R)-B^\tau_\varepsilon(n,R)
\le n\bigl[A(d_\varepsilon)+2\bigr]+d_\varepsilon.
\]
\end{theorem}

\begin{proof}
Apply Lemma~\ref{fr:coarse-modulus} to each nonempty batch under the
reweighted law $V$ of the protocol. Along each execution, the batch
sizes satisfy $\sum_tb_t=n$ and there are at most $R$ nonempty
batches. Hence
\[
\sum_tb_t\log(m_t/b_t)
\le n\log n-\sum_tb_t\log b_t\le n\log R.
\]
The number of labels available to a batch satisfies $m_t\le n$, and
the total correlations of the batches have expected sum at most $d$,
so the total private conditional entropy is at most
$n\log R+n(td+\Psi(t))$.
The elementary bound
\[
x^2e^{-tx/(2k_0)}\le\frac{16k_0^2}{e^2t^2}
\]
and the choice $t=1/(1+d)$ give
\[
td+\Psi(t)
\le1+\log(1+K_0)+2\log(1+d).
\]
Combining this with $H_V(X)\ge\log n!-d$,
$\log n!\ge n\log n-n$, and
the shared-entropy bounds of Lemma~\ref{fr:overlap} proves
\eqref{fr:eq-sparse-finite}.

For exact achievability, write $n=kR+s$, $0\le s<R$, and use
Lemma~\ref{ex:group-sampler} with the group
$(S_R)^k\times S_s$, omitting the last factor when $s=0$.
Each factor has a base of length at most $R-1$, and a trivial factor
requires no base point. The number of values of the uniform shared
seed, and its logarithm, satisfy
\[
M=\frac{n!}{(R!)^ks!},\qquad
\log M\le n\log(n/R)+n+s\log(R/s)
\le n\log(n/R)+2n,
\]
where the term involving $s$ is zero when $s=0$.
For $n\ge2$, abort this sampler: run it with probability
$p=1-\varepsilon$ and output a fixed word outside $S_n$ otherwise.
The resulting error is $\varepsilon$, and the seed entropy is at most
$h_2(p)+p\log M$.
Comparison with \eqref{fr:eq-sparse-finite} proves both asymptotic
formulas, since $\log(n/R_n)\to\infty$.

Finally, a protocol feasible at error $\varepsilon$ has overlap at least
$1-\varepsilon$. The function $A(d)$ increases with $d$, so the support
cost of such a protocol is at least
$n\log(n/R)-nA(d_\varepsilon)-d_\varepsilon$.
Subtracting this lower bound from the upper bound on $\log M$ for the
exact sampler proves the uniform estimate for the support saving,
which is nonnegative because the exact protocols are among the
feasible ones.
\end{proof}

\section{Information, collisions, and the support frontier}
\label{app:frontiers}

This appendix proves Theorem~\ref{thm:frontiers} on the support cost
with linearly many rounds and determines the static entropy cost of
Theorem~\ref{thm:shannon-frontiers}. It begins with three tools: the
one-batch variational problem, its comparison with the two-coordinate
problem, and the realization of profiles by protocols in both models.

\subsection{The one-batch variational problem}
The one-batch function $K_b$ satisfies elementary bounds and reduces
to a convex envelope in one variable. The reduction also shows that
bounded density laws with finite support suffice, and the profiles
defined below are built from such laws.

For each positive integer $b$, let $\mathcal M_b$ be the set of laws
of $a=(a_1,\ldots,a_b)\in[0,\infty)^b$ satisfying $\E a_i=1$ for
every $i$; its elements are called density laws. We
use the information and collision quantities $d_b,c_b,K_b$ from
\eqref{eq:local-action}, with $0\log0=0$. Jensen's inequality gives
the lower bound below, while uniform densities and disjoint supports
give the two upper bounds:
\begin{equation}\label{front:elementary-K}
 0\le K_b(t)\le\min\{\tbinom b2t,b\log b\};
\end{equation}
moreover, $K_b$ is nondecreasing and concave, and $K_1=0$. For a
product law with marginals $p_1,\ldots,p_b$ on $m$ labels, its
empirical density law is the law of the vector of densities at a
uniformly random label, and it satisfies
\begin{equation}\label{front:empirical}
 a_i(y)=mp_i(y),\quad d_b=\sum_iD(p_i\Vert u_m),\quad
 c_b=m\sum_{i<j}\langle p_i,p_j\rangle.
\end{equation}
When the same marginals arise from a law supported on injections,
Lemma~\ref{fr:coarse-modulus} gives $c_b\le m\TC$.

\begin{lemma}[Scalar convexification]\label{front:convexification}
Let $\Delta_b=\{v\in[0,1]^b:\sum_i v_i=1\}$.
For $t>0$, put
$f_{b,t}(s)=s\log s+s\min_{v\in\Delta_b}
\{\sum_iv_i\log v_i+ts(1-\sum_iv_i^2)/2\}$. Then
\begin{equation}\label{front:scalar-K}
 K_b(t)=(\conv f_{b,t})(b)
       =(\conv f_{b,1})(bt)/t-b\log t.
\end{equation}
The infimum can be approximated by laws with two values of the total
density $s$. At each positive value, the normalized density vector is
uniform or has the form $(x,y,\ldots,y)$, up to permutation, with $x>y$.
\end{lemma}
\begin{proof}
Symmetrize the density law over its $b$ coordinates. The individual
mean constraints then reduce to $\E s=b$, where $s=\sum_i a_i$.
Conditioning on $s$ gives $f_{b,t}(s)$, and minimization under this
single mean constraint gives the lower convex envelope. Two-point
laws approximate its value at $b$. The identity
$f_{b,t}(s)=f_{b,1}(ts)/t-s\log t$ gives the scaling formula.

For $s>0$, every minimizer over the simplex lies in its interior:
moving a small mass $e$ to a zero coordinate changes the entropy term
by $e\log e+O(e)$, while the interaction term changes by $O(e)$. The
Lagrange multiplier equation is $\log v_i-tsv_i=\mathrm{constant}$
and has at most two positive solutions. If the larger solution
appeared twice, transferring mass between those two coordinates would
have a negative second derivative, contradicting minimality. Thus a
nonuniform minimizer has one large coordinate and $b-1$ equal smaller
coordinates. Finally, $f_{b,t}(s)\ge s\log(s/b)$ is coercive, which
justifies approximation on bounded intervals of total density.
\end{proof}
The minimization at a fixed total density is the mean-field Potts
variational problem, and convexification over the total density has
the same variational structure as the mean-field Potts gas
\cite{EllisWang1990,GeorgiiEtAl2006}. Bounded
density laws with finite support suffice in the infimum defining
$K_b$. To see this, truncate each coordinate at a level $M$ and
divide it by its mean. Pair products converge by
monotone convergence before normalization, and the entropy terms
converge because the negative part of $a\log a$ is bounded.
Discretization on a compact set, followed by rational approximation
of the probabilities and mean normalization, preserves the limiting
information and collision quantities.

\subsection{Comparison with the two-coordinate problem}

The two-coordinate problem has an explicit solution, and this
solution bounds $K_b$ from below for every batch size $b$. For a
range of multipliers the bound reduces the infimum over batch sizes
in Theorem~\ref{thm:frontiers} to singletons and pairs. Set
\[
 I(\theta)=\frac{(1+\theta)\log(1+\theta)+(1-\theta)\log(1-\theta)}2,
 \qquad
 K(t)=\min_{0\le\theta\le1}\{2I(\theta)+t(1-\theta^2)\}.
\]
The minimizing $\theta$ is zero for $t\le1$ and is the unique
positive solution of $\operatorname{arctanh}\theta=t\theta$ for
$t>1$.

\begin{lemma}[Comparison across batch sizes]
\label{pair:cross-batch}
For every $b\ge2$ and $t\ge0$,
\begin{equation}
 K_2(t)=K(t),\qquad
 K_b(t)\ge(b-1)K_2(bt/2).
 \label{pair:comparison}
\end{equation}
Consequently, for $b\ge3$,
\begin{equation}
 2K_b(t)\ge bK_2(t)+(b-2)K_2(bt)
          \ge bK_2(t)+(b-2)K_2(3t).
 \label{pair:three}
\end{equation}
Also $K_b(t)=\binom b2t$ for $0\le t\le2/b$. Consequently every
admissible density law satisfies $c_b\ge\binom b2-\frac b2d_b$.
For \(0\le\eta\le K_2(3t)\), these inequalities give
\begin{equation}
 \inf_{b\ge1}\frac{K_b(t)+\eta}{b}
 =\min\left\{\eta,\frac{K_2(t)+\eta}{2}\right\}.
 \label{pair:binary-reduction}
\end{equation}
\end{lemma}

\begin{proof}
We first compare entropy at a fixed value of $\sum_i v_i^2$ and then
average over the total density. For $v\in\Delta_b$, put
$\theta^2=(b\sum_i v_i^2-1)/(b-1)$,
$x=[1+(b-1)\theta]/b$, $y=(1-\theta)/b$.
Cauchy--Schwarz gives
$\sum_jv_j^2\ge v_i^2+(1-v_i)^2/(b-1)$, hence $v_i\le x$.
For $F(z)=z\log z$, the
quadratic interpolating $F(y),F'(y),F(x)$ has Hermite remainder
$F^{(3)}(\xi)(z-y)^2(z-x)/6\ge0$ on $(0,x]$.
Matching the first two moments shows that $(x,y,\ldots,y)$ maximizes
entropy at this value of $\sum_i v_i^2$
\cite{HarremoesTopsoe2001}; continuity handles
zeros and endpoints. Convexity gives
$F(1+(b-1)\theta)\ge(b-1)F(1+\theta)$, hence
\begin{equation}
 \frac b{b-1}D(v\Vert u_b)\ge2I(\theta).
 \label{pair:purity}
\end{equation}
To average over the total density, we use the convexity of
$G_u(f)=f\log f+fK(uf)$ for each $u\ge0$.
On the uniform branch, where the minimizing $\theta$ is zero,
$fG_u''=1+2uf>0$. On the nonuniform branch, put $z=uf$ and
$q=1-\theta^2$. Differentiating the stationarity equation gives
$K'=q,\quad K''=-\frac{2\theta^2q}{1-zq},\quad fG_u''=\frac{1-zq(2z-1)}{1-zq}>0.$
The denominator is positive by stationarity. With
$w=\operatorname{arctanh}\theta$, positivity of the numerator is
equivalent to $\sinh^2w+w\tanh w>2w^2$, the hyperbolic Wilker
inequality \cite[Theorem 1]{Zhu2007}. It also follows directly by
multiplying the difference by $\cosh^2w$: the coefficients of
$w^2$ and $w^4$ vanish, and the coefficient of $w^{2k}$ for $k\ge3$ is
\[
 \frac{2^{2k-2}}{(2k)!}\bigl(2^{2k-1}-4k(k-1)\bigr)>0.
\]
The first derivative matches at the transition, so $G_u$ is convex
on the whole nonnegative half-line.

For a law with $\E a_i=1$, put $f=b^{-1}\sum_i a_i$, so $\E f=1$,
and $v_i=a_i/(bf)$ when $f>0$. On the event $f=0$, set $v=u_b$;
all corresponding terms below are zero, with $0\log0=0$.
The identities $d_b=b\E[f\log f+fD(v\Vert u_b)]$ and
$c_b=\binom b2\E[f^2(1-\theta^2)]$ separate the density $f$
from the normalized shape $v$. Using \eqref{pair:purity} and
$\E f\log f\ge0$ gives
\[
 \frac{d_b+tc_b}{b-1}
 \ge\E\left[f\log f+f\left(2I(\theta)
       +\frac{btf}{2}(1-\theta^2)\right)\right]
 \ge\E G_{bt/2}(f)\ge G_{bt/2}(1)=K(bt/2),
\]
where the last step is Jensen's inequality.
For $b=2$, uniformly permuting $(1+\theta,1-\theta)$ attains $K$.
Concavity of $K_2$ with weights $b/[2(b-1)]$ and $(b-2)/[2(b-1)]$
gives \eqref{pair:three}, and the uniform density law attains
equality for $bt/2\le1$. Two cases give the last identity. If
$\eta\le K_2(t)$, \eqref{pair:comparison} gives
$K_b(t)\ge(b-1)\eta$, so singletons are optimal. If
$K_2(t)\le\eta\le K_2(3t)$, \eqref{pair:three} gives
$2K_b(t)\ge bK_2(t)+(b-2)\eta$, so pairs are optimal.
\end{proof}

\subsection{Finite profiles and their realization}
With linearly many rounds, the limiting behavior of a protocol is
described by a profile. Every finite profile is realized by protocols
in both models, which gives the upper bounds in
Theorems~\ref{thm:frontiers} and~\ref{thm:shannon-frontiers}.

A density law for a batch of size $b$ specifies an action
$(r,d,c)=(1,d_b,c_b)/b$, whose three coordinates are the round use,
the information, and the collision coefficient per output coordinate.
A finite profile is a piecewise constant choice, at each fraction of
unused labels, of a finite mixture of such actions, drawn from a
finite collection of bounded density laws with finite support. The
mixture averages $(r,d,c)$ at each remaining fraction $u$. By
definition, uniform singletons occupy an interval
$[0,\varepsilon]$ for some $\varepsilon>0$, so every integral below
is finite. A profile is collision-free when $c=0$ throughout. The
round density, the information, and the collision exponent of a
profile are
\begin{equation}\label{front:profile}
 R=\int_0^1r,\qquad B=\int_0^1d,\qquad
 D=\int_0^1 c(u)/u\,du,
\end{equation}
and its survival function, its limiting collision-free probability,
and its survival-weighted information are
\begin{equation}\label{ad:profile}
 s_\pi(u)=\exp\left[-\int_u^1 c(v)/v\,dv\right],\qquad
 p_\pi=e^{-D},\qquad A_\pi=\int_0^1s_\pi d.
\end{equation}
Here $s_\pi(u)$ is the limiting probability that the prefix is
collision-free when the fraction $u$ of labels remains unused;
execution proceeds from $u=1$ toward $u=0$. Let $\mathcal R$ denote the
closure of the set of these triples $(R,D,B)$, enlarged by adding the
nonnegative orthant. The lower boundary of this region at round
budget $\rho$ and collision budget $D$ is denoted
$\mathcal B_\rho(D)$, and we put
$B_\rho(p)=\mathcal B_\rho(-\log p)$, with $B_\rho(0)=0$.
For a normalized support budget $c\ge0$, write
$P_\rho(c)=\sup\{p\in[0,1]:B_\rho(p)\le c\}$ for the maximal overlap.

\begin{lemma}[Adding a collision-free component]\label{front:profile-slack}
Mix a finite profile $\pi$ pointwise with a collision-free profile $h$
in proportion $\theta\in[0,1]$. Then $R$ and $B$ are affine in
$\theta$, $D_\theta=(1-\theta)D$, $s_\theta=s^{1-\theta}$, and
$A_\theta\le e^{\theta D}A+\theta B(h)$.
Collision-free profiles with arbitrarily small positive $R(h)$ exist.
\end{lemma}
\begin{proof}
The identities follow by integrating the pointwise mixture. For the
inequality, use $s^{1-\theta}\le e^{\theta D}s$ on the information of
the given profile and $s^{1-\theta}\le1$ on that of the collision-free
profile. For the last assertion, take batches of one large fixed size
with disjoint supports, followed by a short interval of singletons.
Consequently, a mixture of vanishing weight with a collision-free
profile of smaller round density makes the round constraint strict
while preserving the limiting cost and overlap.
\end{proof}

The next lemma realizes one block of a profile by a codebook, that
is, a finite family of private-coin protocols for the block. The uniform
mixture of their output laws, each conditioned on a collision-free
output, is close to the uniform law on the collision-free outputs of
the block.

\begin{lemma}[Finite block codebooks]\label{ad:block-codebook}
Fix a finite family of density laws with finite support and
$\varepsilon,\xi>0$.
For all sufficiently large $n$ (at fixed $\xi$) and any deterministic
action sequence producing at most $n$ outputs,
with $m_t\ge\varepsilon n$ labels at every stage, put
$\mathsf h=\sum_t d_{b_t}$. There is a codebook of
$M=\lceil e^{\mathsf h+n\xi}\rceil$ private-coin protocols, each with
collision-free probability
\begin{equation}\label{ad:block-survival}
 q=\prod_tq_t,\qquad q_t=1-c_{b_t}/m_t+O(m_t^{-2}),
\end{equation}
whose output laws conditioned on a collision-free output, denoted
$W_i$, satisfy
\begin{equation}\label{ad:block-cover}
 \TV\left(P,M^{-1}\sum_iW_i\right)
 \le C/(n\xi^2)+\tfrac12e^{-n\xi/4},
\end{equation}
where $P$ is uniform on the ordered injections for this block.
The constants depend only on the family and $\varepsilon$.
Relabeling transports the same family of protocols to every
collision-free history at the start of the block.
\end{lemma}
\begin{proof}
The proof realizes one density law by product marginals, assembles a
random private-coin protocol from such realizations, and selects the
codebook by a resolvability bound.

For a density law $\mu=\sum_\kappa w_\kappa\delta_{a^{(\kappa)}}$,
give $n_\kappa=mw_\kappa+O(1)$ of the available labels type $\kappa$,
and use the marginals $p_i(y)=a_i^{(\sigma(y))}/Z_i$, where the
superscript is the type of the label and $Z_i=m+O(1)$ normalizes. We
call such a choice of types a density assignment. All assignments are
relabelings of one another, so their collision-free probabilities
agree. A pair of coordinates collides with probability
$\E a_i a_j/m+O(m^{-2})$, and the intersection of two distinct pair
events has probability $O(m^{-2})$, also when the pairs share a
coordinate. This proves \eqref{ad:block-survival}.

Each private-coin protocol specifies a density assignment at every
collision-free history and uses deterministic output kernels after a
collision. Choosing these assignments independently and uniformly
defines a random private-coin protocol $T$. The survival probability of a
stage, that is, the probability that it produces no collision, is
independent of the assignment and of the preceding collision-free
history, so every protocol has collision-free probability $q$.
Permutation symmetry gives $\E W_T=P$. Under the joint law
conditioned on a collision-free block, each newly encountered
assignment remains uniform, because both its own survival probability
and the survival probability of the remaining stages are constant.
For the sum $g_t=\sum_i\log a_i^{(\sigma_t(Y_i))}$ of logarithmic
densities at the outputs of a stage, rounding and conditioning on an
event of probability $1-O(m_t^{-1})$ give
$\E[g_t\mid\text{past}]=d_{b_t}+O(m_t^{-1})$, a deterministic
quantity that depends only on the stage. Every positive density
belongs to a fixed finite set, so $g_t$ is bounded. Therefore the
centered sums are martingales. The information density
$\imath(T;Y)=\log[W_T(Y)/P(Y)]$ therefore satisfies
\begin{equation}\label{ad:block-spectrum}
 \imath(T;Y)=\sum_tg_t+
 \sum_t\{\log(m_t)_{b_t}-\sum_i\log Z_{t,i}-\log q_t\},\quad
 |\E\imath-\mathsf h|=O(1),\quad\Var(\imath)=O(n).
\end{equation}
The second sum is deterministic and $O(1)$.

For independent copies of the random protocol, the finite-channel
resolvability bound is
\begin{equation}\label{front:soft-cover}
 \E\TV(P,M^{-1}\sum_jW_{T_j})
 \le\Pr\{\imath>\log M-\gamma\}+\tfrac12e^{-\gamma/2}.
\end{equation}
Indeed, truncate the density $L_T=dW_T/dP$ at $Me^{-\gamma}$.
Independence and the bound $(L_T^-)^2\le Me^{-\gamma}L_T^-$ for the
truncated density bound the expected $L^1(P)$ norm of the centered
average by $e^{-\gamma/2}$, while the discarded mass is the tail
probability in the display \cite{HanVerdu1993,Cuff2013}. Chebyshev's
inequality with $\gamma=n\xi/2$ then proves \eqref{ad:block-cover}
for some deterministic codebook.
\end{proof}

The proposition below concatenates such codebooks along a profile,
with the codebook indices drawn either all at the start or one block
at a time.

\begin{proposition}[Static and adaptive realization]\label{ad:deferred-profile}
For a finite profile and a partition
$1=u_0>\cdots>u_J=\varepsilon$ refining its intervals of constancy
above the final singleton interval, put
$h_j=\int_{u_j}^{u_{j-1}}d$ and
$A_{\pi,\mathcal P}=\sum_js_\pi(u_{j-1})h_j$.
For every $\delta>0$ there are sequences of static and adaptive
protocols on $[n]$ with the same output law, deterministic round
counts with $R_n/n\to R(\pi)$, overlap tending to $p_\pi$, and
\begin{equation}\label{ad:deferred-limits}
 \limsup B^{\rm st}/n=\limsup C_H^{\rm st}/n\le B(\pi)+\delta,
 \quad\limsup B^{\rm ad}/n\le B(\pi)+\delta,
 \quad\limsup C_H^{\rm ad}/n\le A_{\pi,\mathcal P}+\delta.
\end{equation}
The static seed is uniform, and the adaptive protocol makes at most
$J$ shared draws.
\end{proposition}
\begin{proof}
On intervals of $v=\lceil\sqrt n\rceil$ coordinates, realize each
action weight $w_a$ by $\lfloor w_av/b_a\rfloor$ batches and fill the
remaining coordinates with uniform singletons. There are
$O_\pi(\sqrt n)$ such singletons, and their contribution vanishes in
all Riemann sums. Apply the preceding lemma to each block of the
partition, with $\xi=\delta/J$ and $\varepsilon/2$. Uniformly over
the stages, $m\ge\varepsilon n-O_\pi(1)$. Hence the information sum,
the collision-free probability, and the total variation error of
each block satisfy
\begin{equation}\label{ad:deferred-riemann}
 \mathsf h_{n,j}/n\to h_j,\quad-\log q_{n,j}\to
 D_j:=\int_{u_j}^{u_{j-1}}c(u)/u\,du,\quad e_{n,j}\to0.
\end{equation}
Transport the codebooks using the increasing bijection between the
available label sets. The codebook indices are independent and
uniform; the static protocol draws all of them initially, and the
adaptive protocol draws each index only when its block is entered at
a collision-free history. Indices of blocks still to be entered are
independent of the preceding outputs and of the used indices, so the
two protocols have the same output law. After a collision, both
protocols retain the deterministic schedule and use fixed kernels.
Constant survival probabilities give the entry probability
$r_{n,j}=\prod_{i<j}q_{n,i}$ and the collision-free probability
$p_n=\prod_jq_{n,j}$. At each entry history of the adaptive protocol,
before its shared draw, conditioning on the survival of the block
preserves the uniform law of the index. We analyze the adaptive
protocol; the static protocol has the same output law. Contraction
of total variation under the subsequent sampling kernels, followed
by completion with uniform singletons, gives, for the conditioned
output law and for the overlap,
\begin{equation}\label{ad:deferred-overlap}
 \TV(W_n,P_n)\le\sum_je_{n,j},\qquad
 p_n(1-\sum_je_{n,j})\le1-\TV(Q_n,P_n)\le p_n.
\end{equation}
The static protocol has support cost and entropy cost both equal to
$\sum_j\log M_{n,j}$, and the adaptive protocol has
\begin{equation}\label{ad:deferred-cost}
 C_H^{\rm ad}=\sum_jr_{n,j}\log M_{n,j},\qquad
 B^{\rm ad}\le\sum_j\log M_{n,j}.
\end{equation}
The stated limits follow from
$\log M_{n,j}=\mathsf h_{n,j}+n\delta/J+O(1)$ and
$r_{n,j}\to s_\pi(u_{j-1})$.
\end{proof}

\subsection{Support and static entropy frontiers}
The first result is the full form of Theorem~\ref{thm:frontiers}.
With linearly many rounds, the optimal support cost per coordinate
is, in both models, the lower boundary of the region of profile
triples, and this boundary has a dual formula that decouples into
one-batch problems. The accompanying lower bound on the entropy cost
is the input for the static entropy cost determined later in the
subsection.

\begin{theorem}[Support cost]\label{front:support}
Let $n\to\infty$ and $R/n\to\rho\in(0,1)$. In both models the
optimal support cost per coordinate is $B_\rho(p)$, and it is
attained in the closed rate region by protocols with a uniform
initial seed. Moreover,
\begin{equation}\label{front:dual}
 \mathcal B_\rho(D)=\sup_{\eta,\xi\ge0}
 \left\{\int_0^1\inf_{b\ge1}
 \frac{K_b(\xi/u)+\eta}{b}\,du
 -\eta\rho-\xi D\right\},\qquad B_\rho(1)=\Phi(\rho).
\end{equation}
In either model, every sequence of protocols whose bounds on the
number of rounds satisfy $R_n/n\to\rho$ and whose overlaps satisfy
$\liminf p(Q_n)\ge p>0$ obeys the entropy-cost bound
\[
 \liminf_{n\to\infty}C_H(Q_n)/n\ge pB_\rho(p).
\]
\end{theorem}
\begin{proof}
The proof has five parts. The first two establish the dual formula,
away from the zero-collision boundary and then on it. The third is
an entropy inequality for a protocol with bounded batch sizes, the
fourth removes that bound and yields both lower bounds, and the
fifth constructs protocols that realize the closed region.

Throughout, for $\eta,\xi\ge0$ we write
\[
 \ell_{\eta,\xi}(u)
 =\inf_{b\ge1}\frac{K_b(\xi/u)+\eta}{b},
 \qquad 0<u\le1.
\]
The singleton action gives $0\le\ell_{\eta,\xi}\le\eta$.
Since $K_b$ is nondecreasing, $\ell_{\eta,\xi}$ is nonincreasing
in $u$. We extend it to $u=0$ by its bounded one-sided limit.

\emph{The closed profile region and its dual.}
Let $\mathcal R$ be the upward closed region of profile triples
defined above. Pointwise mixing of two finite profiles makes
$(R,D,B)$ affine, so $\mathcal R$ is convex. It is closed as well:
if a convergent sequence of triples is written as a sum of a triple
in the closed profile set and a nonnegative vector, both summands
are bounded coordinatewise and have convergent subsequences.
For every $\rho>0$ there is a finite collision-free profile with
$R<\rho$ and finite $B$, by Lemma~\ref{front:profile-slack}.
Consequently $\mathcal B_\rho(D)$ is finite for $\rho>0$, $D\ge0$,
and its defining infimum is attained in $\mathcal R$.
Upward closure makes $\mathcal B_\rho(D)$ nonincreasing in each
budget, and hence makes $B_\rho(p)$ nondecreasing in $p$.

Every finite profile satisfies the pointwise bound
$d(u)+\xi c(u)/u+\eta r(u)\ge\ell_{\eta,\xi}(u)$.
Integration, followed by closure, therefore gives
\[
 \mathcal B_\rho(D)\ge
 \int_0^1\ell_{\eta,\xi}(u)\,du-\eta\rho-\xi D.
\]
To identify the converse inequality, we first establish
\begin{equation}\label{front:profile-lagrangian}
 \inf_{\pi\ {\rm finite}}
 \{B(\pi)+\eta R(\pi)+\xi D(\pi)\}
 =\int_0^1\ell_{\eta,\xi}(u)\,du.
\end{equation}
The pointwise bound shows that the left side is at least the right
side. If $\xi=0$, then $K_b(0)=0$ and $\ell_{\eta,0}=0$.
A profile that uses uniform marginals in batches of a fixed size
$b$, followed by singletons on $[0,a]$, has $B=0$ and
$R=a+(1-a)/b$. Letting $b\to\infty$ and then $a\downarrow0$
proves \eqref{front:profile-lagrangian} in this case.

Suppose now that $\xi>0$. The uniform density bound yields
\begin{equation}\label{front:batch-cutoff}
 \frac{K_b(\xi/u)+\eta}{b}
 \ge \log b-\Psi(\xi),
 \qquad 0<u\le1.
\end{equation}
Choose an integer $M$ such that
$\log b-\Psi(\xi)>\eta+1$ for $b>M$.
The infimum defining $\ell_{\eta,\xi}$ may then be restricted to
$1\le b\le M$, uniformly in $u$. Fix $a>0$.
Each $K_b$ is finite and continuous on $(0,\infty)$, so
$\ell_{\eta,\xi}$ is continuous on $[a,1]$.
For any $\zeta>0$ and each $u_0\in[a,1]$, choose a batch size
$b\le M$ and a bounded finite density law whose local objective at
$u_0$ is at most $\ell_{\eta,\xi}(u_0)+\zeta$.
Such laws suffice by the approximation following
Lemma~\ref{front:convexification}. The chosen objective is continuous
in $u$, as it has the form $d+\eta/b+\xi c/u$.
Throughout some neighborhood of $u_0$ the error is therefore at
most $2\zeta$. Compactness supplies finitely many such
neighborhoods and density laws. A sufficiently fine partition into
intervals, with one of these laws used on each interval, has local
objective at most $\ell_{\eta,\xi}+3\zeta$ throughout $[a,1]$.
On $[0,a]$ we use uniform singletons. The resulting finite profile
has objective at most
\[
 \int_a^1\ell_{\eta,\xi}(u)\,du+3\zeta+\eta a
 \le \int_0^1\ell_{\eta,\xi}(u)\,du+3\zeta+\eta a.
\]
Letting first $\zeta\downarrow0$ and then $a\downarrow0$ proves
\eqref{front:profile-lagrangian}, by approximation with finite
action sets on compact intervals.

Now fix $\rho>0$ and $D>0$, and put $B_*=\mathcal B_\rho(D)$.
A supporting hyperplane to $\mathcal R$ at $(\rho,D,B_*)$ has
nonnegative coefficients because $\mathcal R$ is upward closed.
Its coefficient on $B$ is positive. Indeed, a hyperplane with zero
$B$ coefficient would be contradicted by a collision-free profile
with $R<\rho$ and $D=0<D$. Normalize the coefficient on $B$ to one
and denote the other coefficients by $\eta,\xi\ge0$.
Closure preserves the infimum of this continuous linear functional,
and adding a nonnegative vector can only increase it. Its infimum
over $\mathcal R$ is therefore the infimum over finite profiles.
The supporting identity and \eqref{front:profile-lagrangian} give
\[
 B_*+\eta\rho+\xi D
 =\inf_{\pi\ {\rm finite}}
   \{B(\pi)+\eta R(\pi)+\xi D(\pi)\}
 =\int_0^1\ell_{\eta,\xi}(u)\,du.
\]
Together with the preceding lower bound, this proves
\eqref{front:dual} for $D>0$.

\emph{The zero-collision boundary.}
For fixed $\eta\ge0$ and $\xi>0$, the density bound also gives
\[
 \ell_{\eta,\xi}(u)\ge
 \inf_{b\ge1}\left(\log b+\frac{\eta}{b}\right)-\Psi(\xi).
\]
At $D=0$, first let $\xi\to\infty$ and then take the supremum over
$\eta\ge0$. The supporting-line formula for the lower convex
envelope of $(1/b,\log b)$ yields
\[
 \mathcal B_\rho(0)\ge
 \sup_{\eta\ge0}\left\{
   \inf_{b\ge1}\left(\log b+\frac{\eta}{b}\right)-\eta\rho
 \right\}=\Phi(\rho).
\]
For the reverse inequality, mix the disjoint-support actions of
the two adjacent sizes whose reciprocal sizes bracket $\rho$.
Their per-coordinate information is $\log b$ and their collision
coefficient is zero, so this constant mixture has round density
$\rho$ and information $\Phi(\rho)$.
Replacing its restriction to $[0,a]$ by singletons gives finite
profiles whose triples converge to $(\rho,0,\Phi(\rho))$ as
$a\downarrow0$. Hence
$\mathcal B_\rho(0)=\Phi(\rho)$.
The weak dual inequality and the same limit of multipliers prove
\eqref{front:dual} also at $D=0$.

We record the continuity needed below. The function
$(\rho,D)\mapsto\mathcal B_\rho(D)$ is finite and convex on
$(0,1)\times(0,\infty)$, hence continuous there.
Closedness of $\mathcal R$ gives lower semicontinuity at $D=0$.
The bound $\mathcal B_\rho(D)\le\Phi(\rho)$ and continuity of $\Phi$
give the matching upper limit as $(\rho',D')\to(\rho,0)$.
Thus the profile value is jointly continuous for
$0<\rho<1$, $D\ge0$.

\emph{A finite entropy inequality.}
Consider a protocol $Q$ with overlap $p>0$ and batches of size at
most $L$. Let $V$ be its reweighted law on transcripts, as in
Lemma~\ref{lem:overlap-main}. Write $J_V(Q)$ for the
shared entropy under $V$, $D_p$ for the negative logarithm of the
overlap, and $\mathsf H_n$ for the $n$th harmonic number:
\[
 J_V(Q)=\sum_t H_V(Z_t\mid\mathcal H_t),\qquad
 D_p=\log(1/p),\qquad \mathsf H_n=\sum_{j=1}^n\frac1j.
\]
At a collision-free history immediately before a private batch, let
$m$ be the number of available labels, $b$ the batch size, and $\nu$
the conditional law of the batch under $V$. Write
\[
 d=\sum_{i=1}^bD(\nu_i\Vert u_m),\qquad
 \delta=\TC(\nu),\qquad
 \gamma(m,b)=b\log m-\log(m)_b.
\]
Then
$H(\nu)=b\log m-d-\delta
       =\log(m)_b+\gamma(m,b)-d-\delta$.
The falling factorials telescope on every collision-free execution:
$\sum_t\log(m_t)_{b_t}=\log n!$.
Moreover,
\[
 \gamma(m,b)
 =\sum_{i=0}^{b-1}\log\left(1+\frac{i}{m-i}\right)
 \le (L-1)\sum_{j=m-b+1}^{m}\frac1j.
\]
The integer intervals in this last sum are disjoint along an
execution and together cover $[n]$. Thus
$\sum_t\gamma(m_t,b_t)\le(L-1)\mathsf H_n$ pathwise.

The entropy chain rule bounds $H_V(X)$ by the shared conditional
entropies plus the private batch entropies. Substituting the
preceding identities and using Lemma~\ref{fr:overlap} gives
\begin{equation}\label{front:finite-support}
 \begin{aligned}
 B(Q)\ge J_V(Q)
 &\ge \log p+\E_V\sum_t(d_t+\delta_t)
             -\E_V\sum_t\gamma(m_t,b_t)\\
 &\ge \log p+\E_V\sum_t d_t-(L-1)\mathsf H_n,
 \qquad
 \E_V\sum_t\delta_t\le D_p.
 \end{aligned}
\end{equation}
The conditional marginals of $\nu$ form an admissible density law,
with $c_b\le m\delta$. Therefore, at any fixed $\xi\ge0$,
\[
 d+\xi n\delta
 \ge K_b(\xi n/m).
\]
If $N$ is the number of nonempty output rounds, then for any
$\eta\ge0$,
\[
 \E_V\sum_t d_t
 \ge \E_V\sum_t b_t\ell_{\eta,\xi}(m_t/n)
       -\eta\,\E_V N-\xi nD_p.
\]
On each collision-free execution, the intervals
$[(m_t-b_t)/n,m_t/n]$ partition $[0,1]$ and have length at most
$L/n$. The right-endpoint Riemann sum of the bounded nonincreasing
function $\ell_{\eta,\xi}$ over this partition satisfies
\[
 \frac1n\sum_t b_t\ell_{\eta,\xi}(m_t/n)
 \ge \int_0^1\ell_{\eta,\xi}(u)\,du-\frac{L\eta}{n}.
\]
Since $N\le R$, we obtain the finite inequality
\begin{equation}\label{front:finite-dual-bound}
 \frac{J_V(Q)}n\ge
 \int_0^1\ell_{\eta,\xi}(u)\,du
 -\eta\frac Rn-\xi\log(1/p)
 +\frac{\log p}{n}
 -\frac{(L-1)\mathsf H_n+L\eta}{n}.
\end{equation}

\emph{Arbitrary batch sizes and the order of limits.}
To remove the restriction on batch sizes, divide each batch into
groups of at most $L$ coordinates, retaining all its marginal
kernels and ignoring the intermediate outputs until that batch is
complete. The output law, the shared costs, and the histories at
shared draws are preserved. In particular, $J_V(Q)$ is preserved.
On every execution, the number of additional rounds is at most
\[
 \sum_t\bigl(\lceil b_t/L\rceil-1\bigr)\le\frac nL.
\]

Consider any protocol sequence with $R_n/n\to\rho$ and overlap
$p_n$. Along a subsequence with $p_n\to q>0$, apply
\eqref{front:finite-dual-bound} to these subdivided protocols.
For fixed $L,\eta,\xi$, first let $n\to\infty$.
The terms $\mathsf H_n/n$ and $\log p_n/n$ vanish, giving
\[
 \liminf_n\frac{J_{V_n}(Q_n)}n
 \ge
 \int_0^1\ell_{\eta,\xi}(u)\,du
 -\eta(\rho+1/L)-\xi\log(1/q).
\]
Next let $L\to\infty$, keeping $\eta,\xi$ fixed.
The resulting inequality holds for every pair of nonnegative
multipliers, so taking the supremum over them and using
\eqref{front:dual} gives
\[
 \liminf_n\frac{J_{V_n}(Q_n)}n
 \ge B_\rho(q).
\]
If $\liminf_n p_n\ge p>0$, every subsequence has a further one
with overlap limit $q\ge p$.
The inequality $B(Q_n)\ge J_{V_n}(Q_n)$ and monotonicity of
$B_\rho$ give the converse for the support cost.
Likewise, Lemma~\ref{fr:overlap} gives
$C_H(Q_n)\ge p_nJ_{V_n}(Q_n)$, so the same subsequence has
\[
 \liminf_n\frac{C_H(Q_n)}n
 \ge qB_\rho(q)\ge pB_\rho(p).
\]
Applying this to a subsequence along which the cost in question
attains its lower limit proves the claimed bound for the whole
sequence.

\emph{Realizing the closed region.}
Fix $(\rho,D,B_*)\in\mathcal R$ with $0<\rho<1$ and finite $D$.
By the definition of this region, there are finite profiles $\pi_j$
and numbers $\zeta_j\downarrow0$ with
$R(\pi_j)\le\rho+\zeta_j$,
$D(\pi_j)\le D+\zeta_j$, and
$B(\pi_j)\le B_*+\zeta_j$.
Mix each with a fixed collision-free profile of round density
strictly below $\rho$, using weights tending to zero sufficiently
slowly. Lemma~\ref{front:profile-slack} gives profiles with
$R<\rho$, information at most $B_*+o(1)$, and overlap at least
$e^{-D-o(1)}$.
For each fixed profile, Proposition~\ref{ad:deferred-profile}
constructs protocols with a uniform initial seed.
Because its round density is strictly below $\rho$, these protocols
respect any prescribed bounds on the number of rounds with
$R_n/n\to\rho$, for all sufficiently large $n$. Choose successively
larger thresholds for the profile index, the coding accuracy, and
the rounding error. A diagonal sequence then has limiting support
cost at most $B_*$ and lower limiting overlap at least $e^{-D}$.
Static protocols are also admissible in the adaptive model, so this
proves achievability in both models and completes the proof.
\end{proof}

The remaining results of this subsection determine the overlap
attainable at zero rate, the regularity of the optimal support cost,
the static entropy cost, and a bound on the gain from approximation
that is uniform in the round density. Together they prove the cost
comparisons stated in Theorem~\ref{thm:shannon-frontiers}.

\begin{proposition}[Zero-rate overlap]\label{front:zero}
For permutations, if $2/(j+1)\le\rho\le2/j$ with an integer $j\ge2$,
the greatest overlap attainable at zero support rate, and likewise at
zero entropy rate, is
\begin{equation}\label{front:pzero}
 p_0(\rho)=\sqrt j\,(j-1)!
 \left(\frac{\rho-1/j}{j-1}\right)^{(j-1)/2}.
\end{equation}
\end{proposition}
\begin{proof}
The optimization over uniform marginals agrees with the joint-scaling
frontier and with the exact characterization under a hard bound on
the number of rounds in
\cite[Section~4.2 and Appendix~C]{WenEtAl2026}.
The relation between the distance conventions is explicit for the
schedules with fixed batch sizes that attain this value.
Let $b_1,\ldots,b_R$ be the positive batch sizes of such a schedule,
with $\sum_t b_t=n$, and let $m_t=n-\sum_{s<t}b_s$ count the labels
still available.
For independent uniform draws among the available labels, every
permutation $\sigma$ has probability
\[
 Q(\sigma)=\prod_{t=1}^R m_t^{-b_t}=\frac{q}{n!},
 \qquad
 q=\prod_{t=1}^R\frac{(m_t)_{b_t}}{m_t^{b_t}}.
\]
Thus $q$ is both the collision-free probability and the overlap, and
\[
 \TV(P_n,Q)=1-q,\qquad
 D(P_n\Vert Q)=-\log q,\qquad
 D_{\mathrm{bits}}(P_n\Vert Q)=-\log_2 q.
\]
Here $D_{\mathrm{bits}}$ is KL divergence measured in bits, so the
collision exponent in this paper is
$D=(\log2)D_{\mathrm{bits}}$.
The number of rounds of a fixed schedule is deterministic, so its
mean number of rounds equals its hard bound on the number of rounds.

For an arbitrary output law $Q$ with $q=Q(S_n)>0$, the law $Q_S$
obtained by conditioning on a collision-free output satisfies
$D(P_n\Vert Q)=-\log q+D(P_n\Vert Q_S)$, and $p(Q)\le q$.
For a deterministic schedule whose batch sizes adapt to the values
already written and which samples from conditional marginals, write
$G(\sigma)=\sum_t\log[m_t(\sigma)^{b_t(\sigma)}/(m_t(\sigma))_{b_t(\sigma)}]$.
Then $Q(\sigma)=P_n(\sigma)e^{-G(\sigma)}$, whence
$p(Q)=q=\E_{P_n}e^{-G}$ and $D(P_n\Vert Q)=\E_{P_n}G\ge-\log p(Q)$.
Fixed batch sizes make $G$ constant and attain equality. When words
with a repeated label have positive probability,
$D(Q\Vert P_n)=+\infty$; all KL quantities in this comparison
have the forward orientation.

We now show that uniform laws also determine the zero-cost boundary
for arbitrary private kernels.
An action has zero information precisely when its densities are
uniform, and its per-coordinate collision coefficient is then
$c=(b-1)/2$. These actions also determine the zero-cost boundary
of the closed region. To prove this, first divide each action of a
profile into groups of at most $L$ coordinates, adding at most $1/L$
to the round density. Then replace the actions on the interval $u<a$
by singletons, adding at most $a$ to the round density, and replace
the remaining density laws by uniform laws.
Lemma~\ref{pair:cross-batch} gives
$c_b\ge\binom b2-(b/2)d_b$, so this last step increases the
integrated collision exponent by at most $LB/(2a)$.
Let the information tend to zero, $B\to0$, and then let
$L\to\infty$ and $a\downarrow0$.
The optimum over uniform actions is continuous and minimizes
$(b-1)/(2u)+\eta/b$ pointwise over the batch size.
Writing $A=2\eta$, the size $b$ is optimal when
$(b-1)b\le Au\le b(b+1)$. Integrating in the case where the largest
size used is $j$ gives
$\rho=1/j+(j-1)/A$ and
$D=(j-1)\log A/2-\log j/2-\log((j-1)!)$, which proves the formula.
The bound $C_H/n\ge pB(p)+o(1)$ shows that the entropy cost has the
same zero-rate boundary, and the endpoint $\rho=1$ is attained by
singletons.
\end{proof}

\begin{lemma}[Regularity and finite-size support limits]
\label{front:regularity}
The map $(\rho,p)\mapsto B_\rho(p)$ is continuous on
$(0,1)\times[0,1]$, nondecreasing in $p$, and nonincreasing in $\rho$.
The map $(\rho,D)\mapsto\mathcal B_\rho(D)$ is convex.
Furthermore,
\[
 B_\rho(p)=0\ \Longleftrightarrow\ p\le p_0(\rho),
 \qquad B_\rho(1)=\Phi(\rho).
\]
For each fixed $\rho\in(0,1)$ there are $c_\rho>0$ and
$\varepsilon_\rho>0$ such that
\begin{equation}\label{front:endpoint-softening}
 B_\rho(1-\varepsilon)\le\Phi(\rho)
       -c_\rho\varepsilon\log(1/\varepsilon),
 \qquad 0<\varepsilon<\varepsilon_\rho.
\end{equation}
For every integer sequence $1\le R_n\le n$ with $R_n/n\to\rho$,
every fixed $0\le\varepsilon<1$, and either model,
\begin{equation}\label{front:finite-limit}
 \lim_{n\to\infty}\frac{B^\tau_\varepsilon(n,R_n)}n
       =B_\rho(1-\varepsilon).
\end{equation}
In particular, for fixed $\rho\in(0,1)$ and $0<\varepsilon<1$, the
support saving relative to the exact optimum is
$\Theta_{\rho,\varepsilon}(n)$.
\end{lemma}
\begin{proof}
Convexity of the profile region gives the qualitative claims, a
perturbation of an exact profile gives the endpoint estimate, and
the support converse together with the profile construction gives
the finite-size limit.

The closed upward set of profile triples is convex. Its lower
boundary is finite for $\rho>0,D\ge0$, since collision-free profiles
have arbitrarily small round density and finite cost. This proves
monotonicity, convexity, and joint continuity for $D>0$. Closedness
gives lower semicontinuity at $D=0$ as well: for a bounded sequence
of boundary costs, pass to the limit of the corresponding triples.
The upper bound
$\mathcal B_\rho(D)\le\Phi(\rho)$ and continuity of $\Phi$ give the
matching upper limit at $D=0$. Proposition~\ref{front:zero} gives the
zero set, and since its threshold is continuous and positive, it
also gives continuity at $p=0$.

For the quantitative improvement at overlap one, take an exact
collision-free profile on two adjacent batch sizes, with round
density $\rho$ and cost $\Phi(\rho)$.
Because $\rho<1$, on some interval bounded away from zero an action
of size $b\ge2$ has positive weight $w$. On that interval, replace
this action by the uniformly permuted density vector with entries
$b(1-\alpha),b\alpha/(b-1),\ldots,b\alpha/(b-1)$ for small
$\alpha>0$. Its per-coordinate information and
collision coefficients are
\[
 d_\alpha=\log b-h_2(\alpha)-\alpha\log(b-1),\qquad
 c_\alpha=b\alpha-\frac{b^2\alpha^2}{2(b-1)}.
\]
The round density is unchanged. On an interval $[a,z]\subset(0,1)$,
the added collision exponent is
$w c_\alpha\log(z/a)=\kappa\alpha+O(\alpha^2)$ with
$\kappa>0$, and the information saving is
$w(z-a)[h_2(\alpha)+\alpha\log(b-1)]$.
Choose the small solution for which the added collision exponent
equals $-\log(1-\varepsilon)$. Then
$\alpha=\Theta_\rho(\varepsilon)$, which proves
\eqref{front:endpoint-softening}. The perturbed profiles belong to
the closed region: replace a vanishing terminal interval by
singletons and mix with a collision-free profile of strictly smaller
round density. Both corrections vanish in the limit, and the mixing
only decreases the collision rate.

For \eqref{front:finite-limit}, the support converse applies to every
sequence with these round caps and overlap at least $1-\varepsilon$,
which gives the lower limit. For the upper limit when
$0<\varepsilon<1$, first fix
$\rho'<\rho$ and $1-\varepsilon<p'<1$. The profile construction and
mixing with a collision-free profile give static protocols with
limiting overlap at least $p'$, round density at most $\rho'$, and
normalized support cost arbitrarily close to $B_{\rho'}(p')$. They
satisfy the actual cap $R_n$ and the required overlap for all
sufficiently large $n$.
Send $\rho'\uparrow\rho$ and $p'\downarrow1-\varepsilon$, using joint
continuity, to obtain the upper limit in both models.
For $\varepsilon=0$, Theorem~\ref{ex:linear-density} gives the limit.
Finally, \eqref{front:endpoint-softening} and monotonicity imply
$B_\rho(1-\varepsilon)<\Phi(\rho)$ for every fixed positive error;
\eqref{front:finite-limit} and the exact rate prove the stated saving.
\end{proof}

\begin{theorem}[Static entropy cost]\label{front:static-entropy}
For $\rho\in(0,1)$ and $p\in[0,1]$,
\begin{equation}\label{front:static-convex}
 C^{\mathrm{st}}_\rho(p)=
 \inf\left\{\sum_jw_jB_\rho(p_j):
 \sum_jw_jp_j\ge p,\ \sum_jw_j=1\right\}.
\end{equation}
The infimum ranges over finite probability vectors $(w_j)$ and
overlap levels $p_j\in[0,1]$.
Two mixture components suffice. Equivalently, the largest overlap
$P_\rho$ attainable at a given normalized support cost is replaced
by its least concave majorant $\cav P_\rho$.
\end{theorem}
\begin{proof}
Write $G_\rho(p)$ for the right side of
\eqref{front:static-convex}. The upper bound
$C^{\rm st}_\rho(p)\le G_\rho(p)$ comes from mixing, and the
converse from grouping the seed values by their probabilities.

An initial mixture of the uniform-seed constructions in
Theorem~\ref{front:support} has seed entropy
$H(w)+\sum_jw_jH(Z_j)$, and its overlap is at least the weighted
average of the component overlaps. The components obey common
prescribed round caps, and a component with $p_j=0$ outputs a fixed
word with a repeated label.
Since $H(w)=O(1)$, the upper bound follows.

For the converse, first extend the support bound to capped overlap.
For $A\ge1$, put
\[
 p_A(Q)=\sum_{\sigma\in S_n}\min\{A/n!,Q(\sigma)\},
 \qquad
 V_X^A(\sigma)=\frac{\min\{A/n!,Q(\sigma)\}}{p_A(Q)}
\]
when $p_A(Q)>0$, and extend $V_X^A$ to transcripts through the
conditional law $Q(T\mid X)$.
Then
\[
 H_{V^A}(X)\ge\log n!+\log p_A-\log A,\qquad
 D(V_T^A\Vert Q_T)\le\log(1/p_A),\qquad
 J_{V^A}(Q)\le B(Q).
\]
Thus the proof of \eqref{front:finite-dual-bound} applies with
$p_A$ in place of $p$ and the extra term $-\log A/n$.
Subdivision into batches of size at most $L$ preserves the output
and support cost and adds at most $n/L$ rounds.
For fixed $L,\eta,\xi$, let $n\to\infty$, then $L\to\infty$,
and finally take the supremum over the multipliers. We obtain
\begin{equation}\label{front:capped-support}
 \begin{gathered}
 R_n/n\to\rho,\quad \log A_n=o(n),\quad
 p_{A_n}(Q_n)\to q>0\quad\Longrightarrow\\
 \liminf_n B(Q_n)/n\ge B_\rho(q).
 \end{gathered}
\end{equation}
The case $q=1$ is covered by the dual formula at $D=0$.

Now consider static protocols with seeds $Z_n$, round caps
$R_n/n\to\rho$, and lower limiting overlap at least $p$.
If $c=\liminf_nH(Z_n)/n<\infty$, pass to a subsequence attaining
this limit. Fix $\delta>0$ and an integer $K$ with
$K\delta\ge\Phi(\rho)$. Group the seed values of positive
probability according to
\[
 j\delta\le -n^{-1}\log\Pr(Z_n=z)<(j+1)\delta,
 \qquad 0\le j<K,
\]
and place all remaining values in a tail group.
Let $w_{j,n},Q_{j,n}$ be the probabilities of the groups and the
output laws conditioned on them. Each ordinary group contains fewer
than $e^{n(j+1)\delta}$ values, so its conditional support cost is at
most $n(j+1)\delta$. Groups of probability zero are omitted.

The inequality
$\min\{x,\sum_j a_j\}\le\sum_j\min\{x,a_j\}$ gives
\begin{equation}\label{front:grouped-overlap}
 p(Q_n)\le
 \sum_{j=0}^{K-1}w_{j,n}p_{1/w_{j,n}}(Q_{j,n})
 +w_{{\rm tail},n}.
\end{equation}
Assign cost $(j+1)\delta$ to group $j$ and $\Phi(\rho)$ to the tail.
By the definitions of the groups,
\begin{equation}\label{front:grouped-entropy}
 \sum_{j=0}^{K-1}w_{j,n}(j+1)\delta
 +w_{{\rm tail},n}\Phi(\rho)
 \le H(Z_n)/n+\delta.
\end{equation}

Pass to a further subsequence on which all group probabilities
converge and, for each group with limit $w_j>0$, also
$p_{1/w_{j,n}}(Q_{j,n})\to q_j$.
For such a group $\log(1/w_{j,n})=O(1)$, so
\eqref{front:capped-support} implies
$B_\rho(q_j)\le(j+1)\delta$; this also holds at $q_j=0$.
Groups whose probability tends to zero contribute nothing, since
capped overlap is at most one and the assigned rates are fixed.
With overlap one assigned to the tail,
\eqref{front:grouped-overlap} shows that the limiting weights and
overlaps form a feasible mixture for $G_\rho(p)$.
Using \eqref{front:grouped-entropy} and
$B_\rho(1)=\Phi(\rho)$ therefore yields
\[
 G_\rho(p)\le
 \sum_{j:w_j>0}w_jB_\rho(q_j)+w_{\rm tail}\Phi(\rho)
 \le c+\delta.
\]
Let $\delta\downarrow0$. The case $c=\infty$ is immediate,
which completes the converse.

Finally, by Lemma~\ref{front:regularity} the graph of $B_\rho$ is
compact, and so is its convex hull in the plane. The least cost
subject to mean overlap at least $p$ is therefore attained.
Because $B_\rho$ is nondecreasing, lowering component overlaps
can make their mean exactly $p$ without increasing cost.
For $0<p<1$, a supporting line to this nondecreasing lower boundary
has nonnegative slope, and the minimum is attained by a mixture of
its two extreme contact points with the graph.
At $p=0$ one zero-cost component suffices, and at $p=1$ all
components of positive weight have overlap one. Hence two components
suffice; equivalently, at fixed cost the attainable overlap is the
least concave majorant $\cav P_\rho$.
\end{proof}

\begin{corollary}[An approximation gain bounded uniformly in the
round density]
\label{front:bounded-gap}
For $p\in(0,1]$, set
\[
 c_p=\inf_{t>0}\{t\log(1/p)+\Psi(t)\}.
\]
Then, for every $\rho\in(0,1)$,
\[
0\le\Phi(\rho)-B_\rho(p)\le c_p.
\]
\end{corollary}
\begin{proof}
The exact profile gives the first inequality. For any feasible
permutation profile, the coarse density bound gives
$d(u)\ge\log b-\Psi(t)-tc(u)$ after averaging over the mixture of
actions.
Integer convexity and the round constraint imply
$\int\log b\ge\Phi(\rho)$, while
$\int c(u)\,du\le\int c(u)/u\,du\le\log(1/p)$.
Integrating the pointwise bound and minimizing over $t$ proves the
second inequality, and passing to the closure preserves it.
\end{proof}

\begin{proof}[Proof of the bounds in Section~\ref{sec:sharing-costs}]
The overlap converse gives the lower bound, and aborting an exact
protocol gives the upper bound, in
\[
 pB_\rho(p)\le C^\tau_\rho(p)\le p\Phi(\rho).
\]
Together with Corollary~\ref{front:bounded-gap}, these imply
\[
 C^\tau_\rho(p)-pB_\rho(p)\le pc_p,\qquad
 B_\rho(p)-C^{\rm st}_\rho(p)
 \ge(1-p)\Phi(\rho)-c_p.
\]
For fixed $p<1$, the latter bound is positive at sufficiently small
$\rho$, since $\Phi(\rho)\to\infty$ and $c_p$ is independent of $\rho$.

By Proposition~\ref{front:zero} and
Theorem~\ref{front:static-entropy}, the static cost vanishes on
$[0,p_0]$ and lies below the chord from $(p_0,0)$ to
$(1,\Phi(\rho))$. This chord lies strictly below the cost
$p\Phi(\rho)$ of aborting, because for $p_0\le p<1$,
\[
 p-\frac{p-p_0}{1-p_0}
 =\frac{p_0(1-p)}{1-p_0}>0.
\]
For $0<p<p_0$, zero cost is likewise below $p\Phi(\rho)$.
\end{proof}
\section{Adaptive entropy cost}
\label{app:adaptive}

This appendix determines the trade-off between entropy cost and
overlap in the adaptive model with linearly many rounds, as stated
in Theorem~\ref{thm:shannon-frontiers}. Profiles generate the
achievable region, and a finite dynamic program gives the matching
converse.

For a finite profile $\pi$, recall from \eqref{ad:profile} the
survival function $s_\pi$, the limiting collision-free probability
$p_\pi$, and the survival-weighted information $A_\pi$; the last two
serve as the overlap and the adaptive information cost of the
profile. Let $\mathcal A_\rho$ be the closed convex hull of the
pairs $(A_\pi,p_\pi)$ with $R(\pi)\le\rho$, together with all pairs
obtained by increasing the cost or decreasing the overlap. The round constraint applies to each profile that enters a
convex combination. Define
\[
 \mathcal V_\rho(\lambda)
 =\sup_{\pi:R(\pi)\le\rho}\{p_\pi-\lambda A_\pi\},
 \qquad \lambda\ge0,
\]
where the supremum includes limits of profile triples $(R,A,p)$.
Also put
\[
 P^{\mathrm{ad},H}_\rho(c)
 =\sup\{p:(c,p)\in\mathcal A_\rho\}.
\]
Thus $\mathcal V_\rho$ is the value of a linear objective, whereas
$P^{\mathrm{ad},H}_\rho$ gives the largest overlap at a prescribed
normalized entropy cost.

\begin{theorem}[Adaptive entropy frontier]\label{ad:frontier}
The closed adaptive entropy cost--overlap region is $\mathcal A_\rho$.
Writing its least cost at a given overlap as
$C^{\mathrm{ad}}_\rho(p)$, one has
\begin{equation}\label{ad:dual}
 P^{\mathrm{ad},H}_\rho(c)
 =\inf_{\lambda\ge0}
   \{\mathcal V_\rho(\lambda)+\lambda c\}.
\end{equation}
\end{theorem}

The theorem is proved at the end of this appendix. Its converse is
established first for protocols whose collision-free probability
already equals their overlap with the target. The following
reduction brings an arbitrary protocol into this form, using one
additional output round and private randomness only.

\begin{lemma}[Realizing the overlap as collision-free probability]
\label{ad:calibration}
Let the target be the uniform law $P_{n,k}$ on injections, with
$n\ge k\ge2$. One additional output round suffices to replace the
collision-free part of $Q$ by $\min(P_{n,k},Q)$. Once shared draws
that follow the completed output have been removed, the shared
choices and their costs are preserved.
\end{lemma}
\begin{proof}
First remove the shared draws made after all coordinates have been
written. When a batch is about to complete the output, withhold one
of its coordinates and retain the marginal distribution that the
protocol $Q$ assigned to that coordinate. Generate the other
coordinates of the batch, then use the retained marginal in a final
singleton round. Conditional independence within the batch under $Q$
ensures that the resulting candidate output has law $Q$.

Given a collision-free candidate $x$ with $Q(x)>0$, retain the last
coordinate with private probability
\[
 \min\{1,P_{n,k}(x)/Q(x)\}.
\]
On rejection, output a label that has already been written. Such a
label exists because $k\ge2$. For a candidate with a repeated label,
output the candidate last coordinate unchanged. The candidate and the
acceptance decision are both generated inside the final singleton
kernel. No additional shared choice is required, and no earlier
shared choice changes its law. Every rejection produces a collision,
while the probability of each collision-free output $x$ becomes
$\min\{P_{n,k}(x),Q(x)\}$.
\end{proof}

\subsection{A finite adaptive converse}

The converse rests on a finite dynamic program. Its value function
bounds, up to explicit error terms, the linear objective of a finite
protocol with bounded batch size.

Fix the number of labels $n$, the number of output coordinates
$k\le n$, a maximum batch size $L$, and an entropy multiplier
$\lambda>0$. A state $(m,\ell,r)$ records the number of available
labels, the number of unfinished coordinates, and the number of
remaining output rounds. Along a collision-free execution
starting at $(n,k,R)$, one has $m-\ell=n-k$; the permutation case has
$k=n$.

The value function has the boundary values $W(m,0,r)=1$ and
$W(m,\ell,0)=0$ for $\ell>0$.
At a state with $r\ge1$ and $\ell>0$, write
$W_b^+=W(m-b,\ell-b,r-1)$ and define
\begin{equation}\label{ad:recursion}
 W(m,\ell,r)=\max_{1\le b\le\min(L,\ell)}w_b,
 \qquad
 w_b+\frac\lambda n
 K_b\left(\frac{nw_b}{\lambda m}\right)=W_b^+.
\end{equation}
The left side of the scalar equation is continuous and strictly
increasing in $w_b$, equals zero at $w_b=0$, and is at least $W_b^+$
at $w_b=W_b^+$. Hence the equation has a unique solution in
$[0,W_b^+]$. Induction on the number of remaining rounds gives
$0\le W\le1$.

Written as an optimization over density laws, the solution is
\begin{equation}\label{ad:rational-control}
 w_b=\sup_{\mu\in\mathcal M_b}
       \frac{W_b^+-\lambda d_b(\mu)/n}{1+c_b(\mu)/m}.
\end{equation}
Indeed, the defining equation says that the infimum over $\mu$ of
$w_b(1+c_b/m)+\lambda d_b/n$ is $W_b^+$. Every ratio in
\eqref{ad:rational-control} is therefore at most $w_b$, and laws
approaching that infimum make the ratios approach $w_b$.

Two elementary comparisons bound the actions needed in this
optimization. The uniform action has $d_b=0$ and
$c_b=\binom b2$, so it gives a nonnegative ratio and dominates all
actions with a negative numerator. Among actions with nonnegative
numerator, it also dominates any action with $c_b>\binom b2$.
The disjoint-support action, with $d_b=b\log b$ and $c_b=0$,
dominates any remaining action with $d_b>b\log b$. Thus we may use
only actions satisfying
\[
 0\le d_b\le b\log b,\qquad 0\le c_b\le\binom b2.
\]
Bounded finite laws suffice by the approximation established in
Appendix~\ref{app:frontiers}. If an approximating law crosses one of
these bounds, the corresponding uniform or disjoint-support action
can be used in its place without decreasing the ratio.

Since $W\ge w_b$ and the left side of the scalar equation is
increasing, every allowed batch size satisfies
\begin{equation}\label{ad:bellman-inequality}
 W_b^+-W\le\frac\lambda n
 K_b\left(\frac{nW}{\lambda m}\right).
\end{equation}
This dynamic-programming inequality applies to every batch of an
arbitrary protocol, whether or not the action of that batch attains
the finite value function.

\begin{lemma}[A finite adaptive converse]\label{ad:finite}
Suppose that a finite protocol outputs $k\le n$ coordinates in at
most $R$ rounds, uses batches of size at most $L$, and assigns
probability at most $1/(n)_k$ to every injection. If its output is
collision-free with probability $p$, then, with
$\mathsf H_m=\sum_{j=1}^m j^{-1}$ and $\mathsf H_0=0$,
\begin{equation}\label{ad:finite-bound}
 p-\lambda C_H/n\le W(n,k,R)
 +\frac\lambda n\{a+1+(L-1)(\mathsf H_n-\mathsf H_{n-k})\}+e^{-a}
 \qquad(a>0).
\end{equation}
\end{lemma}
\begin{proof}
The proof follows the protocol tree with a potential built from the
value function and from the largest density of the collision-free
completions, bounds its expected increase at each step, and sums
these bounds over the tree.

\emph{Collision-free completions.}
Let $E$ be the event that the completed output is an injection.
At a collision-free history $\mathcal H$ of positive probability,
let $S(\mathcal H)$ be the set of unused labels and
$I(\mathcal H)$ the set of unfinished coordinates, with
cardinalities $m$ and $\ell$. The collision-free completions are the
injections from $I(\mathcal H)$ to $S(\mathcal H)$; their uniform
law, denoted by $U_\mathcal H$, assigns probability $1/(m)_\ell$ to
each completion. Define the subprobability measure
\[
 L_\mathcal H(x)=Q(E,X_{I(\mathcal H)}=x\mid \mathcal H)
\]
and its largest logarithmic density relative to $U_\mathcal H$ by
\[
 A(\mathcal H)=\log\max_x\frac{L_\mathcal H(x)}{U_\mathcal H(x)}.
\]
If $Q(E\mid \mathcal H)=0$, set $A(\mathcal H)=-\infty$.
Summing the pointwise bound
$L_\mathcal H\le e^{A(\mathcal H)}U_\mathcal H$ gives
$Q(E\mid \mathcal H)\le e^{A(\mathcal H)}$. At the root, the assumed
bound on the probability of each injection gives $A\le0$; at every
collision-free leaf, $A=0$.

We next compare these densities across the two kinds of steps.
For a shared draw $z$ of positive conditional probability, the
mixture identity
$L_\mathcal H(x)=\sum_zQ(z\mid \mathcal H)L_{\mathcal H,z}(x)$ gives
the first inequality below.
For a private batch of size $b$, its coordinate set is known from
the history and the preceding shared choice. For a collision-free
batch output $y$, the factorization
$(m)_\ell=(m)_b(m-b)_{\ell-b}$ and conditional probability give the
second inequality:
\begin{equation}\label{ad:density-updates}
 \begin{aligned}
 A(\mathcal H,z)&\le A(\mathcal H)-\log Q(z\mid \mathcal H),\\
 A(\mathcal H,y)&\le A(\mathcal H)-\log Q(y\mid \mathcal H)
                   -\log(m)_b.
 \end{aligned}
\end{equation}
To verify the second inequality, observe that if $x'$ completes $y$,
then
\[
 \frac{L_{\mathcal H,y}(x')}{U_{\mathcal H,y}(x')}
 =\frac{L_\mathcal H(y,x')/U_\mathcal H(y,x')}
       {Q(y\mid \mathcal H)(m)_b},
\]
which proves the private-step bound after maximizing over $x'$.
The argument allows the coordinate set of the batch to depend on the
history, because this set is fixed before the private batch is
generated.

\emph{Stopping and the potential.}
Stop the protocol tree at its first collision or at its first
history with $A(\mathcal H)<-a$. The cylinder events associated with
these first stopping histories are disjoint. At a stopping history of
the second kind, the conditional probability of eventually producing
an injection is at most $e^{-a}$; at a collision it is zero. Stopping
therefore removes collision-free probability at most $e^{-a}$.

At a retained history put $J(\mathcal H)=A(\mathcal H)+a\ge0$ and
define
\[
 \mathcal F(\mathcal H)=
 \min\{1,W(m,\ell,r)+\lambda J(\mathcal H)/n\}.
\]
Set $\mathcal F=0$ at a stopped node. At the root this potential is
at most $W(n,k,R)+\lambda a/n$, including the case that the root is
stopped. Every retained collision-free leaf has potential one.

A shared draw leaves $(m,\ell,r)$ unchanged. By
\eqref{ad:density-updates}, each retained child satisfies
$J(\mathcal H,z)\le J(\mathcal H)-\log Q(z\mid \mathcal H)$.
The function
$x\mapsto\min\{1,x\}$ is increasing and $1$-Lipschitz on the
nonnegative half-line; stopped children have potential zero.
Thus the conditional expected increase of $\mathcal F$ at this
draw, which we call its drift, is at most
\[
 \frac\lambda n H_Q(Z\mid \mathcal H).
\]
Here the entropy is taken under the protocol law $Q$ at the current
history.

\emph{A private batch conditioned on retention.}
Now consider a retained history immediately before a private batch
of size $b$, and identify its available labels with $[m]$.
Let $\mathcal G_\mathcal H$ be the set of outputs that are
injections into those labels and lead to a child with
$A(\mathcal H,y)\ge-a$. The conditional batch law under $Q$ is a
product, say $\bigotimes_i Q_i$.
Write
\[
 q=Q(X_{A_t}\in\mathcal G_\mathcal H\mid \mathcal H).
\]
If $q=0$, all children are stopped and the drift is nonpositive.
For $q>0$, let $\nu$ be the conditional batch law
\[
 \nu(y)=\frac{\mathbf1_{\mathcal G_\mathcal H}(y)}q
              \prod_iQ_i(y_i).
\]
This is a probability law on injections.
Put
\[
 h=-\log q,\qquad
 d=\sum_iD(\nu_i\Vert u_m),\qquad
 \delta=\TC(\nu),\qquad
 \gamma=b\log m-\log(m)_b.
\]
The event defining $\mathcal G_\mathcal H$ is a function of the current
batch output and history. Conditioning the product law on that event
therefore gives the exact identities
\[
 h=D\!\left(\nu\middle\Vert\bigotimes_iQ_i\right)
   =\delta+\sum_iD(\nu_i\Vert Q_i),
 \qquad
 H(\nu)=b\log m-d-\delta.
\]
In particular, $0\le\delta\le h$.
Write $J'$ for the value of $J$ at the child reached by the batch
output. Averaging the private-step density update under $\nu$ yields
\[
 \mathbb E_\nu J'
 \le J+H(\nu)+h-\log(m)_b
 =J-d-\delta+h+\gamma.
\]

\emph{The finite drift inequality.}
If the parent potential equals one, its drift is nonpositive.
Otherwise $\mathcal F(\mathcal H)=W+\lambda J/n$. Bounding the
potential of each retained child by $W_b^++\lambda J'/n$ and assigning
zero to the other children gives
\[
 \begin{aligned}
 \mathbb E_Q[\mathcal F(\mathcal H,Y)\mid \mathcal H]
 -\mathcal F(\mathcal H)
 \le{}&q(W_b^+-W)-(1-q)W-\frac\lambda n qd\\
 &+\frac\lambda n q(h-\delta+\gamma)
       -\frac\lambda n(1-q)J.
 \end{aligned}
\]
The injection collision bound, applied to the marginals of $\nu$,
gives $c_b\le m\delta$. Consequently,
\[
 K_b\left(\frac{nW}{\lambda m}\right)
 \le d+\frac{nW}{\lambda}\delta.
\]
Together with \eqref{ad:bellman-inequality}, this bounds the first
line of the drift estimate by
\[
 W\{q\delta-(1-q)\}\le0.
\]
Here we used $\delta\le h$ and $q(-\log q)\le1-q$.
Since $J,\delta\ge0$, the second line is at most
\[
 \frac\lambda n\{(1-q)+q\gamma\}.
\]
Thus this last expression bounds the private-step drift in all cases.

\emph{Summing over the stopped tree.}
Weight each conditional drift by the probability under $Q$ of reaching
its retained parent. The terms from shared draws sum to at most
$\lambda C_H/n$, since stopping only removes nonnegative conditional
entropy contributions. The weighted terms $1-q$ are probabilities
of first stopping immediately after a private batch. These events
are disjoint, so their sum is at most one.

For a collision-free private batch,
\[
 \begin{aligned}
 \gamma
 &=\sum_{i=0}^{b-1}-\log(1-i/m)\\
 &\le\sum_{i=0}^{b-1}\frac{i}{m-i}
 \le(L-1)\sum_{j=m-b+1}^{m}\frac1j.
 \end{aligned}
\]
Along any retained path, the intervals of remaining-label counts in
this sum are disjoint. Hence their total contribution is at most
$(L-1)(\mathsf H_n-\mathsf H_{n-k})$. The same bound holds after
averaging over retained private edges, which supplies the factors
$q$ above.

The expected terminal potential equals the retained collision-free
probability. Telescoping the drifts over the finite stopped tree,
using the root bound, and adding back the removed probability, which
is at most $e^{-a}$, gives \eqref{ad:finite-bound}.
\end{proof}

\subsection{Continuum limit and realization}

Two lemmas prepare the proof of the frontier theorem, which closes
this subsection. The first compares the rational survival factor of
the finite recursion with the exponential survival of a continuous
profile. All estimates in the comparison are uniform over the actions
selected by the recursion once their batch size is bounded by $L$.
The second lemma shows that $\mathcal V_\rho(\lambda)$ is
right-continuous in the round density $\rho$.

\begin{lemma}[Comparison with the profile optimization]\label{ad:bridge}
Fix $L\ge2$, $\lambda>0$, and $0<\varepsilon<1/2$. If
$\varepsilon n\ge2L$ and $R/n+L/n<1$, then
\begin{equation}\label{ad:bridge-bound}
 W(n,n,R)\le\mathcal V_{R/n+L/n}(\lambda)
 +\lambda(\varepsilon+L/n)\log L
 +O_{L,\lambda}\left(\frac1{\varepsilon^2n}\right).
\end{equation}
\end{lemma}
\begin{proof}
The proof extracts a deterministic sequence of actions from the
recursion, compares it with a finite profile as long as at least
$\varepsilon n$ coordinates remain, and completes this profile on
the short terminal interval.

\emph{A deterministic sequence from the recursion.}
If $W(n,n,R)=0$, the assertion follows from
$\mathcal V_{R/n+L/n}(\lambda)\ge0$. Otherwise, at each visited
state choose a batch size attaining the maximum in
\eqref{ad:recursion} and a bounded finite density law approaching
the corresponding ratio in \eqref{ad:rational-control}.
Take its error smaller than half the positive value at that state,
so that the chosen ratio and its continuation value are positive.
The resulting sequence of states is deterministic. It has at most
$n$ steps, and the approximation errors can be chosen with total
at most an arbitrary $\zeta>0$.

Write $(b_t,d_t,c_t)$ for the batch size of a chosen action and for
its unnormalized information and collision quantities, and put
\[
 q_t=(1+c_t/m_t)^{-1},\qquad
 Q_t=\prod_{i\le t}q_i,\qquad Q_0=1.
\]
The action bounds give $d_t\le b_t\log L$ and
$c_t\le\binom L2$. Repeated substitution of the chosen ratios gives,
for every prefix of $T$ steps,
\[
 W(n,n,R)
 \le Q_T W(m_{T+1},m_{T+1},R-T)
       -\frac\lambda n\sum_{t\le T}Q_t d_t+\zeta,
\]
where $m_{t+1}=m_t-b_t$ is the number of coordinates that remain
after step $t$. The total approximation error is at most
$\zeta$ because all the factors $q_t$ are at most one.
The remaining value $W(m_{T+1},m_{T+1},R-T)$ is at most one.

Stop before the first chosen batch that would leave fewer than
$\varepsilon n$ coordinates. If $m_*$ denotes the number of
coordinates remaining at that point, then
\[
 \varepsilon n\le m_*<\varepsilon n+L.
\]
A state has positive value precisely when its unfinished coordinates
fit into its remaining batches, that is, when $\ell\le Lr$.
This follows from the recursion by induction on the number of
remaining rounds; the uniform action shows that the value is positive
whenever the condition holds. Thus the remaining state can be
completed with batches of size at most $L$.

\emph{Uniform comparison on the retained interval.}
Represent step $t$ by its constant action on
$[(m_t-b_t)/n,m_t/n]$. Since $m_t=\ell_t$ along the sequence, the
integrated collision exponent on this interval is
\[
 x_t=\frac{c_t}{b_t}\log\frac{m_t}{m_t-b_t}.
\]
The rational factor of the recursion has the exponent
$y_t=-\log q_t=\log(1+c_t/m_t)$. Both exponents differ from
$c_t/m_t$ only by terms of second order in $1/m_t$, because $b_t$ and
$c_t$ are bounded in terms of $L$. For $m_t\ge\varepsilon n\ge2L$, the
logarithm estimates
\[
 \begin{aligned}
 0\le x_t-\frac{c_t}{m_t}
 &\le\frac{c_tb_t}{2m_t(m_t-b_t)},\\
 0\le\frac{c_t}{m_t}-y_t
 &\le\frac{c_t^2}{2m_t^2}
 \end{aligned}
\]
show that $|x_t-y_t|\le C_L/m_t^2$ and
$x_t\le c_t/(m_t-b_t)\le C_L/(\varepsilon n)$, where the constant
depends only on $L$. Since there are at most $n$ steps,
$\sum_{t\le T}|x_t-y_t|\le C_L/(\varepsilon^2n)$. The map $x\mapsto e^{-x}$ is $1$-Lipschitz for $x\ge0$.
It follows that the discrete survival $Q_t$ and the continuous
survival at the corresponding step endpoint differ by at most
$C_L/(\varepsilon^2n)$, uniformly over prefixes.

Let $A_{\mathrm{pre}}$ be the continuous information cost on these
intervals. The contribution of step $t$ lies between the continuous
survival probabilities at its two endpoints, each multiplied by
$d_t/n$. Since
$\sum_t d_t/n\le\log L$, the preceding survival estimate and the
within-step bound on $x_t$ imply
\[
 \left|A_{\mathrm{pre}}-\frac1n\sum_{t\le T}Q_t d_t\right|
 \le\frac{C_L\log L}{\varepsilon^2n}
     +\frac{C_L\log L}{\varepsilon n}.
\]
For fixed $L$ and $\varepsilon<1$, the right side is
$O_L((\varepsilon^2n)^{-1})$.
Writing $s_*$ for the continuous survival at $u=m_*/n$, we therefore
have
\[
 W(n,n,R)
 \le s_*-\lambda A_{\mathrm{pre}}
       +O_{L,\lambda}((\varepsilon^2n)^{-1})+\zeta.
\]

\emph{Completing the terminal interval.}
If $r_*$ rounds remain, positivity of the continuation value gives
$m_*\le Lr_*$. Partition the remaining coordinates into at most
$r_*$ batches of size at most $L$, and use the disjoint-support
action in each batch.
This adds zero collision exponent and information cost at most
\[
 \frac{m_*}{n}\log L
 \le(\varepsilon+L/n)\log L.
\]
Replace the last of these batches by uniform singletons. This creates
a terminal singleton interval and increases the round density by at
most $L/n$. All action laws are bounded and finite, so the completed
object is an admissible finite profile with round density at most
$R/n+L/n$. The final overlap of this profile remains $s_*$, and its
adaptive cost is at most
$A_{\mathrm{pre}}+(\varepsilon+L/n)\log L$, because survival is at
most one throughout the terminal interval. The definition of
$\mathcal V$, followed by the limit $\zeta\downarrow0$, gives
\eqref{ad:bridge-bound}.
\end{proof}

\begin{lemma}[Round continuity]\label{ad:round-continuity}
For fixed $\rho\in(0,1)$ and $\lambda>0$,
\[
 \mathcal V_{\rho+\delta}(\lambda)
 \longrightarrow\mathcal V_\rho(\lambda)
 \qquad\text{as }\delta\downarrow0.
\]
\end{lemma}
\begin{proof}
Since $\mathcal V_\rho(\lambda)$ is nondecreasing in $\rho$, it
suffices to bound $\mathcal V_{\rho+\delta}(\lambda)$ from above.
The bound is obtained by mixing a near-maximizing profile with a
fixed profile of smaller round density.

Fix a collision-free finite profile $h$ with $R(h)<\rho$ and
$B(h)<\infty$. Such a profile exists by
Lemma~\ref{front:profile-slack}.
Fix also a zero-information profile with round density below
$\rho$ and positive overlap $v_0$. For example, uniform densities
at a sufficiently large fixed batch size, followed by a sufficiently
short singleton interval, have these properties. Consequently,
$\mathcal V_{\rho+\delta}(\lambda)\ge v_0$ for all small
$\delta\ge0$.

It therefore suffices to consider near-maximizing finite profiles
$\pi$ with
\[
 p_\pi-\lambda A_\pi\ge v_0/2.
\]
These profiles satisfy, uniformly in $\pi$ and small $\delta$,
\[
 D(\pi)\le\log(2/v_0),\qquad \lambda A_\pi\le1.
\]
For $R(\pi)\le\rho+\delta$, mix $\pi$ pointwise with $h$, giving
$h$ the weight
\[
 \theta=\frac{\delta}{\rho+\delta-R(h)}.
\]
The resulting profile has round density at most $\rho$.
For its overlap and adaptive cost, Lemma~\ref{front:profile-slack}
gives $p_\theta\ge p_\pi$ and
$A_\theta\le e^{\theta D(\pi)}A_\pi+\theta B(h)$.
Its loss in the linear objective is therefore at most
\[
 \bigl(e^{\theta D(\pi)}-1\bigr)\lambda A_\pi
       +\lambda\theta B(h)=O_{\rho,\lambda,h,v_0}(\delta).
\]
This estimate is uniform over all near-maximizing profiles.
For a point in the closure, apply it to finite approximants, with
$\delta+o(1)$ in the mixing weight, and then pass to the limit.
We obtain
\[
 \mathcal V_\rho(\lambda)
 \le\mathcal V_{\rho+\delta}(\lambda)
 \le\mathcal V_\rho(\lambda)+O(\delta),
\]
which proves right continuity.
\end{proof}

\begin{proof}[Proof of Theorem~\ref{ad:frontier}]
For the converse, fix $\lambda>0$ and an achievable pair $(c,p)$
at round density $\rho$. Fix $L$ with $\rho+1/L<1$.
By Lemma~\ref{ad:calibration}, one additional round makes the
collision-free probability of every protocol in an achieving
sequence equal to its overlap with the target.
Subdivide each private batch into groups of at most $L$ coordinates,
retaining its kernels and ignoring the intermediate feedback.
The output and the shared cost are unchanged, and the new number of
rounds satisfies $\widehat R_n\le R_n+n/L+1$.

Apply Lemma~\ref{ad:finite} with $a=\log n$ and then
Lemma~\ref{ad:bridge}. For fixed $L,\lambda$ and
$0<\varepsilon<1/2$, this gives
\[
 p(Q_n)-\lambda C_H(Q_n)/n
 \le\mathcal V_{\widehat R_n/n+L/n}(\lambda)
   +\lambda(\varepsilon+L/n)\log L+o(1).
\]
Monotonicity and Lemma~\ref{ad:round-continuity} give, successively
as $n\to\infty$, $\varepsilon\downarrow0$, and $L\to\infty$,
\[
 p-\lambda c\le\mathcal V_\rho(\lambda).
\]
For $\lambda=0$ the corresponding inequality is $p\le1$.
These inequalities suffice. Because $\mathcal A_\rho$ is closed
under increasing the cost, every separating linear functional that
is bounded above on it has a nonpositive cost coefficient.
Since $(0,0)\in\mathcal A_\rho$, a functional whose
overlap coefficient is also nonpositive cannot separate a point
with nonnegative cost and overlap. Every nontrivial separating
functional can therefore be normalized to $p-\lambda c$ with
$\lambda\ge0$. The separation theorem for closed convex sets now
places every achievable pair in $\mathcal A_\rho$.

For achievability, Proposition~\ref{ad:deferred-profile} realizes
a finite profile $\pi$ with adaptive normalized cost at most
$A_{\pi,\mathcal P}$. Continuity of its survival function gives
\[
 0\le A_{\pi,\mathcal P}-A_\pi
 \le\omega_{s_\pi}(|\mathcal P|)B(\pi)\to0,
\]
where $\omega_{s_\pi}$ is the modulus of continuity of $s_\pi$ and
the limit is taken as the mesh $|\mathcal P|$ of the partition tends
to zero.
Choose the partition and the coding accuracy before taking
$n\to\infty$, then refine the partition and send that accuracy to
zero. Mixing with a collision-free profile of smaller round density
at vanishing weight, as in Lemma~\ref{front:profile-slack}, ensures
that the prescribed hard bound on the number of rounds holds after
rounding.

Mixing finitely many of these protocols at the start adds only
$O(1)$ entropy and preserves the common bound on the number of
rounds. The collision-free parts of their output laws, once
normalized, approach $P_n$, so the overlap of the mixture approaches
the weighted average of the collision-free probabilities.
A diagonal sequence realizes the closed convex hull, including all
larger cost budgets and smaller overlap requirements.
This proves the equality of the two regions, and \eqref{ad:dual} is
its expression through supporting lines.
\end{proof}

\section{Strict advantage of adaptive sharing}
\label{app:strictness}

This appendix proves the strict separation between the two sharing
models stated in Theorem~\ref{thm:shannon-frontiers}: with linearly
many rounds, the adaptive model has strictly smaller entropy cost
than the static model at every overlap strictly between
$p_0(\rho)$, the largest overlap attainable at zero cost, and one.

\begin{theorem}\label{strict:universal}
For $\rho\in(0,1)$ and $p_0(\rho)<p<1$,
\begin{equation}\label{strict:chain}
 C^{\mathrm{ad}}_\rho(p)
 <C^{\mathrm{st}}_\rho(p)\le B_\rho(p).
\end{equation}
All three costs vanish at $p_0$ and equal $\Phi(\rho)$ at one.
\end{theorem}
\begin{proof}
The first part of the proof fixes an interior overlap and shows
that, for profiles whose information cost approaches the support
cost, the adaptive cost is smaller than the information cost by a
constant independent of the profile. The second
part applies the resulting strict inequality to one component of an
optimal mixture representing the static entropy cost.

\emph{The dual maximum is attained at positive multipliers.}
Fix $p_0(\rho)<p<1$ and put
\[
 D_*=-\log p>0,\qquad B_*=\mathcal B_\rho(D_*)>0.
\]
Choose a collision-free finite profile $h$ with
$r_h=R(h)<\rho$ and $B_h=B(h)<\infty$.
Testing the dual objective in \eqref{front:dual} against $h$ shows
that its value at the multipliers $(\eta,\xi)$ is at most
\[
 B_h-\eta(\rho-r_h)-\xi D_*.
\]
Hence any maximizing sequence with positive objective remains in a
bounded subset of the nonnegative quadrant of multipliers. The dual
objective is an infimum, over finite profiles, of continuous affine
functions of the multipliers, and is therefore upper semicontinuous.
Its positive maximum is consequently attained at finite multipliers
$(\eta,\xi)$. Both multipliers are positive. Indeed, if $\eta=0$,
the singleton action makes the local infimum zero and the dual
objective is $-\xi D_*\le0$. If $\xi=0$, arbitrarily large batch
sizes make that infimum zero, and the objective is $-\eta\rho\le0$.
Either case contradicts the positivity of the maximum.

Fix such maximizing multipliers and define, as functions of the
remaining fraction $u$, the collision multiplier and the local
infimum
\[
 t(u)=\xi/u,\qquad
 \ell(u)=\inf_{b\ge1}\frac{K_b(t(u))+\eta}{b}.
\]
The singleton action gives $0\le\ell(u)\le\eta$.
For a particular action, given by a density law for a batch of size
$b$, write $d=d_b/b$ and $c=c_b/b$ for the batch information $d_b$
and the collision coefficient $c_b$ per output coordinate, and
define its nonnegative dual gap by
\[
 g(u;b,\mu)=d+t(u)c+\eta/b-\ell(u)\ge0.
\]

\emph{Feasible profiles approaching the boundary.}
There are feasible finite profiles $\pi_j$ whose round density,
collision exponent, and information cost satisfy
\[
 R_j\le\rho,\qquad D_j\le D_*,\qquad B_j\longrightarrow B_*.
\]
To obtain them directly from the closed profile region, start with
finite approximants, whose quantities are marked by primes and whose
constraint excesses tend to zero. Mix each approximant pointwise
with $h$. The mixing weight can be chosen as the larger of
\[
 \frac{(R_j'-\rho)_+}{R_j'-r_h}
 \quad\text{when }R_j'>\rho,
 \qquad
 \frac{(D_j'-D_*)_+}{D_j'}
 \quad\text{when }D_j'>D_*,
\]
with zero used when the corresponding excess is absent.
Both weights tend to zero because $\rho-r_h>0$ and $D_*>0$.
By Lemma~\ref{front:profile-slack}, the mixture is feasible and its
information cost still tends to $B_*$.

At each $u$, average action quantities using the finite mixture
specified by $\pi_j$. Since the chosen multipliers attain the dual
maximum $B_*$, the integrated dual gap satisfies
\[
 \begin{aligned}
 0\le G_j:=\int_0^1\mathbb E_{\pi_j(u)}g\,du
 &=B_j-B_*+\eta(R_j-\rho)+\xi(D_j-D_*)\\
 &\le B_j-B_*\longrightarrow0.
 \end{aligned}
\]
For every action, $\ell\le\eta$ also gives
\[
 d\le\eta+g.
\]
This bound holds for every profile, and it controls the information
carried by actions with a large dual gap and by a short terminal
interval.

\emph{A finite set of batch sizes away from the endpoint.}
Choose $a\in(0,1)$ such that $\eta a<B_*/4$.
The uniform batch bound \eqref{front:batch-cutoff} gives
\[
 g\ge\log b-\Psi(\xi)-\eta.
\]
Thus a fixed integer $M\ge2$ can be chosen so that $b>M$ implies
$g\ge1$, uniformly in $u$ and in the density law of the action.
On $u\in[a,1]$, the collision multiplier is bounded by
$T:=\xi/a$.

For each $2\le b\le M$, perturb the density law with disjoint
supports by taking the law of a uniformly random permutation of the
density vector
\[
 \bigl(b(1-\alpha),b\alpha/(b-1),\ldots,
              b\alpha/(b-1)\bigr).
\]
Its per-coordinate information and collision quantities are
\[
 d_\alpha=\log b-h_2(\alpha)-\alpha\log(b-1),
 \qquad
 c_\alpha=b\alpha-\frac{b^2\alpha^2}{2(b-1)}.
\]
Choose $\alpha=\exp(-MT-1)$. Since
$h_2(\alpha)\ge\alpha\log(1/\alpha)$ and $c_\alpha\le M\alpha$,
this single choice gives, simultaneously for $2\le b\le M$ and
$0\le t\le T$,
\[
 \frac{K_b(t)}b\le d_\alpha+tc_\alpha\le\log b-\alpha.
\]

\emph{A small dual gap forces a positive collision rate.}
Put $\delta_0=\alpha/4$, and choose $T'>0$ such that
$\Psi(T')\le\alpha/4$. For an action with
$u\ge a$, $2\le b\le M$, and $g\le\delta_0$, the definition of
$\ell$ and the preceding display give
\[
 d+t(u)c
 =g+\ell(u)-\eta/b
 \le\frac{K_b(t(u))}{b}+\delta_0
 \le\log b-3\alpha/4.
\]
On the other hand, the uniform entropy inequality gives
\[
 d\ge\log b-\Psi(T')-T'c.
\]
If $c\le\kappa:=\alpha/(4T')$, this last bound would give
$d\ge\log b-\alpha/2$, contradicting the preceding upper bound.
Every such action therefore satisfies
\[
 c>\kappa,
 \qquad d\le\eta+\delta_0.
\]
All constants $a,M,\alpha,T',\kappa,\delta_0$ depend only on the
fixed overlap, the round density, and the chosen dual multipliers;
they are independent of $j$.

Retain exactly these actions from each profile and let
\[
 \iota_j(u)=\mathbb E_{\pi_j(u)}
    \bigl[d\,\mathbf1_{\{u\ge a,\ 2\le b\le M,\ g\le\delta_0\}}\bigr]
\]
be their information-cost density. The discarded actions carry little
information. The omitted terminal interval carries at most
$\eta a+G_j$. Actions with $b>M$ carry at most $(\eta+1)G_j$, by
$g\ge1$ and $d\le\eta+g$. Among the remaining actions, those with
$g>\delta_0$ carry at most $(\eta/\delta_0+1)G_j$.
For singletons, $g=d+\eta-\ell\ge d$, so their information is at
most $G_j$.
Consequently,
\[
 \mathcal I_j:=\int_0^1\iota_j(u)\,du
 \ge B_j-\eta a-(\eta+\eta/\delta_0+4)G_j
 \ge B_*/2
\]
for all sufficiently large $j$.

\emph{A uniform reduction in adaptive cost.}
The retained actions have $c>\kappa$ and
$d\le\eta+\delta_0$. Since $u\le1$, their contribution alone bounds
the full collision rate of the profile from below:
\[
 \frac{c_j(u)}u\ge\kappa_*\iota_j(u),
 \qquad \kappa_*:=\frac{\kappa}{\eta+\delta_0}>0.
\]
Thus the survival function of $\pi_j$ satisfies
\[
 s_j(u)
 \le\exp\left[-\kappa_*\int_u^1\iota_j(v)\,dv\right].
\]
Using the identity $B_j-A_j=\int_0^1(1-s_j)d_j$ for the difference
between the information cost and the adaptive cost of the profile,
and then changing variable to $\int_u^1\iota_j$, we obtain
\begin{equation}\label{strict:discount}
 \begin{aligned}
 B_j-A_j
 &\ge\int_0^1(1-s_j(u))\iota_j(u)\,du\\
 &\ge\mathcal I_j-
        \frac{1-e^{-\kappa_*\mathcal I_j}}{\kappa_*}\\
 &\ge\frac{B_*}{2}
        -\frac{1-e^{-\kappa_*B_*/2}}{\kappa_*}>0.
 \end{aligned}
\end{equation}
The last positive constant is independent of $j$. Each profile has
$p_j=e^{-D_j}\ge p$ and $R_j\le\rho$, so
Theorem~\ref{ad:frontier} gives $C^{\mathrm{ad}}_\rho(p)\le A_j$.
Letting $j\to\infty$ in \eqref{strict:discount} proves
\[
 C^{\mathrm{ad}}_\rho(p)<B_\rho(p)
 \qquad\text{for every }p_0(\rho)<p<1.
\]
Because the constant in \eqref{strict:discount} is uniform, the
inequality stays strict when the support infimum is approached by a
sequence of profiles.

\emph{Passing from support cost to the static convex envelope.}
Write $p_0=p_0(\rho)$ and $\Phi=\Phi(\rho)$ in the remainder of
the proof. By Lemma~\ref{front:regularity}, $B_\rho$ is continuous
on $[0,1]$, vanishes precisely on $[0,p_0]$, and takes value $\Phi$
at one. Theorem~\ref{front:static-entropy} represents
$C^{\mathrm{st}}_\rho(p)$ as a mixture of at most two points of
this curve, and the infimum over such mixtures is attained. Points
of the curve are written as pairs of cost and overlap.

At $p>p_0$, the static cost is positive, since the support converse
gives $C^{\mathrm{st}}_\rho(p)\ge pB_\rho(p)>0$.
An optimal representation has mean overlap exactly $p$.
Indeed, if its mean were $\bar p>p$, retain that mixture with
weight $(p-p_0)/(\bar p-p_0)$ and assign the remaining weight to
the zero-cost point $(0,p_0)$. This gives mean $p$ and strictly
lowers its positive cost.
The same argument shows that a positive-weight zero-cost component
must have overlap $p_0$: raising any smaller zero-cost overlap to
$p_0$ and then restoring the mean would strictly reduce the cost.

We next exclude the representation whose only components have
overlaps $p_0$ and one.
The endpoint perturbation in
\eqref{front:endpoint-softening} gives
\[
 B_\rho(1-\varepsilon)
 \le\Phi-c_\rho\varepsilon\log(1/\varepsilon).
\]
Choose $\varepsilon>0$ small enough that $q=1-\varepsilon>p$ and
$c_\rho(1-p_0)\log(1/\varepsilon)>\Phi$.
Then
\[
 \frac{B_\rho(q)}{q-p_0}<\frac{\Phi}{1-p_0}.
\]
The mixture of $(0,p_0)$ and $(B_\rho(q),q)$ with mean overlap $p$
therefore has strictly smaller cost than the mixture supported at
$p_0$ and one. Every optimal static representation at $p$ must
accordingly contain a positive-weight component with overlap in
$(p_0,1)$.

For that component, the strict inequality of the first part
supplies an adaptive realization whose cost is below its support
cost. Realize the other component by its support construction, and
mix the two with one initial shared draw that selects the component.
The mixture respects the bound on the number of rounds of each
component and adds only $O(1)$ shared entropy. The resulting
normalized adaptive cost is strictly below the optimal static
mixture cost, proving \eqref{strict:chain}.

Finally, Proposition~\ref{front:zero} gives $0<p_0<1$ and zero cost
at $p_0$. At overlap one, the rate lower bound
$C^{\mathrm{ad}}_\rho(p)\ge pB_\rho(p)$ and the exact profile give
both entropy costs equal to $B_\rho(1)=\Phi(\rho)$.
\end{proof}
\section{Explicit frontiers for pairs and singletons}
\label{app:pair}

When the round density is at least two thirds, the cost--overlap
curves of Theorems~\ref{thm:frontiers}
and~\ref{thm:shannon-frontiers} are attained by profiles that use
only batches of size two (pairs) and of size one (singletons), and
the curves can be described explicitly. The explicit description
also shows that the adaptive entropy cost lies strictly above its
general lower bound.

\subsection{The support and static entropy curves}

The support curve and the static entropy curve coincide, and both
are given by a one-dimensional variational problem.
For $a\in[0,1)$ and $\beta\ge0$, define
\begin{equation}
 J_a(\beta)=\inf_{\int_a^1 I(\theta(x))\,dx\le\beta}
            \frac12\int_a^1\frac{1-\theta(x)^2}{x}\,dx.
 \label{pair:J}
\end{equation}
The infimum ranges over measurable functions
$\theta:[a,1]\to[0,1]$, and the integrals are improper when $a=0$.
The nontrivial budget interval is $0\le\beta\le(1-a)\log2$: one has
$J_a(\beta)=0$ for larger budgets, and $J_0(0)=+\infty$.
Let $\vartheta(t)$ denote the minimizing parameter in $K_2(t)$:
it is zero for $t\le1$ and is the positive solution of
$\operatorname{arctanh}\vartheta(t)=t\vartheta(t)$ for $t>1$.
For an interior budget, the minimizing profile in \eqref{pair:J} is
$\theta(x)=\vartheta(v/x)$, with $v>0$ chosen to satisfy the budget
constraint with equality.

For a budget $\beta\ge0$, the optimal errors under support cost and
under static entropy cost are
\[
 E^{\mathrm{supp}}_\rho(\beta)=1-P_\rho(\beta),\qquad
 E^{\mathrm{st},H}_\rho(\beta)
 =1-\sup\{p:C^{\mathrm{st}}_\rho(p)\le\beta\}.
\]

\begin{theorem}[Frontiers attained by pairs and singletons]
\label{pair:domain}
Let $\rho\in[2/3,1)$ and $a=2\rho-1$. The unrestricted support
and static entropy error frontiers, for every $\beta\ge0$, are both
\begin{equation}
 E^{\mathrm{supp}}_\rho(\beta)
 =E^{\mathrm{st},H}_\rho(\beta)
 =1-e^{-J_a(\beta)}.
 \label{pair:full-curve}
\end{equation}
\end{theorem}

\begin{proof}
The proof exhibits a profile, verifies its optimality by duality,
and passes to static entropy cost by concavity.
Use pairs while the remaining fraction $x$ of coordinates lies in
$[a,1]$ and singletons afterward. This profile has round density
$(1-a)/2+a=\rho$, information cost
$\int_a^1 I(\theta(x))\,dx$, and the collision exponent appearing in
\eqref{pair:J}. In the support dual, choose the collision
multiplier $v$ and the round multiplier $\eta=K_2(v/a)$, so the
local collision multiplier is $t=v/x$.
For $x\ge a$, the inequality $a\ge1/3$ gives
$K_2(t)\le\eta\le K_2(3t)$; for $x<a$, one has
$\eta\le K_2(t)$. Lemma~\ref{pair:cross-batch} therefore proves
optimality over every batch size. The endpoints of zero cost and of
perfect overlap follow by continuity, because the profile region is
closed.

For static entropy cost, put $q(t)=(1-\vartheta(t)^2)/2$ and define
the information budget, the collision exponent, and the overlap by
\[
 B(v)=\int_a^1I\bigl(\vartheta(v/x)\bigr)\,dx,\qquad
 D(v)=\int_v^{v/a}\frac{q(t)}t\,dt,\qquad
 p(v)=e^{-D(v)}.
\]
Positive budgets correspond to $v>a$. Differentiation with respect
to $v$ gives
\[
 B'(v)=-vD'(v)=q(v)-q(v/a)\in(0,1/2],\qquad
 \frac{dp}{dB}=\frac pv.
\]
Since $\frac{d}{dv}(p/v)=p(B'-1)/v^2<0$, the overlap is concave
as a function of the budget $B$.
For $0<v\le a$, the profile is uniform and $B=0$. At perfect overlap
the limiting slope of the curve is zero, so its continuation by a
constant is again concave. Theorem~\ref{front:static-entropy}
then shows that the static entropy curve equals the support curve.
\end{proof}

\subsection{The adaptive curve in one dimension}

For the same round densities, the adaptive entropy curve is
parametrized by the solution of a scalar differential equation.
The parametrization uses the functions
\[
 \kappa(v)=\frac{K_2(v)}2,\qquad
 q(v)=\kappa'(v)=\frac{1-\vartheta(v)^2}{2},\qquad
 g(v)=v+\kappa(v).
\]
For $a\in(0,1)$ and $u>1$, define $z=z_a(u)$ by
\begin{equation}
 \int_z^u\frac{dv}{g(v)}=\log(1/a).
 \label{pair:scalar-ode}
\end{equation}
Since $v\le g(v)\le3v/2$, this equation has a unique solution with
$a^{3/2}u\le z_a(u)\le au$.

\begin{theorem}[Explicit adaptive frontier]
\label{pair:adaptive}
For $\rho\in[2/3,1)$ and $a=2\rho-1$, the
unrestricted adaptive entropy frontier is parametrized by
\begin{equation}
 p=\frac{g(z_a(u))}{a g(u)},\qquad
 c=u\frac{g(z_a(u))}{g(u)}-z_a(u),
 \qquad 1<u<\infty.
 \label{pair:adaptive-parametric}
\end{equation}
The optimal overlap is strictly concave as a function of cost for
$0<c<\Phi(\rho)$, with
$dp/dc=1/(au)$, and its endpoints are
$(c,p)=(0,\sqrt a)$ and $(\Phi(\rho),1)$. For
$c\ge\Phi(\rho)$ the optimal overlap is one.
\end{theorem}

\begin{proof}
We first optimize among profiles that use pairs on $[a,1]$ and
singletons on $[0,a]$, which yields the parametrization; we then
show that these profiles are optimal among all batch sizes.
The variable $x$ denotes the remaining fraction of coordinates,
so an execution traverses the interval $[a,1]$ from $1$ toward $a$.
For a multiplier $\lambda>0$ on the entropy cost, define the
continuation value by
\[
 W(a)=1,\qquad
 W'(x)=-\lambda\kappa\!\left(\frac{W(x)}{\lambda x}\right).
\]
For a pair control $\theta(x)$, write
$q_\theta(x)=(1-\theta(x)^2)/2$. Its survival function satisfies
$s'=s q_\theta/x$ when differentiated with respect to the remaining
fraction. The inequality
$\kappa(v)\le I(\theta)+v q_\theta$ gives
$(sW)'\ge-\lambda s I(\theta)$. Integrating from $a$ to $1$ yields
$p-\lambda c\le W(1)$, and the minimizing control in $\kappa$
attains equality.
With $v(x)=W(x)/(\lambda x)$, the equation becomes
$xv'=-g(v)$. Its boundary values are
$u=v(a)=1/(\lambda a)$ and $z=v(1)$, and integration gives
\eqref{pair:scalar-ode}.

The envelope identity $g'=1+q$ now gives
\[
 p=\frac{g(z)}{ag(u)},\qquad c=aup-z.
\]
Differentiating these identities with respect to the parameter $u$,
we obtain
\[
 z'=ap,\qquad
 \frac{p'}p=\frac{q(z)-q(u)}{g(u)}>0,\qquad
 c'=aup'.
\]
Thus $dp/dc=1/(au)$, and the overlap is strictly concave in cost.
As $u\downarrow1$, the identity $\kappa(v)=v/2$ on $[0,1]$ gives
$z=a^{3/2}$ and $(c,p)=(0,\sqrt a)$.
As $u\to\infty$, the bounds on $z$ imply $z\to\infty$;
using $\kappa(v)\to\log2$ in \eqref{pair:scalar-ode} gives $z/u\to a$.
Consequently,
\[
 c=\frac{u\kappa(z)-z\kappa(u)}{g(u)}
   \longrightarrow(1-a)\log2=\Phi(\rho),\qquad p\longrightarrow1.
\]

We establish optimality over all batch sizes in two steps. First,
singletons can be placed after all larger batches. Fix the sequence
of nonsingleton sizes $b(t)$, where $t$ counts the fraction of
coordinates emitted in those batches alone. If singletons are
interspersed, the remaining fraction satisfies $x(t)\le1-t$. The
optimized continuation value along this schedule solves
\[
 U'(t)=\frac{\lambda}{b(t)}
 K_{b(t)}\!\left(\frac{U(t)}{\lambda x(t)}\right),
 \qquad U(T)=1.
\]
Monotonicity of $K_b$ and comparison backward from $T$ show that
replacing $x(t)$ by $1-t$ increases the value and preserves the
number of rounds used. For a finite profile, $x$ stays positive and
the right-hand side is bounded and locally Lipschitz in the positive
continuation value. Deterministic interleavings therefore
approximate mixtures of actions, and taking the profile closure
extends the comparison.

Second, the value of the pair profiles satisfies the
dynamic-programming inequality for every nonsingleton size. At a
state with remaining coordinate mass $x$ and round budget $r$,
assume $2x/3\le r\le x$ and put
\[
 A=2r-x,\qquad a=A/x,\qquad u=1/(\lambda A),\qquad
 W(x,r)=\lambda xz,
\]
where $\int_z^u dv/g(v)=\log(x/A)$. The same equation gives
$z=ua^{3/2}$ when $u\le1$, and the boundary condition is
$W(x,x)=1$. Here and below,
$p=g(z)/(ag(u))$. Differentiation gives
\[
 \eta:=\frac{W_r}{\lambda}=pK_2(u),\qquad
 W_x=-\frac{W_r+\lambda K_2(z)}2.
\]
To bound the multiplier $\eta$ from both sides, define
\[
 k_z(v)=\exp\!\left(\int_z^v\frac{dt}{g(t)}\right),\qquad
 p_z(v)=\frac{k_z(v)g(z)}{g(v)}.
\]
The identity
\[
 \log\frac{p\kappa(u)}{\kappa(z)}
 =\int_z^u q(v)\left(\frac1{\kappa(v)}-\frac1{g(v)}\right)dv\ge0
\]
gives $K_2(z)\le\eta$.
For the upper bound, consider the elasticity
$\chi(v)=v\kappa'(v)/\kappa(v)$. It equals one on $v\le1$ and is
nonincreasing on $v>1$. Indeed, with
$w=\operatorname{arctanh}\vartheta(v)$, one has
$v\kappa'(v)=w/\sinh(2w)$, which decreases, while $\kappa(v)$
increases. Since $w_z(v):=zk_z(v)\le v$,
\[
 \frac{d}{dv}\log\frac{\kappa(w_z(v))}{p_z(v)\kappa(v)}
 =\frac{\chi(w_z(v))-\chi(v)}{g(v)}\ge0.
\]
This ratio starts at one. At $v=u$, the identities
$k_z(u)=1/a$ and $p_z(u)=p$ give $\eta\le K_2(z/a)$.
Thus $a\ge1/3$ implies
$K_2(z)\le\eta\le K_2(3z)$, and
\eqref{pair:binary-reduction} yields the dynamic-programming
equation
\[
 W_x+\inf_{b\ge1}
 \frac{W_r+\lambda K_b(W/(\lambda x))}{b}=0.
\]

Parametrize an execution by its total emitted coordinate mass.
While it uses nonsingleton batches, $x'=-1$ and $r'=-1/b$, so
$A'=1-2/b\ge0$. Hence $A/x\ge1/3$ persists until the boundary
$r=x$ is reached, where uniform singletons have continuation value
one. For an action $(b,d_b,c_b)$ before that boundary, the survival
function satisfies
$s'=-s c_b/(bx)$. The dynamic-programming inequality gives
$(sW)'\le\lambda s d_b/b$.
Integration therefore proves $p-\lambda c\le W(1,\rho)$, and the
pair profile attains equality.
At the initial state, multipliers $\lambda\ge1/a$ give the zero-cost
uniform branch, while
$\lambda=0$ gives perfect overlap. Thus the pair profiles attain
every supporting line of the closed convex region.
Theorem~\ref{ad:frontier} identifies this region with the one
achievable by unrestricted adaptive protocols.
\end{proof}

When $z_a(u)<1$, the optimal profile initially uses uniform pair
marginals. Its information-cost density is zero for $x\in[x_0,1]$ and
positive for $x\in[a,x_0)$, where
\[
 x_0=a\exp\!\left(\int_1^u\frac{dv}{g(v)}\right).
\]
This follows from $\vartheta(v)=0$ for $v\le1$ and the monotone
increase of $v(x)$ as the remaining fraction decreases.

The explicit profile also shows that the general lower bound on the
adaptive entropy cost is strict.

\begin{corollary}[Strictness of the overlap lower bound]
\label{pair:strict-lower}
For $\rho\in[2/3,1)$ and $p_0(\rho)<p<1$,
\[
 pB_\rho(p)<C^{\rm ad}_\rho(p)
 <C^{\rm st}_\rho(p)=B_\rho(p).
\]
For $0<p\le p_0(\rho)$ all these quantities vanish; at $p=1$
they all equal $\Phi(\rho)$.
\end{corollary}
\begin{proof}
Put $a=2\rho-1>0$.
Theorem~\ref{pair:adaptive} provides a finite parameter $U>1$
for every $p\in(\sqrt a,1)$ and an optimizing pair profile $\pi$ on
$[a,1]$, followed by singletons on $[0,a]$.
Its control $v(x)$ is determined by
\[
 \int_{v(x)}^U\frac{dt}{g(t)}=\log(x/a),
 \qquad v(a)=U,
\]
and its density law is the pair
$(1+\vartheta(v(x)),1-\vartheta(v(x)))$ in uniformly random order.
Write
\[
 i(x)=I(\vartheta(v(x))),\qquad
 q_\pi(x)=\tfrac12(1-\vartheta(v(x))^2),\qquad
 s(x)=\exp\!\left[-\int_x^1\frac{q_\pi(y)}y\,dy\right].
\]
Approximation by finitely many steps shows that these bounded
continuous controls, together with their singleton suffix, belong
to the profile closure.
Their round density is $(1-a)/2+a=\rho$, their final overlap is
$s(a)=p$, and the attaining profile in Theorem~\ref{pair:adaptive}
has adaptive cost
$C^{\rm ad}_\rho(p)=\int_a^1s(x)i(x)\,dx$.
The same profile is feasible for the support variational problem,
so $\int_a^1i(x)\,dx\ge B_\rho(p)$.
Since $v(x)\le U<\infty$, one has $\vartheta(v(x))<1$ and
$q_\pi(x)>0$ throughout $[a,1]$. Hence $s(x)>s(a)=p$ for $x>a$.
Also $v(a)=U>1$ and continuity imply $i(x)>0$ on a nonempty
interval immediately to the right of $a$. Therefore
\[
 C^{\rm ad}_\rho(p)-p\int_a^1i(x)\,dx
 =\int_a^1(s(x)-p)i(x)\,dx>0,
\]
which proves the first inequality. The others follow from
Theorems~\ref{strict:universal} and~\ref{pair:domain};
the remaining cases follow from the endpoint formulas for zero cost
and for exact sampling.
\end{proof}
\section{Randomness representations and costs under the reweighted law}
\label{app:information}

This appendix collects the general facts about shared randomness
behind the comparisons of Section~\ref{sec:model}: protocols may be
taken to be finite trees, static costs are costs of mixtures of
private-coin protocols, and a uniform initial seed approximates any
finite protocol (Proposition~\ref{prop:resource-comparisons}). Its
last subsection measures the shared randomness under the law
reweighted by the common part of the output and target distributions.

Fix a finite product space $\mathcal X=\prod_{i=1}^n\mathcal X_i$.
Protocols that terminate almost surely have the same cost infima as
finite trees. For the adaptive entropy cost, truncate the protocol
after $T$ stages and, at each unfinished history, make one shared
draw from the prescribed conditional law of the remaining tuple. An
unfinished history has an output round remaining, so the output law
and the hard bound on the number of rounds are preserved. Writing
$T_{\rm end}$ for the completion time, the extra expected entropy is
at most $\Pr(T_{\rm end}>T)\log|\mathcal X|\to0$.
For a static protocol, remove the deterministic empty stages; the
initial seed and the output histories are finite.

For a support cost $B<\infty$, every prefix of positive probability
has a terminating extension, and hence its cumulative shared
log-support is at most $B$. A path therefore has at most $B/\log2$
nontrivial shared choices and $n$ output stages. After the
deterministic empty stages are deleted, finite branching and this
uniform bound on the length of a path give a finite tree;
almost-sure termination excludes a deterministic infinite wait.

Let $\mathcal Q_R$ be the set of output laws of protocols with no
shared randomness and at most $R$ output rounds, including schedules
that depend on previous outputs. For a target $P$ on $\mathcal X$,
we write $P$ in place of $n$ in the notation for the finite sampling
costs. Let $\mathcal D_\varepsilon(P,R)$ be the set of all finite
representations
\[
 Q=\sum_{j=1}^M w_jQ_j,\qquad
 Q_j\in\mathcal Q_R,\quad w_j\ge0,\quad \sum_jw_j=1,
 \qquad \TV(P,Q)\le\varepsilon.
\]

\begin{lemma}[One output round]\label{info:one-round-timing}
For $R=1$, the static and adaptive optima coincide for both costs.
\end{lemma}
\begin{proof}
In an adaptive protocol, discard any shared draws made after the
output batch. Before that batch the history contains no private
outputs, so the remaining shared draws can be combined into their
joint transcript $T$ and drawn at the start as an initial seed. The
output law and the bound on the number of rounds are unchanged. By
the chain rule, the entropy cost of these draws equals $H(T)$, while
the number of leaves of the shared tree is at most the maximum
product of its branching numbers along a path. Thus the initial seed
costs no more in either measure. The reverse inequalities are
immediate.
\end{proof}

\begin{proposition}[Static mixture representation]
\label{info:mixture}
For $1\le R\le n$ and $0\le\varepsilon<1$,
\begin{equation}
 B^{\mathrm{st}}_\varepsilon(P,R)
   =\min_{\mathcal D_\varepsilon(P,R)}\log|\supp w|,
 \qquad
 C^{\mathrm{st}}_\varepsilon(P,R)
   =\min_{\mathcal D_\varepsilon(P,R)}H(w).
 \label{info:mixture-costs}
\end{equation}
Each minimum is attained by a representation with at most
$|\mathcal X|$ positive weights. At $R=1$, $\mathcal Q_1$ is the
family of product laws $\bigotimes_{i=1}^n q_i$. In particular, on
$[n]^n$,
\[
 B^{\mathrm{st}}_\varepsilon(P,1)=\log r_\varepsilon(P),\qquad
 C^{\mathrm{st}}_\varepsilon(P,1)=G_\varepsilon(P),
\]
where $r_\varepsilon$ is the nonnegative tensor rank minimized over
probability distributions within total variation $\varepsilon$ of $P$,
and $G_\varepsilon$ is the mixing entropy of Appendix~\ref{rank:section}. The $n$ tensor
factors are the output coordinates, or $n$ parties with one
coordinate each.
\end{proposition}

\begin{proof}
Conditioning a static protocol on its shared seed gives the required
mixture. Conversely, an initial draw of the mixture index realizes
any such representation. The family $\mathcal Q_R$ is compact,
because there are finitely many possible scheduling trees and each
fixed tree gives a continuous image of a product of probability
simplexes.
If a representation has more than $|\mathcal X|$ positive weights,
then its component laws are affinely dependent, and moving the
weights along this dependence until one of them vanishes preserves
the output law. By concavity of entropy, one endpoint of this
interval has entropy no larger than that of the original weights.
Iterating this step reduces both objectives to at most
$|\mathcal X|$ components. Compactness then shows that the entropy
minimum is attained, and the support minimum is attained because
the support cardinality is integer-valued.
With one round, the private laws are exactly the product laws, and
Lemma~\ref{info:one-round-timing} gives the corresponding adaptive
optima.
\end{proof}

For two coordinates, these quantities recover the nonnegative rank
of classical correlation generation \cite{JainEtAl2013} and, at zero
error, the common entropy
$G(X_1;X_2)=\min_{X_1\to W\to X_2}H(W)$
\cite{KumarLiElGamal2014}. Wyner's common information minimizes
$I(W;X_1,X_2)$ under the same conditional-independence constraint
\cite{Wyner1975}. Allowing several rounds extends the mixture
components from product distributions to private-coin protocols.
For general finite joint distributions, computing the common entropy
is NP-hard \cite{BrunoVaccaro2026}; here we study the structured
family of permutations and its extensions with a bounded number of
rounds.

\begin{proposition}[Finite static approximation]
\label{info:static-simulation}
For any finite protocol in the model and any integer $M\ge1$, there
is a protocol with the same bound on the number of output rounds and
one uniform $M$-state initial seed such that
\[
 \TV(Q,Q_M)\le e^B/M.
\]
More generally, if $S=\sum_t-\log Q(Z_t\mid \mathcal H_t)$ is its
realized shared surprisal, then for every $L>0$,
\[
 \TV(Q,Q_M)\le\Pr_Q(S>L)+e^L/M\le C_H/L+e^L/M.
\]
For $0<\varepsilon\le1$, total-variation error at most $\varepsilon$
is attained with support cost at most
$B+\log(1/\varepsilon)+\log2$.
\end{proposition}

\begin{proof}
Generate the private draws from fresh independent uniforms and the
shared choices by arithmetic decoding from one independent
$U\sim\Unif[0,1)$ \cite{HanHoshi1997}: the current interval of this
variable is subdivided according to each conditional shared law and
stays unchanged during private draws. Conditional rescaling
preserves the required uniform law.

Fix the private uniforms. The leaves of the remaining shared-choice
tree are half-open intervals of lengths $e^{-S}$. Coupling $U$ with
$\lfloor MU\rfloor/M$ changes its leaf only within distance $1/M$
of the left boundary of a leaf. At most $e^L$ leaves have length at
least $e^{-L}$, so averaging over the private uniforms gives the
first surprisal bound. Markov's inequality and $\E S=C_H$ give the
second.
A tree whose branching numbers have product at most $C$ along every
path has at most $C$ leaves: if the root has $j$ children, then each
subtree has the bound $C/j$, and induction on the height applies.
Taking $C=e^B$ proves the first bound.

At every history, the resulting protocol uses fresh private uniforms
and retains its initial seed, so each batch still has a product law
conditional on the public history and the seed. Finally, the choice
$M=\lceil e^B/\varepsilon\rceil$ gives the assertion on the support
cost.
\end{proof}

\subsection{Information under the reweighted law}

The shared randomness of a protocol is measured here under its
reweighted law, by expected surprisal and by entropy. The two
measures differ by at most $\log(1/p)$, and for permutations with
linearly many rounds both have optimal rate $B_\rho(p)$;
Section~\ref{sec:sharing-costs} uses this fact.

For a uniform target on a finite set $\Omega$ of collision-free
words, a protocol $Q$, and its reweighted law $V$, constructed as in
Lemma~\ref{lem:overlap-main}, define
\[
 C_{\rm ov}(Q)=\E_V\sum_t-\log Q(Z_t\mid \mathcal H_t),\qquad
 C_H(V)=\sum_tH_V(Z_t\mid \mathcal H_t).
\]

\begin{proposition}[Cost under the reweighted law]
\label{info:overlap-cost}
If the overlap is $p>0$, then
\begin{equation}
 0\le C_{\rm ov}(Q)-C_H(V)
 =\sum_t\E_VD(V_{Z_t\mid \mathcal H_t}\Vert Q_{Z_t\mid \mathcal H_t})
 \le\log(1/p).
 \label{info:overlap-cost-difference}
\end{equation}
For permutations, fix $\rho\in(0,1)$ and $p\in(0,1]$. Among sequences
of protocols $Q_n$ with $R_n/n\to\rho$ and $\liminf_n p(Q_n)\ge p$,
the optimal normalized cost for either quantity is $B_\rho(p)$.
\end{proposition}

\begin{proof}
The equality is the decomposition of each conditional cross-entropy
into a conditional entropy and a relative entropy. The sum is part
of the chain rule for the relative entropy $D(V_T\Vert Q_T)$, which
is at most $\log(1/p)$ because the density of the reweighted law
with respect to the protocol law is at most $1/p$.

For the converse, take a subsequence along which the overlaps
converge, $q_n\to q\ge p$. The bound \eqref{front:finite-support} on
the shared entropy under the reweighted law and its dual limit in
Theorem~\ref{front:support} give
$\liminf C_H(V_n)/n\ge B_\rho(q)\ge B_\rho(p)$.
The subdivision into smaller batches retains the prescribed marginal
kernels and preserves the histories at all shared draws.
Every subsequence has a further subsequence with such a limit, which
proves the converse; \eqref{info:overlap-cost-difference} transfers
it to $C_{\rm ov}$.
For achievability, a uniform initial codebook has constant surprisal
$\log M$, and its entropy under the reweighted law differs from
this value by at most $\log(1/q_n)=O(1)$.
\end{proof}

\section{Extension to injections}
\label{app:partial-injection-frontiers}

This appendix proves Theorem~\ref{thm:injections}. The arguments
for permutations track only the fraction of unused labels, so they
carry over to sampling without replacement. The first five
subsections treat linearly many rounds, and the last treats exact
sampling and sublinearly many rounds.

Let $\Omega_{N,k}$ be the set of injections from $[k]$ to $[N]$, and
let $P_{N,k}=\Unif(\Omega_{N,k})$. With linearly many rounds, all
costs and round densities are normalized by $N$, and we assume
\[
 k/N\longrightarrow\alpha\in(0,1),\qquad
 R/N\longrightarrow\rho\in(0,\alpha),\qquad a=1-\alpha>0.
\]
The protocol model is otherwise unchanged; in particular, the bound
on the number of rounds holds on every execution. We write
$B_{\alpha,\rho}(p)$ and $C^\tau_{\alpha,\rho}(p)$ for the boundaries
of the closed sets of achievable support and entropy cost rates
when the overlap has lower limit at least $p$.
If rates are instead normalized by $k$ and $R/k\to r$, divide all
cost rates below by $\alpha$ and substitute $\rho=\alpha r$.

\subsection{Profiles}
Profiles, which parametrize all limiting costs below, are defined as
for permutations, except that the fraction of unused labels stays
bounded away from zero. The one-batch function
$K_b$ of \eqref{eq:local-action} is unchanged. The coordinates of a
density law satisfy $\E a_i=1$, its information and its collision
coefficient are
\[
 d_b=\E\sum_i a_i\log a_i,\qquad
 c_b=\E\sum_{i<j}a_i a_j,
\]
and its action per output coordinate is $(r,d,c)=(1,d_b,c_b)/b$.
A finite profile on $[a,1]$ is a piecewise constant mixture of finitely
many bounded density laws with finite support. Its round density,
information, and collision exponent, and its survival function,
limiting collision-free probability, and survival-weighted
information are
\begin{align}
 R(\pi)&=\int_a^1r(u)\,du,& B(\pi)&=\int_a^1d(u)\,du,
 &D(\pi)&=\int_a^1\frac{c(u)}u\,du,\label{inj:profiles}\\
 s_\pi(u)&=\exp\left[-\int_u^1\frac{c(v)}v\,dv\right],&
 p_\pi&=s_\pi(a),& A(\pi)&=\int_a^1s_\pi(u)d(u)\,du.
 \nonumber
\end{align}
Because $a>0$, the definition needs no terminal interval of
singletons; requiring one of vanishing length gives the same closed
region. Let $\mathcal R_\alpha$ be the closure of the set of profile
triples $(R,D,B)$, enlarged by adding the nonnegative orthant, and let
$\mathcal B_{\alpha,\rho}(D)$ be its lower boundary in the $B$
coordinate at budgets $(\rho,D)$. Let $\mathcal A_{\alpha,\rho}$ be
the closed convex hull of the pairs $(A(\pi),p_\pi)$ with
$R(\pi)\le\rho$, enlarged by all pairs of larger cost and smaller
overlap. The round constraint is imposed on each component.

\subsection{Finite support converse and realization}
The limit of the support cost in Theorem~\ref{thm:injections} is
proved in three steps: a lower bound for finite protocols, its
identification with the boundary of the profile region, and a
matching construction.

Consider a protocol with overlap $p=p(Q)>0$, and let $V$ be its
reweighted law. The change of measure for a finite uniform target
gives
\[
 H_V(X)\ge\log(N)_k+\log p,\qquad
 D(V\Vert Q)\le-\log p,\qquad
 J_V(Q)\le\min\{B(Q),C_H(Q)/p\},
\]
where $J_V(Q)=\sum_tH_V(Z_t\mid \mathcal H_t)$. The entropy argument
in the proof of Theorem~\ref{front:support} now telescopes over the
interval of unused labels. On collision-free paths whose batches have
size at most $L$, the numbers of unused labels and the batch sizes
satisfy
\[
 \sum_t\log(m_t)_{b_t}=\log(N)_k,\qquad
 \sum_t\gamma(m_t,b_t)\le(L-1)(\mathsf H_N-\mathsf H_{N-k}).
\]
The collision bound $c_b\le m\,\mathrm{TC}(\nu)$ for injections and a
Riemann sum with right endpoints give
\begin{align}
 \frac{J_V(Q)}N\ge{}&
 \int_{1-k/N}^1\ell_{\eta,\xi}(u)\,du
 -\eta\frac RN-\xi\log(1/p)\nonumber\\
 &+\frac{\log p-(L-1)(\mathsf H_N-\mathsf H_{N-k})-L\eta}{N},
 \label{inj:finite-dual}
\end{align}
where $\ell_{\eta,\xi}(u)=\inf_b(K_b(\xi/u)+\eta)/b$.
A protocol with arbitrary batches is reduced to this case by
splitting each batch into groups of at most $L$ coordinates,
retaining its marginal kernels, and ignoring intermediate outputs
until the batch ends. The output law, the shared histories, and the
shared costs are unchanged, and at most $k/L$ rounds are added.
For fixed $L,\eta,\xi$, first let $N\to\infty$ along a subsequence
with overlap tending to $q>0$; the harmonic term is
$O_{\alpha,L}(1)/N$. Then let
$L\to\infty$, and finally optimize over $\eta,\xi$. This proves the
lower bounds once the supremum is identified with the boundary of the
profile region.

The region $\mathcal R_\alpha$ is closed, convex, and upward closed,
and collision-free constant actions
of sufficiently large fixed batch size have round density below any
prescribed $\rho>0$. For fixed multipliers, the infimum of the
Lagrangian over profiles equals $\int_a^1\ell_{\eta,\xi}$. The lower
inequality holds pointwise. For the upper inequality when $\xi>0$,
the bound
$K_b(\xi/u)/b\ge\log b-\Psi(\xi)$ restricts the minimizing batch
sizes to a finite set, and compactness of $[a,1]$ yields finite
profiles whose Lagrangian approximates the integral. When $\xi=0$,
uniform actions with $b\to\infty$ make the infimum zero. A supporting
hyperplane at $D>0$ has a positive coefficient on $B$, because a
collision-free profile has strictly smaller round and collision
budgets. Normalizing that coefficient proves the formula for
$B_{\alpha,\rho}$ in Theorem~\ref{thm:injections}. At
$D=0$, letting the collision multiplier tend to infinity gives
\[
 \mathcal B_{\alpha,\rho}(0)
 =\sup_{\eta\ge0}\left\{
 \alpha\inf_b(\log b+\eta/b)-\eta\rho\right\}
 =\alpha\Phi(\rho/\alpha),
\]
and equality is attained by mixing disjoint-support actions with the
two neighboring reciprocal batch sizes. Convexity gives continuity
of the boundary for $D>0$, and closedness together with this
collision-free upper bound gives continuity at $D=0$.

For the construction, Lemma~\ref{ad:block-codebook} already treats
ordered injections; we apply it with ambient size $N$ and stop after
$k$ outputs. Since the number of unused labels satisfies
$m\ge N-k\ge aN/2$, its estimates are uniform throughout the profile.
Rounding and block composition as in
Proposition~\ref{ad:deferred-profile} give, for each block of the
profile,
\[
 \frac{h_{N,j}}N\longrightarrow\int_{u_j}^{u_{j-1}}d(u)\,du,\qquad
 -\log q_{N,j}\longrightarrow
 \int_{u_j}^{u_{j-1}}\frac{c(u)}u\,du,
\]
and the normalized collision-free part of each block law approaches
the uniform law on injections. Codebooks drawn uniformly at the start
therefore attain cost $B(\pi)$ and overlap $p_\pi$. The deviation of
$k/N$ from its limit changes only a vanishing terminal interval of
the profile. Clipping this interval or extending it with singletons,
and mixing in a vanishing weight of a collision-free profile of
smaller round density, makes the round constraint strict. A diagonal
argument then realizes every point of the closed region. For fixed
$0<p<1$, the lower bound follows by taking a subsequence with overlaps
tending to $q\ge p$ and using monotonicity of $B_{\alpha,\rho}$. For
the upper bound, choose $p'>p$ and a near-optimal finite profile at
$p'$; the strict round constraint and the terminal adjustment meet
the prescribed integer bounds on the number of rounds, with overlap
at least $p$ for all large $N$. Letting $p'\downarrow p$ and using
continuity shows that the optimal support costs of finite protocols
converge to the stated rate in either model.

\subsection{Static entropy converse}
The static part of Theorem~\ref{thm:injections} follows the proof for
permutations: the seeds are grouped by surprisal, and each group is
controlled by a support bound in which the target probabilities are
capped. For $A\ge1$, define the capped overlap
\[
 p_A(Q)=\sum_{x\in\Omega_{N,k}}
 \min\{A/(N)_k,Q(x)\}.
\]
The corresponding reweighted law has entropy at least
$\log(N)_k+\log p_A-\log A$ and divergence at most
$\log(1/p_A)$. Thus \eqref{inj:finite-dual} holds with the additional
term $-\log A/N$. If $\log A_N=o(N)$ and $p_{A_N}(Q_N)\to q>0$,
it yields
$\liminf B(Q_N)/N\ge B_{\alpha,\rho}(q)$.

The argument of Theorem~\ref{front:static-entropy}
applies with $n\mapsto N$, $n!\mapsto(N)_k$,
$B_\rho\mapsto B_{\alpha,\rho}$, and
$\Phi(\rho)\mapsto\alpha\Phi(\rho/\alpha)$.
A surprisal bin of width $\delta N$ contains fewer than
$e^{N(j+1)\delta}$ seeds. For a bin whose limiting probability $w_j$
is positive, the cap $A_N=1/w_{j,N}$ satisfies $\log A_N=O(1)$, so
the capped support bound applies.
Equations~\eqref{front:grouped-overlap} and
\eqref{front:grouped-entropy}, with these substitutions and
$K\delta\ge\alpha\Phi(\rho/\alpha)$, give the lower bound by the
convex envelope after $\delta\downarrow0$. For the reverse
inequality, mix finitely many of the codebook constructions above at
the start, with the same hard bound on the number of rounds in every
component. Continuity of the support curve shows, as in that theorem,
that two components suffice.

\subsection{Adaptive entropy converse}
In the adaptive model, the lower bound runs the dynamic program for
permutations with one more state coordinate and compares its value
with profiles; the construction draws each codebook only while the
output is still collision-free.

Lemma~\ref{ad:calibration} converts overlap into collision-free
probability, using one extra round and no additional shared cost. The
state $(m,\ell,r)$ of the dynamic program from
Appendix~\ref{app:adaptive} now starts at
$(N,k,R)$, maintains $m-\ell=N-k$, and ends at $(N-k,0,r)$.
Lemma~\ref{ad:finite}, in its general form for injections, yields
\begin{equation}\label{inj:adaptive-finite}
 p-\lambda C_H/N\le W(N,k,R)
 +\frac\lambda N\{h+1+(L-1)(\mathsf H_N-\mathsf H_{N-k})\}
 +e^{-h},\qquad h>0.
\end{equation}
To compare this value with profiles, consider first the case
$W(N,k,R)=0$. The comparison is then
immediate: uniform batches of a sufficiently large fixed size have
positive overlap, zero information cost, and arbitrarily small
positive round density, so the profile value is nonnegative at every
positive round budget.
Suppose instead that $W(N,k,R)>0$. Induction along the recursion shows
that $W(m,\ell,r)>0$ precisely when $\ell\le Lr$: the zero values at
$r=0$ give necessity, and a feasible batch followed by the uniform
action in the rational control formula gives sufficiency. At each
state of positive value, choose a maximizing batch size and a bounded
density law with finite support whose ratio differs from the state
value by less than half that value and by less than
$\zeta_N/(k+1)$, where $\zeta_N\downarrow0$. Its ratio is positive,
so its continuation value is positive as well. The resulting sequence
of states is deterministic; $\ell$ decreases by at least one at each
step, and the sequence cannot reach $r=0$ with $\ell>0$ because such
a state has value zero. It therefore reaches $\ell=0$ in at most $k$
steps and within the bound on the number of rounds. The approximation
errors sum to at most $\zeta_N=o(1)$. Put
$q_t=(1+c_t/m_t)^{-1}$ and $Q_t=\prod_{i\le t}q_i$.
Unrolling the recursion, with terminal value one and multipliers at
most one, gives
\[
 W(N,k,R)\le Q_T-\frac\lambda N\sum_{t\le T}Q_td_t+o(1).
\]
Represent step $t$ by its action on
$[(m_t-b_t)/N,m_t/N]$. The collision exponent of this piece of the
profile is $x_t=(c_t/b_t)\log(m_t/(m_t-b_t))$, whereas the
corresponding exponent in the recursion is
$y_t=\log(1+c_t/m_t)$. Since $m_t-b_t\ge N-k\ge aN/2$,
\[
 |x_t-y_t|\le C_L/m_t^2,\quad
 \sum_t|x_t-y_t|=O_{\alpha,L}(N^{-1}),\quad
 \sum_td_t/N\le(k/N)\log L.
\]
Within a step the survival function varies by
$O_{\alpha,L}(N^{-1})$. Therefore the resulting profile satisfies
\[
 W(N,k,R)\le p_\pi-\lambda A(\pi)+O_{\alpha,L,\lambda}(N^{-1})+o(1).
\]
If the interval of this profile differs from $[a,1]$, then clipping
a terminal interval can only increase $p-\lambda A$, while extending
by singletons preserves this objective and adds $o(1)$ to the round
density. Consequently
\[
 W(N,k,R)\le\mathcal V_{\alpha,R/N+o(1)}(\lambda)+o(1),\qquad
 \mathcal V_{\alpha,\rho}(\lambda)
 =\sup_{R(\pi)\le\rho}(p_\pi-\lambda A(\pi)).
\]
This value is right-continuous in $\rho$, and the proof is
independent of the zero-cost boundary characterized below. Fix an
integer $b_*$ with
$\alpha/b_*<\rho$. The constant uniform action of size $b_*$ is a
finite profile with $A=0$, round density $\alpha/b_*$, and overlap
$q_*=a^{(b_*-1)/2}>0$. Thus
$\mathcal V_{\alpha,\rho+\delta}(\lambda)\ge q_*$ for every
$\delta\ge0$. Choose near-maximizers at budgets $\rho+\delta$
whose objective is at least $q_*/2$. They satisfy
$p\ge q_*/2$, $D\le\log(2/q_*)$, and $A\le1/\lambda$.
Fix a collision-free profile $h$ with $R(h)<\rho$ and finite $B(h)$.
Mix each near-maximizer pointwise with $h$ in proportion
$\theta=\delta/(\rho+\delta-R(h))$. The mixed profile has round
density at most $\rho$, its overlap does not decrease, and
$A_\theta\le e^{\theta D}A+\theta B(h)$.
The uniform bounds on $D$ and $A$ make the increase in cost $o(1)$
as $\delta\downarrow0$. Monotonicity in the round budget and this
construction prove right-continuity.

Now apply the calibration lemma and the subdivision of batches, which
together add at most $1+k/L$ rounds. In \eqref{inj:adaptive-finite}
take
$h=\log N$, let $N\to\infty$ at fixed $L$, then let $L\to\infty$.
It follows for every $\lambda>0$ that
\[
 \limsup_N\{p(Q_N)-\lambda C_H(Q_N)/N\}
 \le\mathcal V_{\alpha,\rho}(\lambda).
\]
The case $\lambda=0$ is the trivial bound of the overlap by one. These
multipliers give all separating directions: upward closure in cost
forces a finite supporting functional to have a nonpositive cost
coefficient, and the point $(0,0)$ excludes a nonpositive overlap
coefficient in a strict separation. The remaining normals normalize
to $p-\lambda c$ with $\lambda\ge0$. Separation of closed convex sets
gives the converse for the region.

For the construction, draw the codebook index of a block only if the
output is still collision-free when the block begins. This event has
probability tending to $s_\pi(u_{j-1})$, so the normalized entropy
cost tends to
$\sum_js_\pi(u_{j-1})\int_{u_j}^{u_{j-1}}d$.
Refining the partition gives $A(\pi)$. Mixing finitely many such
protocols at the start and taking closures attains all of
$\mathcal A_{\alpha,\rho}$.

\subsection{Costs under the reweighted law and strict advantage}
Two further statements for permutations carry over: the costs
measured under the reweighted law have the rate of the support
cost, and adaptive sharing is strictly cheaper than static sharing
above the threshold.

For the first, Proposition~\ref{info:overlap-cost} gives
the finite cross-entropy identity for any finite uniform target. Its
rate argument uses a lower bound on $J_V$ and codebooks drawn
uniformly at the start. Equation~\eqref{inj:finite-dual}
and the preceding construction provide both, so both conditioned
costs have rate $B_{\alpha,\rho}(p)$.

For the second, define
\begin{equation}
 D_0(\alpha,\rho)=
 \inf_{\int_a^1\E(1/b)\,du\le\rho}
 \int_a^1\frac{\E(b-1)}{2u}\,du,
 \qquad p_0(\alpha,\rho)=e^{-D_0(\alpha,\rho)}.
 \label{inj:zero}
\end{equation}
These quantities describe the zero-cost boundary. Indeed, a profile
with zero information has uniform marginals. For a sequence of
profiles with information tending to zero, split large actions into
actions of size at most $L$, which adds at most $\alpha/L$ to the
round density. The inequality
$c_b\ge\binom b2-(b/2)d_b$ shows that replacing the actions by
uniform ones increases the collision exponent by at most $LB/(2a)$.
Let $B\to0$ and then $L\to\infty$, using the continuity in the round
budget of the optimum over uniform actions. That optimum selects $b$
when
\[
 b(b-1)\le2\eta u\le b(b+1)
\]
for a round multiplier $\eta$. Uniform actions of a fixed size show
$D_0<\infty$; also $D_0\ge\alpha-\rho>0$, since
$(b-1)/2\ge1-1/b$ and $u\le1$. Thus $0<p_0<1$.
If $\rho\ge\max\{\alpha/2,\alpha-1/3\}$, put
$t=1-2\alpha+2\rho$. Then $t\ge a$ and $t\ge1/3$.
The minimizing profile uses singletons on $[a,t]$ and pairs on
$[t,1]$: their round use is $(t-a)+(1-t)/2=\rho$, and the multiplier
$2\eta=2/t$ makes these actions pointwise optimal. Hence
$D_0=\frac12\log(1/t)$ and $p_0=\sqrt t$.

\begin{proposition}[Strict adaptive advantage for injections]
For $p_0(\alpha,\rho)<p<1$,
\[
 C^{\rm ad}_{\alpha,\rho}(p)<C^{\rm st}_{\alpha,\rho}(p).
\]
\end{proposition}
\begin{proof}
The proof first shows that the adaptive cost lies strictly below the
support cost at every interior point of positive cost, and then
transfers this gap to the static convex envelope.

Write $D_*=-\log p>0$ and $B_*=\mathcal B_{\alpha,\rho}(D_*)>0$.
A collision-free profile $h$ with $R(h)<\rho$ bounds the dual
objective above by
$B(h)-\eta(\rho-R(h))-\xi D_*$, so a maximizing sequence with
positive values has bounded multipliers. By upper semicontinuity the
supremum is attained at multipliers $\eta,\xi>0$, since the dual
objective is nonpositive when either multiplier is zero.

There are finite feasible profiles with $R_j\le\rho$, $D_j\le D_*$,
$B_j\to B_*$. To obtain them, start with profiles approximating a
boundary point of the closure and mix each with $h$, using the larger
of the two vanishing weights that remove its excess round density and
its excess collision exponent. For an action at $u$ put
\[
 g=d+(\xi/u)c+\eta/b-\ell_{\eta,\xi}(u)\ge0.
\]
The integrated mean of this gap satisfies
$0\le G_j\le B_j-B_*\to0$, and each action has $d\le\eta+g$.
The density bound implies $g\ge\log b-\Psi(\xi)-\eta$; hence there is
a fixed $M$ such that $g\ge1$ whenever $b>M$. Because $u\ge a$, all
local collision multipliers are at most $T=\xi/a$.

Uniformly for $2\le b\le M$, perturbing a disjoint-support density
vector to
$(b(1-\tau),b\tau/(b-1),\ldots,b\tau/(b-1))$ and then permuting its
coordinates uniformly at random gives
\[
 d_\tau=\log b-h_2(\tau)-\tau\log(b-1),\qquad
 c_\tau=b\tau-\frac{b^2\tau^2}{2(b-1)}.
\]
Choose $\tau=e^{-MT-1}$. For $0\le t\le T$ this gives
$K_b(t)/b\le\log b-\tau$. Put $\delta_0=\tau/4$ and choose $T'$ so
that $\Psi(T')\le\tau/4$. An action of size $b$ with
$2\le b\le M$ and $g\le\delta_0$ then satisfies
\[
 d+(\xi/u)c\le\log b-3\tau/4,
 \qquad d\ge\log b-\Psi(T')-T'c,
\]
so $c>\kappa:=\tau/(4T')$ and $d\le\eta+\delta_0$.
Retain exactly these actions and call their mean information-cost density
$\iota_j(u)$. Of the discarded information, actions with large
batches contribute at most $(\eta+1)G_j$, actions with large gap at
most $(\eta/\delta_0+1)G_j$, and singletons at most $G_j$. Therefore
$\mathcal I_j:=\int_a^1\iota_j\ge B_*/2$ eventually.
With $\kappa_*:=\kappa/(\eta+\delta_0)$ we have
$c_j(u)/u\ge\kappa_*\iota_j(u)$, whence
\[
 B_j-A_j\ge\int_a^1(1-s_j)\iota_j
 \ge\mathcal I_j-\frac{1-e^{-\kappa_*\mathcal I_j}}{\kappa_*}
 \ge\frac{B_*}2-\frac{1-e^{-\kappa_*B_*/2}}{\kappa_*}>0.
\]
This uniform positive gap proves $C^{\rm ad}(p)<B(p)$ at every
interior point of positive cost.

It remains to pass to the static convex envelope. Its optimal
two-point mixture at $p>p_0$ has mean overlap exactly $p$, and any
zero-cost component can be taken at $p_0$. A mixture supported only
at $p_0$ and one is excluded as follows. The optimal constant
collision-free profile at $\rho$ assigns positive weight $w$ to some
$b\ge2$. Perturbing this action by $\tau$ throughout $[a,1]$
preserves the round density and gives
\[
 D_\tau=w\left(b\tau-\frac{b^2\tau^2}{2(b-1)}\right)\log(1/a),
\quad
 B_\tau=\alpha\Phi(\rho/\alpha)
 -\alpha w\{h_2(\tau)+\tau\log(b-1)\}.
\]
Thus for small $\varepsilon>0$,
$B(1-\varepsilon)\le\alpha\Phi(\rho/\alpha)
-c\varepsilon\log(1/\varepsilon)$ for some $c>0$.
At the target $p$, the chord from $p_0$ to an interior point
sufficiently close to one therefore lies strictly below the chord
from $p_0$ to one. Hence an optimal static mixture has an interior
component of positive cost and positive weight. Replace that
component by its strictly cheaper adaptive realization and leave the
other component unchanged. The initial choice of the component has
negligible normalized entropy, which proves the claim.
\end{proof}

\subsection{Exact and sublinear costs}
The remaining claims of Theorem~\ref{thm:injections} concern exact
sampling and sublinearly many rounds; they are proved in the two
propositions below.

\begin{proposition}[Maximal private support and exact bounds]
Let $1\le k\le N$ and $1\le R\le k$. Define
\[
 F_{N,k,R}=\max_{1\le b\le k}
       L(N,b)F_{N-b,k-b,R-1},\qquad
 F_{N,0,R}=1,\qquad F_{N,k,0}=0\quad(k>0).
\]
Then $F_{N,k,R}$ is the largest support size of a private-coin protocol
whose output is always collision-free, and the maximal output entropy
of such a protocol is $\log F_{N,k,R}$. Define the deficit
$\Delta_{N,k,R}=\log (N)_k-\log F_{N,k,R}$. Each of the four
exact costs is at least $\Delta_{N,k,R}$, and each support cost
is at least $\log\lceil\exp(\Delta_{N,k,R})\rceil$. Moreover,
\[
 B^{\rm st}_0(N,k,R)
 <\log\binom Nk+\Delta(k,R)+38.
\]
Here $\Delta(k,R)$ is the deficit for permutations defined in
Theorem~\ref{thm:exact}. All four costs are exactly zero for $R\ge k$.
\end{proposition}

\begin{proof}
At a collision-free history, the coordinates of a product batch that
is almost surely collision-free have pairwise disjoint supports among
the unused labels. A batch of size $b$ therefore has at most
$L(N,b)$ outcomes, each leaving the state $(N-b,k-b,R-1)$.
This proves the recursion for the support, also when the schedule is
adaptive. Independent uniform draws from balanced supports, followed
recursively by uniform maximizing continuations, attain the recursion
and are uniform on their support. The same recursion and the chain
rule for conditional entropy give the maximal private entropy
$\log F_{N,k,R}$.

In an exact protocol, every private batch of positive probability has
such disjoint supports. Along every execution, the logarithms of the
numbers of outcomes of its batches therefore sum to at most
$\log F_{N,k,R}$, and the chain rule for entropy gives
\[
 \log(N)_k\le C_H+\log F_{N,k,R}.
\]
Since $C_H\le B$, this proves the four lower bounds. For a finite
protocol, $\exp B$ is an integer, which gives the bound with the
ceiling.

For the upper bound, the initial seed chooses a uniform $k$-subset
of $[N]$ and an independent seed for the permutation sampler of
Theorem~\ref{thm:exact} on $k$ labels. Relabeling its outputs by the
chosen subset gives
probability $[\binom Nk k!]^{-1}=1/(N)_k$ to every injection and uses
at most $R$ rounds. The support costs add. For $R\ge k$, sequential
uniform sampling from the unused labels uses no shared randomness.
\end{proof}

With linearly many rounds, the deficit itself satisfies
\begin{equation}\label{inj:capacity-rate}
 \Delta_{N,k,R}/N\longrightarrow\alpha\Phi(\rho/\alpha).
\end{equation}
For the upper bound, put $q=\lfloor k/R\rfloor$ and partition $k$
outputs into $R$ batches of sizes $q$ and $q+1$. Balanced supports
realize this schedule and give
$\Delta_{N,k,R}\le k\Phi(R/k)+O_{\alpha,\rho}(1)$.
For the lower bound, take a private-coin protocol of maximal support and
split its batches into groups of at most $L$ coordinates, retaining
its kernels and ignoring intermediate feedback. This preserves its
output entropy and adds at most $k/L$ rounds. Each of the resulting
batches has entropy at most $\log L(m,b)$.
For $b\le L$ and $m\ge N-k$,
$\log(m)_b-\log L(m,b)=b\log b+O_L(1/m)$.
The cumulative error is $O_{\alpha,L}(1)$, and convexity of $\Phi$
gives the lower bound
$k\Phi(\min\{1,R/k+1/L\})-O_{\alpha,L}(1)$ for the deficit.
Let $N\to\infty$ and then $L\to\infty$. In the same regime,
$\Delta(k,R)/N\to\alpha\Phi(\rho/\alpha)$ and
$N^{-1}\log\binom Nk\to h_2(\alpha)$. The upper bound from the subset
construction and the lower bound from the private support therefore
leave a gap of $Nh_2(\alpha)+o(N)$ between them.
At $R=1$ the two bounds differ, up to $O(1)$, by
$\log L(N,k)=\Theta(N)$ for fixed $\alpha\in(0,1)$.

A direct comparison with permutations gives the exact limit when
almost all labels are sampled. For
$m\ge b\ge1$, write $m=qb+v$ with $0\le v<b$. The balanced-product
formula gives
\[
 1\le\frac{L(m+1,b)}{L(m,b)}=\frac{q+1}{q}
 \le\frac{m+1}{m+1-b}.
\]
Iterate over the $N-k$ spare labels and multiply along any batch-size
composition of $k$. Taking the maximum over compositions yields
$F(k,R)\le F_{N,k,R}\le\binom Nk F(k,R)$. Consequently, each of the
four exact costs $\mathsf C_0$ satisfies
\[
 \Delta(k,R)\le\mathsf C_0(N,k,R)
 <\Delta(k,R)+\log\binom Nk+38.
\]
If $k/N\to1$ then $\log\binom Nk=o(N)$; for $R/N\to\rho\in(0,1)$,
the exact rate for permutations then proves
$\mathsf C_0/N\to\Phi(\rho)$.

\begin{proposition}[Injection costs with sublinearly many rounds]
Fix $0\le\varepsilon<1$. Suppose $k/N\to\alpha\in(0,1]$ and
$1\le R=o(k)$. In either model,
\[
 B^\tau_\varepsilon(N,k,R)
       \sim k\log(k/R),\qquad
 C^\tau_\varepsilon(N,k,R)
       \sim(1-\varepsilon)k\log(k/R).
\]
More generally, if $k/N\ge a>0$, then uniformly over $1\le R\le k$,
\[
 0\le B^\tau_0(N,k,R)-B^\tau_\varepsilon(N,k,R)
       =O_{a,\varepsilon}(k).
\]
\end{proposition}

\begin{proof}
The lower bounds apply the entropy bound for a single batch under
the reweighted law, and the upper bounds combine the exact
construction with aborting.

The lemma on the reweighted law remains valid after replacing $n!$
by $(N)_k$. Put $p=p(Q)>0$ and $d=-\log p$.
Lemma~\ref{fr:coarse-modulus} applies to each batch: for every $t>0$,
\[
 H(\nu)\le b\log(m/b)+tm\operatorname{TC}(\nu)+b\Psi(t).
\]
Along an execution, $m\le N$, $\sum b=k$, and there are at most $R$
batches. Consequently the total private entropy under the reweighted
law is at most $k\log(NR/k)+tNd+k\Psi(t)$. Hence
\[
 B(Q)\ge \log(N)_k-k\log(NR/k)-tNd-k\Psi(t)-d,
\]
and $C_H(Q)$ is at least $p$ times the same right-hand side.
The elementary integral bound $\log(N)_k\ge k\log N-k$ gives
\[
 B(Q)\ge k\log(k/R)-k-tNd-k\Psi(t)-d.
\]
For $p\ge1-\varepsilon$ and $k/N\ge a$, fix any positive $t$
depending only on $a,\varepsilon$, for example
$t=a/(1-\log(1-\varepsilon))$. The loss is then
$O_{a,\varepsilon}(k)$.
For the entropy lower bound, use $p\ge1-\varepsilon$ and the positivity
of the leading bracket when $R=o(k)$.

The exact construction from a random subset and the bound
$\Delta(k,R)\le k\log(k/R)+O(k)$ for permutations give
$B^{\rm st}_0\le k\log(k/R)+O_a(k)$, since
$\log\binom Nk=O_a(k)$. Aborting, that is, running this exact
protocol with probability
$1-\varepsilon$ and outputting a fixed word with a repeated label
otherwise, gives
$C^{\rm st}_\varepsilon\le(1-\varepsilon)B^{\rm st}_0+h_2(\varepsilon)$
for $k\ge2$. Comparing the bounds proves both asymptotics and the
uniform bound on the saving in support cost. The regime in the
proposition has
$k\to\infty$, so the exceptional case $k=1$ is immaterial.
\end{proof}

\end{document}